%% file: MFG-on-Trees-PGame-2.tex
\documentclass[a4,11pt]{article}
\usepackage{graphicx} 
\usepackage{amsmath}
\usepackage{amsthm}
\usepackage{bm}
\usepackage{amsfonts}
\usepackage{latexsym}
\usepackage{amssymb}
\usepackage{setspace}
\usepackage{comment}
\usepackage{ascmac}
\usepackage{cases}
\usepackage{booktabs}
\usepackage{enumitem}
\usepackage{wrapfig}
\usepackage{subcaption} 
\usepackage{float}   
\usepackage{color}
\makeatletter
 \renewcommand{\theequation}{
   \thesection.\arabic{equation}}
  \@addtoreset{equation}{section}
 \makeatother
\begin{document}
\title{Mean-Field Price Formation on Trees \\
 with a Network of Relative Performance Concerns
}

\author{Masaaki Fujii\footnote{mfujii@e.u-tokyo.ac.jp, Graduate School of Economics, The University of Tokyo, Tokyo, Japan }
\footnote{
The author is not responsible in any manner for any losses caused by the use of any contents in this research.
}
}
\date{  \small
First version: December 25, 2025
}
\maketitle


\input{Fmacro-2015.tex}

\def\calf{{\cal F}}
\def\wt{\widetilde}
\def\mbb{\mathbb}
\def\ol{\overline}
\def\ul{\underline}
\def\sign{{\rm{sign}}}
\def\wh{\widehat}
\def\mg{\mathfrak}
\def\display{\displaystyle}

\def\vr{\varrho}
\def\ep{\epsilon}
\def\mr{\mathring}

\def\del{\delta}
\def\Del{\Delta}

\def\deln{\delta_{\mathfrak{n}}}
\def\oldeln{\overline{\delta}_{\mathfrak{n}}}
\def\vep{\varepsilon}
\def\bS{{\bf{S}}}
\def\bs{{\bf{s}}}
\def\bY{{\bf{Y}}}
\def\by{{\bf{y}}}

\def\red{\textcolor{red}}
\def\Ito{{It\^o}~}
\def\blangle{\bigl\langle}
\def\Blangle{\Bigl\langle}
\def\brangle{\bigr\rangle}
\def\Brangle{\Bigr\rangle}
\def\bi{\begin{itemize}}
\def\ei{\end{itemize}}
\def\ac{\acute}
\def\pr{\prime}
\def\mgn{\mathfrak{n}}
\def\mdp{\mathfrak{p}}
\def\mdq{\mathfrak{q}}

\def\part{\partial}
\def\ul{\underline}
\def\ol{\overline}
\def\vp{\varpi}
\def\nn{\nonumber}
\def\be{\begin{equation}}
\def\ee{\end{equation}}
\def\bea{\begin{eqnarray}}
\def\eea{\end{eqnarray}}
\def\bg{\boldsymbol}
\def\bull{$\bullet~$}
\def\ex{\mbb{E}}
\def\opb{\wh{\beta}}
\def\zo{{0,1}}
\def\gmone{{\gamma^1}}

\newtheorem*{remark*}{Remark}

\vspace{-5mm}


\begin{abstract}
Financial firms and institutional investors are routinely evaluated based on their performance relative to their peers. 
These relative performance concerns significantly influence risk-taking behavior and market dynamics. 
While the literature studying Nash equilibrium under such relative performance competitions is extensive,  
its effect on asset price formation remains largely unexplored. This paper investigates mean-field equilibrium price formation 
of a single risky stock in a discrete-time market where agents exhibit exponential utility and relative performance concerns. 
Unlike existing literature that typically treats asset prices as exogenous, we impose a market-clearing condition to determine 
the price dynamics endogenously within a relative performance equilibrium. Using a binomial tree framework, 
we establish the existence and uniqueness of the market-clearing mean-field equilibrium in both single- and multi-population settings. 
Finally, we provide illustrative numerical examples demonstrating the equilibrium price distributions and agents' optimal position sizes.
\end{abstract}

\vspace{0mm}

\vspace{-2mm}
\section{Introduction}
In this paper, we study the problem of equilibrium price formation for agents with exponential utility.
In addition to a partially hedgeable stochastic terminal liability, each agent is assumed to have concerns 
over the average trading performance of peers. We assume the presence of  multiple heterogeneous populations
forming a network of relative performance concerns and study how it influences the equilibrium 
price distribution under the market-clearing condition. To handle this many-agent problem,
we make use of techniques from mean-field game (MFG) theory.
MFG  was pioneered independently by the seminal works
of Lasry \& Lions~\cite{Lions-1, Lions-2, Lions-3} and by those of Huang et al.~\cite{Huang-1, Huang-2, Huang-3}.
MFG is a powerful tool to analyze symmetric strategic interactions
in large populations when each individual player has a negligible
impact on the collective behavior and the state of the others.
This condition is particularly well suited 
for financial and economic applications, 
and there is an extensive body of literature applying MFG techniques to many-agent problems.
See, for example, \cite{Achdou-1,Achdou-2,  Aid-1, Aid-2, Bayraktar, Gabaix} 
and the references therein.

In recent years, there have also been major advances in MFG theory for applications
in the equilibrium price-formation problem, where the asset price process is endogenously constructed to ensure
that demand and supply always balance among heterogeneous but exchangeable agents, 
rather than being exogenously given.
Gomes \& Sa\'ude~\cite{Gomes-Saude} present a deterministic model
of electricity prices.  Its extension with random supply is given by Gomes et al.~\cite{Gomes-random-supply}.
Ashrafyan et al.~\cite{Gomes-Duality} propose a duality approach transforming these problems
into variational ones. Shrivats et al.~\cite{Shrivats} address a price formation problem for
the solar renewable energy certificate (SREC). F\'eron et al.~\cite{Feron-Tankov} develop a tractable 
equilibrium model for intraday electricity markets.
Sarto et al.~\cite{Giulia} study cap-and-trade pollution regulation.
Regarding price formation in securities markets, Fujii \& Takahashi~\cite{Fujii} show
that the equilibrium price process can be characterized by  FBSDEs of  conditional
McKean-Vlasov type. The strong convergence to the mean-field limit from a finite-agent setting
is proved in \cite{Fujii-SC}, and its extension to the presence of a major player is given in \cite{Fujii-Major} by the same authors. 
Fujii~\cite{Fujii-CP} develops a model that allows the co-presence of cooperative and non-cooperative populations.
By using the martingale optimality principle developed by Hu et al.~\cite{Hu-Imkeller}, 
Fujii \& Sekine~\cite{Fujii-Sekine1, Fujii-Sekine2} solve the mean-field price formation
for agents with exponential utilities,  allowing for general self-financing strategies.
In contrast to these continuous-time models, Fujii~\cite{Fujii-Trees} 
adopts a discrete-time framework with a recombining binomial tree for the stock price process.
By leveraging the simple structure of the binomial lattice, the author provides
the explicit solutions for the equilibrium transition probabilities for exponential utility as well as exponential-type recursive utility
in the presence of multiple populations.

Despite these advancements, the aforementioned models assume that agents are concerned exclusively with
their own trading performance. In practice, however, institutional investors and 
fund managers are often evaluated based on their performance relative to a benchmark or the average 
performance of their peers. Such relative performance concerns, often referred to as "keeping up with the Joneses," 
can lead to significant changes in risk-taking behavior and market dynamics.
This motivation has led to an extensive study of optimal investment under relative performance concerns. 
The foundational work of Espinosa \& Touzi~\cite{Espinosa} investigates the Nash equilibrium in a finite-agent setting 
as well as its large population limit. Specifically, they analyze unconstrained agents with general utilities 
and constrained agents with exponential utilities.
On the other hand,  Frei \& dos Reis~\cite{Frei} explore the existence of equilibrium in similar games
and also construct a counterexample to show that an equilibrium may not exist.
Lacker \& Zariphopoulou~\cite{Lacker-Z} introduce a model where agents trade distinct stocks 
correlated through a common noise.
Later, with the development of MFG theory, the idea of distinct stocks introduced by \cite{Lacker-Z}
has been applied to various models and setups. For instance, Fu~\cite{Fu}, Fu \& Zhou~\cite{Fu-Zhou}, 
and more recently Dianetti et al.~\cite{Dianetti} and 
Fu \& Horst~\cite{Fu-Horst} investigate mean-field 
portfolio games with various preference structures, including consumption and Epstein-Zin preferences. 
The use of forward utilities in relative performance MFGs has also been analyzed by dos Reis \& Platonov~\cite{dosReis-1, dosReis-2}.

However, while these works provide deep insights into optimal strategies and Nash equilibria
for the relative performance competitions, they treat the asset price process as exogenously given.
Hence the problem of how such relative concerns endogenously shape the market stock price remains largely unanswered, in particular, 
for incomplete markets. As observed in Espinosa \& Touzi~\cite{Espinosa}, when the risk-premium of the stock 
is independent of the strength of relative concerns $\theta$,
the effective risk tolerance of each agent and hence their optimal position 
increase as $\theta$ increases. They even diverge in the limit $\theta\rightarrow 1$.
This raises a natural question: what would happen if we impose the market-clearing condition?
By definition, market clearing prevents all agents from simultaneously holding infinitely large positions in the same direction.
It is plausible that the risk premium strongly depends on $\theta$ and adjusts the demand for the stock among  the agents
to clear the market.  
Unfortunately, since the existing literature relies heavily on very complex and delicate mathematics 
of quadratic backward stochastic differential equations (BSDEs) and their mean-field extensions, 
imposing the market-clearing condition on top of the relative performance 
Nash equilibrium presents a formidable technical challenge.

In this paper, we fill this gap by extending the discrete-time framework of Fujii~\cite{Fujii-Trees} to incorporate 
relative performance concerns. We combine MFG theory with the classic idea of binomial trees, initiated by Sharpe~\cite{Sharpe}
and formalized in Cox, Ross \& Rubinstein~\cite{Cox}.
We consider a market populated by agents---financial and investment firms---categorized into distinct populations, indexed by $p=1,2, \ldots, m$,
who engage in self-financing trading with a common risky stock and a risk-free money market account.
Agents in each population are subject to distinct stochastic liabilities $F^p$
and distinct distributions of idiosyncratic factors.
More crucially, we introduce a heterogeneous network of relative performance concerns
represented by $(\theta^i_{p,k},1\leq p,k\leq m, i=1,2,\ldots)$, which denote the sensitivity of agent-$i$ in population $p$
relative to the average performance of population $k$. The aggregate structure of this network is captured by an $m\times m$ interaction matrix 
$\Theta:=\mathbb{E}[\theta^i]$.  
Our main contributions are summarized as follows:
\begin{itemize}[noitemsep]
	\item In the single population case ($m=1$), we prove the existence and uniqueness of the market-clearing mean-field 
	equilibrium (MC-MFE) associated with the relative-performance mean-field equilibrium (RP-MFE) 
	for any sensitivity parameter $\Theta\in\mathbb{R}$.
	\item  In the $m$-population case, we prove that there is a unique MC-MFE provided that $(I-\Theta)$ 
	is invertible or $\mathrm{dim}~\mathrm{Ker}(I-\Theta)=1$. Moreover, utilizing resolvent expansion techniques, 
	we demonstrate that the equilibrium solutions remain continuous even around the singular points where 
	$(I-\Theta)$ exhibits a first-order pole.
	\item We derive all these results  explicitly based on tractable backward induction arguments and provide
	several illustrative numerical examples.
\end{itemize}

The rest of the paper is organized as follows: Section~\ref{sec-single-population} studies the single population case $m=1$ 
with relative performance concerns $(\theta^i, i\in \mbb{N})$;
Section~\ref{sec-multi-population} analyzes an $m$-population model with a network of relative
performance concerns $(\theta^i_{p,k}, 1\leq p,k\leq m, i\in \mbb{N})$. The continuity around the first-order pole of $(I-\Theta)$
is also investigated; Section~\ref{sec-numerical} presents several illustrative numerical examples;
and Section~\ref{sec-conclusion} gives concluding remarks.

\section{Single population with relative performance concerns}
\label{sec-single-population}
\subsection{The setup and notation}
\label{sec-setup-single}
We adopt essentially the same setup and notation used in Fujii~\cite{Fujii-Trees}.
$(\Omega^0,\calf^0, (\calf^0_{t_n})_{n=1}^N, \mbb{P}^0)$ is a complete filtered probability space, where 
$0=t_0<t_1<\ldots<t_N=T$ is an equally spaced time sequence using a time step $\Del:=T/N$ where $T<\infty$
and $N\in \mbb{N}$ are given constants. This space is used to model the common shocks to every agent.
Specifically, the filtration $(\calf_{t_n}^0)_{n=0}^N$ is generated by two stochastic processes:  a strictly positive 
process $(S_n:=S(t_n))_{n=0}^N$ and a $d_Y$-dimensional
process $(Y_n:=Y(t_n))_{n=0}^N$, i.e., $\calf_{t_n}^0:=\sigma\{ S_k, Y_k; 0\leq k\leq n\}$.
We use the process $S:=(S_n)_{n=0}^N$ to represent a risky stock price process and $Y:=(Y_n)_{n=0}^N$ 
standalone non-tradable macroeconomic and/or environmental 
factors that affect all the agents. We set $S_0>0$ and $Y_0\in \mbb{R}^{d_Y}$ as given constants and thus $\calf_{0}^0$ is trivial.
For each $n=1,\ldots, N$, we assume that the $\calf_{t_n}^0$-measurable random variable 
\be
\wt{R}_n:=S_n/S_{n-1} \nn
\ee
takes only the two possible values, either $\wt{u}$ or $\wt{d}$. In other words, we restrict 
the trajectories of $(S_n)_{n=0}^N$ onto a recombining binomial tree.
The set of all possible values taken by $S_n$ is thus given by $\cals_n:=\{S_0\wt{u}^k\wt{d}^{n-k}, 0\leq k\leq n\}$,
which is a finite subset of $(0,\infty)$. We denote by $\cals^n:=\{(s_k)_{k=0}^n\in \mbb{R}^{n+1}:
\mbb{P}^0(S_k=s_k, 0\leq k\leq n)>0\}$ the set of all values taken by the stock price trajectory $(S_k)_{k=0}^n$.
To avoid technical subtleties with conditional probabilities, we also assume that the process $Y$
takes values in a finite set at every $t_n$. We use $\caly_n:=\{y\in \mbb{R}^{d_Y}: \mbb{P}^0(Y_n=y)>0\}$
and $\caly^n:=\{(y_k)_{k=0}^n\in \mbb{R}^{d_Y (n+1)}: \mbb{P}^0(Y_k=y_k, 0\leq k\leq n)>0\}$ to denote the set of 
all values taken by $Y_n$ and $(Y_k)_{k=0}^n$, respectively.

In addition to the above space, we introduce a countably infinite number of complete filtered
probability spaces $(\Omega^i,\calf^i,(\calf^i_{t_n})_{n=0}^N, \mbb{P}^i)$, $i\in \mbb{N}$,
which are used to model idiosyncratic shocks to each agent-$i$, $i=1,2,\ldots$.
For each $i$, $(\Omega^i,\calf^i,(\calf_{t_n}^i)_{n=0}^N, \mbb{P}^i)$ is endowed with 
$\calf_0^i$-measurable random variables $(\xi_i,\gamma_i,\theta_i)$ as well as an $(\calf^i_{t_n})_{n=0}^N$-adapted 
stochastic process  $Z^i:=(Z_n^i:=Z^i(t_n))_{n=0}^N$. Here, $\xi_i,\gamma_i$ and $\theta_i$
are all $\mbb{R}$-valued. $\xi_i$ represents the initial wealth, $\gamma_i$
the coefficient of absolute risk aversion, and $\theta_i$ denotes the strength of the relative performance
concern of agent-$i$, respectively. The $d_Z$-dimensional process $Z^i$ is used to model
idiosyncratic shocks to agent-$i$.  We denote the range of random variable $Z_n^i$ by $\calz_n$.
The fact that $(\xi_i,\gamma_i,\theta_i)$ are $\calf_0^i$-measurable means that agent-$i$ knows 
their initial wealth, the size of risk aversion, and the strength of the relative 
performance concern at time zero. Note that there is no need to impose 
the restriction of a finite state space on variables other than $(S,Y)$.

By the standard procedures (see, for example, Klenke~\cite[Chapter 14]{Klenke}), the complete filtered probability space 
$(\Omega,\calf,(\calf_{t_n})_{n=0}^N,\mbb{P})$ is defined as the product of all the above spaces
\be
(\Omega,\calf,(\calf_{t_n})_{n=0}^N,\mbb{P}):=(\Omega^0,\calf^0,(\calf_{t_n}^0)_{n=0}^N,\mbb{P}^0)
\otimes_{i=1}^\infty (\Omega^i,\calf^i,(\calf_{t_n}^i)_{n=0}^N,\mbb{P}^i) \nn
\ee  
which denotes the full probability space containing the entire environment of our model.
Therefore, by construction, $(S, Y)$ and $(\xi_i, \gamma_i, Z^i),~i\in \mbb{N}$ are mutually independent.
On the other hand, the relevant probability space for each agent-$i$ is the product probability space defined by
\be
(\Omega^{0,i},\calf^{0,i}, (\calf_{t_n}^{0,i})_{n=0}^N,\mbb{P}^{0,i}):=(\Omega^0,\calf^0,(\calf_{t_n}^0)_{n=0}^N,\mbb{P}^0)
\otimes (\Omega^i,\calf^i,(\calf_{t_n}^i)_{n=0}^N,\mbb{P}^i), \nn
\ee 
which reflects our assumption that common shocks are public knowledge, but the idiosyncratic shocks are private to 
each agent. We shall use the same symbols,  such as $(S_n, Y_n, \gamma_i, \cdots)$, if they are 
defined as trivial extensions on larger product probability spaces. 
Expectations with respect to $\mbb{P}^0$, $\mbb{P}^i$, $\mbb{P}^{0,i}$ and $\mbb{P}$ are denoted by $\ex^0[\cdot]$, $\ex^i[\cdot]$, 
$\ex^{0,i}[\cdot]$ and $\ex[\cdot]$, respectively. 

We introduce the symbols $\bS^n:=(S_0, S_1,\ldots, S_n)$
to denote a stock-price trajectory and $\bY^n:=(Y_0, Y_1,\ldots, Y_n)$ as a common-noise trajectory,
$\bs^n:=(s_0,s_1,\ldots, s_n)\in \cals^n$ and $\by^n:=(y_0,y_1,\ldots, y_n)\in \caly^n$ serve as their specific realizations.
For $\bs\in \cals^{n-1}$, we  use the symbols $(\bs\wt{u})^n:=(\bs^{n-1}, s_{n-1}\wt{u})\in \cals^n$
and $(\bs\wt{d})^n:=(\bs^{n-1}, s_{n-1}\wt{d})\in \cals^n$. Moreover, for $\by=(y_0,\ldots, y_n) \in \caly^n$, 
$(y_0, \ldots, y_{n-1})$ is denoted by $\by^{-}$.  For $\bs\in \cals^n$, we denote its $k$-th element by $s_k$
and similarly $y_k$ for $\by\in \caly^n$.
We also introduce an $\calf^i_0$-measurable $2$-tuple
$\vr_i:=(\gamma_i, \theta_i)$ for notational simplicity.
To lighten the notational burden, we shall use expressions such as
$\ex^{0,i}[\cdot|\bs,\by,z^i,\vr_i]$ for $(\bs,\by,z^i)\in \cals^{n-1}\times \caly^{n-1}\times \calz_{n-1}$, 
to denote the conditional expectation $\ex^{0,i}[\cdot|\bS^{n-1}=\bs,
\bY^{n-1}=\by, Z_{n-1}^i=z^i,\vr_i=\vr_i]$. With a slight abuse of notation, we shall use the same symbols for the
realizations of the $\calf^i_0$-measurable random variables (e.g., $\vr_i$).

We assume that the risk-free (nominal) interest rate is given by a constant $r\geq 0$.
The time-$t_n$ value of the money-market account is thus given by $\exp(r n \Del)$.
To make the notation simple, we introduce the constants
\be
u:=\wt{u}-\exp(r\Del), \quad d:=\wt{d}-\exp(r\Del), \nn
\ee
and an $\calf^0_{t_n}$-measurable random variable, which takes values either $u$ or $d$, 
\be
R_n:=\wt{R}_n-\exp(r\Del),  \nn
\ee
for each $1\leq n\leq N$.
We also use the symbol $\beta:=\exp(r\Del)$. For notational simplicity, we also introduce the $\calf^i_0$-measurable process $(\gamma_n^i)_{n=0}^N$
defined by $\gamma_n^i:=(\beta^N/\beta^n)\gamma_i$.

For the above introduced variables and processes, we assume the following:
\begin{assumption}
\label{assumption-single-1}
{\rm (i):} $\wt{u}$ and $\wt{d}$ are real constants satisfying $0<\wt{d}<\exp(r\Del)<\wt{u}<\infty$. \\
{\rm (ii):} The variables $(\xi_i, \gamma_i, \theta_i, Z^i)$ are identically distributed across all the agents $i=1,2,\ldots$. \\
{\rm (iii):} There exist real constants $\ul{\xi},\ol{\xi}$, $\ul{\gamma}, \ol{\gamma}$, and $\ul{\theta},\ol{\theta}$ so that
for every $i\in \mbb{N}$, 
\be
\xi_i\in [\ul{\xi},\ol{\xi}]\subset \mbb{R}, \quad \vr_i:=(\gamma_i, \theta_i)\in \Gamma:=[\ul{\gamma},\ol{\gamma}]
\times [\ul{\theta},\ol{\theta}]\subset (0,\infty)\times \mbb{R}.\nn
\ee
{\rm (iv):} For each $i$, the process $Z^i$ is Markovian, i.e., $\ex^i[f(Z_n^i)|\calf^i_{t_k}]=\ex^i[f(Z_n^i)|Z_k^i]$ for every 
bounded measurable function $f$ on $\calz_n$ and $k\leq n$. \\
{\rm (v):} The process $Y$ is Markovian, i.e., $\ex^0[f(Y_n)|\calf^0_{t_k}]=\ex^0[f(Y_n)|Y_k]$ for every bounded measurable
function $f$ on $\caly_n$ and $k\leq n$.\\
{\rm (vi):} The transition probabilities of $S=(S_n)_{n=0}^N$ satisfy, for every $0\leq n\leq N-1$, a.s., 
\be
\begin{split}
&\mbb{P}^0(S_{n+1}=\wt{u}S_n|\calf_{t_n}^0)=\mbb{P}^0(S_{n+1}=\wt{u}S_n|S_n, Y_n)=:p_n(S_n,Y_n), \\
&\mbb{P}^0(S_{n+1}=\wt{d}S_n|\calf_{t_n}^0)=\mbb{P}^0(S_{n+1}=\wt{d}S_n|S_n,Y_n)=:q_n(S_n,Y_n), \nn
\end{split}
\ee
where $p_n, q_n~(:=1-p_n): \cals_n\times \caly_n\rightarrow \mbb{R}, 0\leq n\leq N-1$ are bounded measurable functions
satisfying
\be
0<p_n(s,y),q_n(s,y)<1 \nn
\ee
for every $(s,y)\in \cals_n\times \caly_n$.
\end{assumption}

We now provide remarks on the assumptions. By (i), we have $d<0<u$.
It is well-known that the transition probabilities under the risk-neutral measure $\mbb{Q}$
are given by $p^{\mbb{Q}}:=(\exp(r\Del)-\wt{d})/(\wt{u}-\wt{d})=(-d)/(u-d)$ for the up-move and 
$q^{\mbb{Q}}:=1-p^{\mbb{Q}}$ for the down-move of the stock price. In this paper, we fix the relative transition size $(\wt{u},\wt{d})$
to be constant across all nodes; however, this is done merely for simplicity.
The analysis of the paper still holds even if $(\wt{u},\wt{d})$ varies from node to node. 
It is known that, by adjusting $(\wt{u},\wt{d})$ node by node according to the result by Derman \& Kani~\cite{Derman-Kani},
we can reproduce the implied volatility surface in the option market, while keeping the recombining property of 
the binomial tree intact. Therefore, if necessary, our discussion below can be constructed 
on a binomial tree whose risk-neutral distribution is consistent with the option market.
The Markovian assumptions of (iv) and (v) can be imposed without loss of generality
by lifting $Y$ and $Z^i$ to higher dimensional processes.
The generalization to the path-dependent transition probabilities, such as $p_{n-1}(\bS^n,Y_n)$,
will be discussed in later sections.
Under the above conditions (v) and (vi), $(S_{n+1}, Y_{n+1})$ satisfy the property:
\be
\ex^0[f(S_{n+1})g(Y_{n+1})|\calf_{t_n}^0]=\ex^0[f(S_{n+1})|S_n,Y_n]\ex^0[g(Y_{n+1})|Y_n] ~{\rm a.s.,} \quad 0\leq n\leq N-1,  \nn
\ee
for any bounded measurable functions $f:\cals_{n+1}\rightarrow \mbb{R}$ and $g:\caly_{n+1}\rightarrow \mbb{R}$.
The process $Y$ can be interpreted as representing some standalone macroeconomic and/or environmental factors
which are not influenced by the agents' trading activities. In regime switching models, $Y$ can be used to 
model the state process of the regimes.

\begin{remark}
The bound for the transition probabilities (vi) guarantees that the probability measures
$\mbb{P}^0\circ S^{-1}$ and $\mbb{Q} \circ S^{-1}$ are equivalent. Hence the system is arbitrage free.
\end{remark}

\subsection{Optimization problem}
Consider a financial market which consists of a large number of agents, agent-$i$, $1\leq i\leq N_p$.
We suppose that the preference of each agent-$i$ is characterized 
by the exponential utility at the terminal time $t=t_N$ with the absolute risk-aversion coefficient $\gamma_i$:
\be
-\exp\Bigl(-\gamma_i \Bigl(X_N^i-\frac{\theta_i}{N_p-1} \sum_{j\neq i} X_N^j-F(S_N, Y_N, Z_N^i)\Bigr)\Bigr). \nn
\ee
Here, $X^i:=(X_n^i:=X^i(t_n))_{n=0}^N$ is the wealth process of agent-$i$, and the term $F$ denotes
a stochastic terminal liability which depends on the realizations of the stock price $S_N$, the 
non-tradable macroeconomic factors $Y_N$,  as well as the idiosyncratic factors $Z^i_N$.
A key extension relative to the previous work Fujii~\cite{Fujii-Trees} is the presence of the relative performance concern
represented by the second term. The coefficient $\theta_i$ gives its strength and direction;
$\theta_i>0$ denotes a situation with the relative performance competition, and $\theta_i<0$ 
corresponds to altruistic preferences (or homophilous interaction as termed by Hu \& Zariphopoulou~\cite{Hu-Z}),
where agents derive positive utility from the wealth of others.
In this section, we are interested in the mean-field limit $N_p\rightarrow \infty$ of the above problem.

We now formulate the optimization problem for each agent. Agent-$i$, for $i=1,2,\ldots$, with an initial wealth $\xi_i$,
engages in self-financing trading with the risk-free money market account and a single risky stock.
They adopt an $(\calf^{0,i}_{t_n})_{n=0}^{N-1}$-adapted trading strategy $(\phi_n^i)_{n=0}^{N-1}$, 
denoting the invested amount of cash in the stock at time $t_n$.
The associated wealth process of agent-$i$, $X^i:=(X_n^i:=X^i(t_n))_{n=0}^N$, follows the dynamics
\be
\begin{split}
X_{n+1}^i &=\exp(r\Del)(X_n^i-\phi^i_n)+\phi^i_n\wt{R}_{n+1}\\
& =\beta X_n^i+\phi^i_n R_{n+1}, \nn
\end{split}
\ee
where $X_0^i=\xi_i$. Recall that $\beta:=\exp(r\Del)$ and $R_{n+1}:=\wt{R}_{n+1}-\exp(r\Del)$.

Each agent-$i$ seeks to solve the optimization problem:
\be
\label{problem-single}
\sup_{(\phi^i_n)_{n=0}^{N-1}\in \mbb{A}^i}\ex^{0,i}\Bigl[
-\exp\Bigl(-\gamma_i \bigl(X_N^i-\theta_i \mu_N(\bS^N, \bY^{N-1})-F(S_N, Y_N, Z_N^i)\Bigr)|\calf_0^{0,i}\Bigr], 
\ee
where 
\be
\mbb{A}^i:=\{(\phi^i_n)_{n=0}^{N-1}: \text{$\phi^i_n$ is an $\calf^{0,i}_{t_n}$-measurable real-valued random variable}\} \nn
\ee
denotes the admissible control space. $\mu_N:\cals^N\times \caly^{N-1}\rightarrow \mbb{R}$
is an appropriate measurable function. We shall search for the fixed point $(\mu_n)_{n=0}^N$ with $\mu_0:=\ex^{i}[\xi_i]$ 
and $\mu_n:\cals^n\times \caly^{n-1}\rightarrow \mbb{R}$, $1\leq n\leq N$,  such that
\be
\mu_n(\bS^n,\bY^{n-1})=\ex^{0,i}\bigl[X_n^i|\calf^0_{t_n}\bigr]~{\rm a.s..}\nn
\ee  
The independence of $\mu_n$ from $Y_n$ follows from the dynamics of wealth $X^i$ and the
condition (vi) in  Assumption~\ref{assumption-single-1}.

\begin{remark}[Economic interpretation of the agents]
Throughout this paper, we interpret the agents not as individual investors, 
but as trading desks of institutional market participants, encompassing both financial 
intermediaries (e.g., dealer banks, market makers) and active asset managers (e.g., hedge funds, insurance firms). 
In this context, the terminal liability $F$ represents inventory risk, derivatives obligations, or specific investment 
mandates arising from client orders and insurance contracts, which are not under the direct control of the trading desks. 
In fact, these institutions are often structurally obligated to manage these risks to maintain client relationships or 
fulfill business duties. The fee income or premiums received from clients for these services are also non-tradable 
and inherent to the business; they are thus embedded in $F$ as a negative contribution to the liability. 
This interpretation naturally justifies the presence of liabilities and sophisticated optimization, 
as well as the introduction of relative performance concerns driven by the highly competitive nature of the financial industry.
\end{remark}

\begin{assumption}
\label{assumption-single-2}
{\rm (i):} $F:\cals_N\times \caly_N\times \calz_N\rightarrow \mbb{R}$ is a bounded measurable function. \\
{\rm (ii):} Every agent is a price-taker in the sense that they consider the stock price process 
(and hence its transition probabilities specified in Assumption~\ref{assumption-single-1} (vi)) to be 
exogenously determined by the collective actions of the others and unaffected by the agent's own trading strategies.\\
{\rm (iii):} Every agent treats the mean-field terms $(\mu_n)_{n=0}^N$, $\mu_n:\cals^n\times \caly^{n-1}\rightarrow \mbb{R}$, as 
exogenous bounded measurable functions, believing they are determined by the collective actions of the others and 
unaffected by the agent's own trading strategies.
\footnote{Since the domain $\cals^N\times \caly^{N-1}$ is finite, the boundedness assumption is redundant. 
However, we make it explicit for clarity. We keep this convention throughout the paper.} 
\end{assumption}

As a preliminary step, we first study, for $1\leq n\leq N$, the one-period problem at $t=t_{n-1}$ for the interval $[t_{n-1},t_n]$:
\be
\label{problem-single-tmp}
\sup_{\phi^i_{n-1}}\ex^{0,i}\Bigl[-\exp\Bigl(-\gamma_n^i \bigl(X_n^i-\theta_i \mu_n(\bS^n, \bY^{n-1})\bigr)\Bigr)V_n(S_n, Y_n,Z_n^i,\vr_i)|\calf_{t_{n-1}}^{0,i}
\Bigr]
\ee
where the supremum is taken over the $\calf^{0,i}_{t_{n-1}}$-measurable real-valued random variables. We recall that $\gamma_n^i$ is defined as
$\gamma_n^i:=(\beta^N/\beta^n) \gamma_i$ and $\vr_i$ is the 2-tuple $\vr_i:=(\gamma_i, \theta_i)$.

\begin{lemma}
\label{lemma-single-tmp}
Let Assumption~\ref{assumption-single-1} and Assumption~\ref{assumption-single-2} (ii) and (iii) be in force. 
Assume, moreover, that $V_n:\cals_n\times \caly_n\times \calz_n\times\Gamma\rightarrow \mbb{R}$ is a measurable function
satisfying the uniform bounds $0<c_n\leq V_n\leq C_n<\infty$ on its domain with some positive constants $c_n$ and $C_n$.
Then the problem $(\ref{problem-single-tmp})$ has a unique optimal solution $\phi^{i,*}_{n-1}$
defined by a bounded measurable function $\phi^{i,*}_{n-1}:\cals^{n-1}\times \caly^{n-1}\times \calz_{n-1}\times \Gamma\rightarrow \mbb{R}$,
such that $\phi_{n-1}^{i,*}:=\phi^{i,*}_{n-1}(\bS^{n-1}, \bY^{n-1}, Z_{n-1}^i, \vr_i)$ a.s., where
\be
\label{single-optimal-tmp}
\begin{split}
&\phi^{i,*}_{n-1}(\bs,\by,z^i,\vr_i):=\frac{\theta_i \Del_n(\bs,\by)}{u-d}+\frac{1}{\gamma_n^i (u-d)}
\Bigl\{\log\Bigl(-\frac{p_{n-1}(s,y) u}{q_{n-1}(s,y)d}\Bigr)+\log f_{n-1}(s, y, z^i, \vr_i)\Bigr\}.
\end{split}
\ee
Here,  $s=s_{n-1}\in \cals_{n-1}$ and $y=y_{n-1}\in \caly_{n-1}$ are the last elements of $\bs\in \cals^{n-1}$ and $\by\in \caly^{n-1}$,
respectively. Furthermore, $f_{n-1}:\cals_{n-1}\times \caly_{n-1}\times \calz_{n-1}\times \Gamma\rightarrow\mbb{R}$ 
is a measurable function satisfying the uniform bounds $0<\ol{c}_n\leq f_{n-1}\leq \ol{C}_n<\infty$ on its domain
with some positive constants $\ol{c}_n$ and $\ol{C}_n$, 
and $\Del_n:\cals^{n-1}\times\caly^{n-1}\rightarrow \mbb{R}$ is a bounded measurable function.
They are defined respectively by
\be
\begin{split}
&f_{n-1}(s,y,z^i,\vr_i):=\frac{\ex^{0,i}[V_n( s\wt{u},Y_n,Z_n^i,\vr_i)|y,z^i,\vr_i]}{\ex^{0,i}[V_n(s\wt{d}, Y_n,Z_n^i,\vr_i)|y,z^i,\vr_i]}, \\
&\Del_n(\bs,\by):=\mu_n((\bs\wt{u})^n, \by)-\mu_n((\bs\wt{d})^n,\by). \nn
\end{split}
\ee
\end{lemma}
\begin{proof}
We solve the problem on each set $\{\omega^{0,i}\in \Omega^{0,i}: (X_{n-1}^i, \bS^{n-1}, \bY^{n-1}, Z_{n-1}^i,\vr_i)=(x^i,\bs,\by,z^i,\vr_i)\}$.
Here, with a slight abuse of notation, we use the same symbols for the realizations of $\calf^i_0$-measurable random variables.
Recall that $s:=s_{n-1}$ and $y:=y_{n-1}$, i.e.,  the last elements of $\bs\in \cals^{n-1}$ and $\by\in \caly^{n-1}$, respectively.
Then the problem $(\ref{problem-single-tmp})$ is equivalent to 
\be
\begin{split}
&\inf_{\phi^i\in \mbb{R}}\ex^{0,i}\Bigl[\exp\Bigl(-\gamma_n^i\bigl(\beta x^i+\phi^i R_n-\theta_i \mu_n(\bS^n, \bY^{n-1})\bigr)\Bigr)
V_n(S_n,Y_n,Z_n^i,\vr_i)|\bs,\by,z^i, \vr_i\Bigr]\\
&=\exp(-\gamma_{n-1}^i x^i)\inf_{\phi^i}\Bigl\{ p_{n-1}(s,y)\exp\Bigl(-\gamma_n^i\bigl(\phi^i u-\theta_i \mu_n((\bs\wt{u})^n,\by)\bigr)\Bigr)
\ex^{0,i}[V_n(s\wt{u}, Y_n, Z_n^i,\vr_i)|y,z^i,\vr_i]\\
&\qquad\qquad+q_{n-1}(s,y)\exp\Bigl(-\gamma_n^i\bigl(\phi^i d-\theta_i \mu_n((\bs\wt{d})^n,\by)\bigr)\Bigr)\ex^{0,i}[V_n(s\wt{d},
Y_n,Z_n^i,\vr_i)|y,z^i,\vr_i]\Bigr\},
\label{middle-valueF}
\end{split}
\ee 
where we have used the property (vi) in Assumption~\ref{assumption-single-1}. Since $\gamma_n^i>0 $ and $d<0<u$, it is easy to see
that the optimal trade position $\phi^{i,*}$ is uniquely characterized by
\be
\begin{split}
&p_{n-1}(s,y)u \exp\Bigl(-\gamma_n^i\bigl(\phi^{i,*} u-\theta_i \mu_n((\bs\wt{u})^n,\by)\bigr)\Bigr)
\ex^{0,i}[V_n(s\wt{u}, Y_n, Z_n^i,\vr_i)|y,z^i,\vr_i] \\
&\quad+q_{n-1}(s,y)d\exp\Bigl(-\gamma_n^i\bigl(\phi^{i,*} d-\theta_i \mu_n((\bs\wt{d})^n,\by)\bigr)\Bigr)\ex^{0,i}[V_n(s\wt{d},
Y_n,Z_n^i,\vr_i)|y,z^i,\vr_i]=0, \nn
\end{split}
\ee
which gives the desired result. The existence of positive constants $\ol{c}_n$ and $\ol{C}_n$ for $f_{n-1}$ follows 
immediately from the boundedness of $V_n$.
\end{proof}

\subsection{Relative performance mean-field equilibrium}

\begin{definition}
\label{def-single-rp-mfe}
We sat that the system is  in the relative performance mean-field equilibrium (RP-MFE)
if the problem $(\ref{problem-single})$ has an optimal solution $(\phi^{i,*}_{n-1})_{n=1}^N, i=1,2,\ldots$, 
for the agents satisfying Assumption~\ref{assumption-single-2} (ii) and (iii), 
such that, 
with $\mu_0:=\ex^i[\xi_i]$, the bounded measurable functions $\mu_n:\cals^n\times \caly^{n-1}\rightarrow \mbb{R}$, $1\leq n\leq N$ satisfy
the fixed point condition:
\be
\label{def-single-mu}
\mu_n(\bS^n, \bY^{n-1})=\ex^{0,i}\bigl[X_n^{i,*}|\calf^{0}_{t_n}\bigr] ~{\rm a.s.,} \quad 1\leq n\leq N, 
\ee
where $X^{i,*}$ denotes the wealth process associated with the optimal control $\phi^{i,*}$.
Moreover, we denote the associated processes in the RP-MFE by $(\wh{\mu}, \wh{\phi}^{i})$, 
and refer to them as the solution pair of the RP-MFE. We also use the symbol $\wh{X}^i$ for $i=1,2,\ldots$
to represent the wealth process of agent-$i$ associated with $\wh{\phi}^i$ and the initial condition $\xi_i$.
\end{definition}
Since we are dealing with a symmetric problem with i.i.d.\! variables and processes $(\xi_i,\gamma_i,\theta_i, Z^i)_{i\geq 1}$, the choice 
of the representative agent is arbitrary. 
Since the filtration $(\calf^0_{t_k})_{k=0}^N$ is generated by $(S,Y)$, the fixed-point condition $(\ref{def-single-mu})$
can be equivalently represented by using agent-$1$ as the representative:
\be
\mu_n(\bs,\by^-)=\ex^1\bigl[X_n^{1,*}|\bs,\by\bigr]=\ex^1\bigl[X_n^{1,*}|\bs,\by^-\bigr], \nn
\ee
for every $(\bs,\by^-)\in \cals^n\times \caly^{n-1}$. Here, $\by=(\by^-,y_n)\in \caly^n$.
In the following, we prove the existence and uniqueness of the RP-MFE. We will show that the path-dependence 
of $\phi^{i,*}_{n-1}$ on $(\bs,\by)\in \cals^{n-1}\times \caly^{n-1}$ reduces to dependence on the current state
 $(s,y)=(s_{n-1},y_{n-1})\in \cals_{n-1}\times \caly_{n-1}$ in the RP-MFE.
It is well known that the RP-MFE constitutes an $\ep$-Nash equilibrium for the 
corresponding finite-agent game; we refer the reader to Appendix~\ref{sec-A} for a brief explanation.

\begin{theorem}
\label{th-single-1}
Let Assumptions~\ref{assumption-single-1} and \ref{assumption-single-2} be in force. Assume further that $\ex^1[\theta_1]\neq 1$.
Then there exists a unique RP-MFE for the problem $(\ref{problem-single})$ with the solution pair $(\wh{\mu}, \wh{\phi}^i)$. The associated
optimal strategy $(\wh{\phi}^i_{n-1})_{n=1}^N$ is given by the bounded measurable function $\wh{\phi}_{n-1}^{i}:\cals_{n-1}\times \caly_{n-1}
\times \calz_{n-1}\times \Gamma\rightarrow \mbb{R}$, $1\leq n\leq N$, such that
\be
\label{single-optimal}
\begin{split}
&\wh{\phi}^{i}_{n-1}(s,y,z^i,\vr_i):=\frac{1}{\gamma_n^i(u-d)}\Bigl\{\log\Bigl(-\frac{p_{n-1}(s,y)u}{q_{n-1}(s,y)d}\Bigr)+\log f_{n-1}(s,y,z^i,\vr_i)\Bigr\}\\
&\qquad+\frac{1}{u-d}\frac{\theta_i}{1-\ex^1[\theta_1]}\Bigl\{
\log\Bigl(-\frac{p_{n-1}(s,y)u}{q_{n-1}(s, y)d}\Bigr)\ex^1\Bigl[\frac{1}{\gamma_n^1}\Bigr]+\ex^1\Bigl[\frac{
\log f_{n-1}(s,y,Z^1_{n-1},\vr_1)}{\gamma_n^1}\Bigr]\Bigr\}
\end{split}
\ee
and the dynamics of the associated mean-field terms $(\wh{\mu}_n)_{n=1}^N$ is given by
\be
\label{single-rp-mfe}
\begin{split}
\wh{\mu}_n((\bs\wt{u})^n, \by )&=\beta \wh{\mu}_{n-1}(\bs, \by^-)+u\ex^1[\wh{\phi}^{1}_{n-1}(s,y,Z_{n-1}^1,\vr_1)], \\
\wh{\mu}_n((\bs\wt{d})^n, \by )&=\beta \wh{\mu}_{n-1}(\bs, \by^-)+d\ex^1[\wh{\phi}^{1}_{n-1}(s,y,Z_{n-1}^1,\vr_1)], 
\end{split}
\ee
with the initial condition $\wh{\mu}_0:=\ex^1[\xi_1]$ and
\be
\label{single-optimal-exp}
\ex^1[\wh{\phi}^{1}_{n-1}(s,y,Z_{n-1}^1,\vr_1)]=\frac{1}{u-d}\frac{1}{1-\ex^1[\theta_1]}
\Bigl\{\log\Bigl(-\frac{p_{n-1}(s,y)u}{q_{n-1}(s,y)d}\Bigr)\ex^1\Bigl[\frac{1}{\gamma_n^1}\Bigr]+
\ex^1\Bigl[\frac{\log f_{n-1}(s,y,Z_{n-1}^1,\vr_1)}{\gamma_n^1}\Bigr]\Bigr\}. 
\ee
Here, $s:=s_{n-1}$ and $y:=y_{n-1}$ are the last elements of $(\bs,\by)\in \cals^{n-1}\times \caly^{n-1}$,
and $\by^{-}$ is such that $(\by^{-},y_{n-1})=\by\in \caly^{n-1}$. 
The function $f_{n-1}:\cals_{n-1}\times\caly_{n-1}\times \calz_{n-1}\times \Gamma\rightarrow \mbb{R}$
is defined as in Lemma~\ref{lemma-single-tmp}. It satisfies the uniform bounds $0<\ol{c}_n\leq f_{n-1}\leq \ol{C}_n<\infty$
on its domain for some positive constants $\ol{c}_n$ and $\ol{C}_n$, and is given by
\be
f_{n-1}(s,y,z^i,\vr_i):=\frac{\ex^{0,i}[V_n(s\wt{u},Y_n,Z_n^i,\vr_i)|y,z^i,\vr_i]}{\ex^{0,i}[V_n(s\wt{d},Y_n,Z_n^i,\vr_i)|y,z^i,\vr_i]}. \nn
\ee
Here, $V_n:\cals_n\times \caly_n\times \calz_n\times \Gamma\rightarrow \mbb{R}$, $0\leq n\leq N$,
are measurable functions satisfying the uniform bounds $0<c_n\leq V_n\leq C_n<\infty$ on their respective domains
for some positive constants $c_n$ and $C_n$. They are defined recursively by 
\be 
V_N(s,y,z^i,\vr_i):=\exp\bigl(\gamma_i F(s, y,z^i)\bigr), \nn
\ee
for each $(s,y,z^i,\vr_i)\in \cals_N\times\caly_N\times \calz_N\times \Gamma$, and 
\be
\label{single-Vnm1}
\begin{split}
&V_{n-1}(s,y,z^i,\vr_i)\\
&=p_{n-1}(s,y)\exp\Bigl(-\gamma_n^i u\bigl(\wh{\phi}^{i}_{n-1}(s,y,z^i,\vr_i)-\theta_i\ex^1[\wh{\phi}^{1}_{n-1}(s,y,Z^1_{n-1},\vr_1)]\bigr)\Bigr)
\ex^{0,i}\bigl[V_n(s\wt{u},Y_n,Z_n^i,\vr_i)|y,z^i,\vr_i\bigr]\\
&+q_{n-1}(s,y)\exp\Bigl(-\gamma_n^i d\bigl(\wh{\phi}^{i}_{n-1}(s,y,z^i,\vr_i)-\theta_i\ex^1[\wh{\phi}^{1}_{n-1}(s,y,Z^1_{n-1},\vr_1)]\bigr)\Bigr)
\ex^{0,i}\bigl[V_n(s\wt{d},Y_n,Z_n^i,\vr_i)|y,z^i,\vr_i\bigr]
\end{split}
\ee
for each $(s,y,z^i,\vr_i)\in \cals_{n-1}\times \caly_{n-1}\times \calz_{n-1}\times \Gamma$, $1\leq n\leq N$.
\end{theorem}
\begin{remark*}
Note that $(1/\gamma_n^i)\log V_n$ represents the effective liability at $t_n$.
\end{remark*}
\begin{proof}
Suppose that the problem for agent-$i$ at $t_{n-1}$ for the period $[t_{n-1},t_n]$ is
given by the form $(\ref{problem-single-tmp})$. To apply Lemma~\ref{lemma-single-tmp}, 
we assume that $V_n:\cals_n\times \caly_n\times \calz_n\times \Gamma \rightarrow \mbb{R}$ 
is a measurable function satisfying the uniform bounds $c_n\leq V_n\leq C_n$ on its domain for some positive constants $c_n$
and $C_n$. This clearly holds at $t=t_{N-1}$ for the last interval $[t_{N-1},t_N]$
with $V_N(s,y,z^i,\vr_i):=\exp\bigl(\gamma_i F(s,y,z^i)\bigr)$.

In view of $(\ref{single-optimal-tmp})$ in Lemma~\ref{lemma-single-tmp}, the evolution of the wealth process of agent-$i$ under the 
optimal strategy $\phi^{i,*}_{n-1}$ is given by 
\be
X_n^{i,*}=\beta X_{n-1}^{i,*}+\phi_{n-1}^{i,*}(\bS^{n-1}, \bY^{n-1},Z_{n-1}^i,\vr_i)R_n. \nn
\ee
Consider the problem conditioned on the event $\{(\bS^{n-1}, \bY^{n-1})=(\bs,\by)\}$ in $\calf^{0}_{t_{n-1}}$. 
Under the induction hypothesis that $\ex^{0,1}[X_{n-1}^{1,*}|\bs,\by]$ 
is of the form $\mu_{n-1}(\bs,\by^-)$ with $(\bs,\by^-)\in \cals^{n-1}\times
\caly^{n-2}$ (when $n=1$, we simply set $\mu_0= \ex^1[\xi_1]$), and utilizing the i.i.d.\! property of $(\xi_i, \gamma_i,\theta_i,Z^i), i=1,2,\ldots$,  
the condition for the RP-MFE for the period $[t_{n-1},t_n]$ is 
equivalently given by  the following evolution equations:
\be
\label{single-RP-MFE-eq}
\begin{split}
\mu_n((\bs\wt{u})^n,\by)&=\beta \mu_{n-1}(\bs,\by^-)+u\ex^1[\phi^{1,*}_{n-1}(\bs,\by,Z_{n-1}^1,\vr_1)],  \\
\mu_n((\bs\wt{d})^n,\by)&=\beta \mu_{n-1}(\bs,\by^-)+d\ex^1[\phi^{1,*}_{n-1}(\bs,\by,Z_{n-1}^1,\vr_1)].
\end{split}
\ee
To solve the above equations, observe that $\phi^{i,*}_{n-1}$ in $(\ref{single-optimal-tmp})$ depends
on the mean-field term $\mu_n$ only through its difference $\Del_n(\bs,\by)$.
By taking the difference in $(\ref{single-RP-MFE-eq})$, we obtain the equation for $\Del_n(\bs,\by)$ as
\be
\label{single-Deln-consistency}
\Del_n(\bs,\by)=\ex^1[\theta_1]\Del_n(\bs,\by)+\Bigl\{\log\Bigl(-\frac{p_{n-1}(s,y)u}{q_{n-1}(s,y)d}\Bigr)\ex^1\Bigl[\frac{1}{\gamma_n^1}\Bigr]
+\ex^1\Bigl[\frac{\log f_{n-1}(s,y,Z^1_{n-1},\vr_1)}{\gamma_n^1}\Bigr]\Bigr\}. 
\ee
Since $\ex^1[\theta_1]\neq 1$ by assumption, this equation determines $\Del_n$ uniquely as
\be
\label{single-Deln}
\Del_n(s,y)=\frac{1}{1-\ex^1[\theta_1]}\Bigl\{\log\Bigl(-\frac{p_{n-1}(s,y)u}{q_{n-1}(s,y)d}\Bigr)\ex^1\Bigl[\frac{1}{\gamma_n^1}\Bigr]
+\ex^1\Bigl[\frac{\log f_{n-1}(s,y,Z^1_{n-1},\vr_1)}{\gamma_n^1}\Bigr]\Bigr\}
\ee
for each $(s,y)=(s_{n-1},y_{n-1})$. Note that $\Del_n$ is seen to be independent of the entire trajectory of $(S,Y)$ up to time $t_{n-2}$,
depending only on the current state $(s_{n-1},y_{n-1})$.
Substituting $\Del_n(s,y)$ in $(\ref{single-Deln})$ for $\Del_n(\bs,\by)$ in $(\ref{single-optimal-tmp})$ yields
the expression $\phi^{i,*}_{n-1}$ as given by $(\ref{single-optimal})$ and hence its expectation as in $(\ref{single-optimal-exp})$.
Moreover, $(\ref{single-RP-MFE-eq})$ is shown to be equivalent to $(\ref{single-rp-mfe})$. 
Since $\phi^{i,*}_{n-1}$ given by $(\ref{single-optimal})$ is a bounded measurable function due to 
the uniform bounds on $f_{n-1}$, i.e., $\ol{c}_n\leq f_{n-1}\leq \ol{C}_n$, the mean-field term $\mu_n$ is also bounded and measurable 
provided that  $\mu_{n-1}$ is.
Thus, under the assumption that $\ex^{0,1}[X_{n-1}^{1,*}|\bs,\by]=\mu_{n-1}(\bs,\by^-)$ with some bounded measurable
function $\mu_{n-1}$, $(\ref{single-optimal})$ and $(\ref{single-rp-mfe})$ 
provide a unique candidate for the solution pair of RP-MFE for the interval $[t_{n-1},t_n]$.

It remains to show that this procedure can be iterated backward from the last interval $[t_{N-1}, t_N]$ to the 
first one $[t_0,t_1]$. From $(\ref{middle-valueF}$), we see that the 
value function at $t_{n-1}$ is given by
\be
\begin{split}
&-\exp(-\gamma_{n-1}^i x^i)\Bigl\{p_{n-1}(s,y)e^{-\gamma_n^i(\phi^{i,*}_{n-1} u-\theta_i \mu_n((\bs\wt{u})^n,\by))}
\ex^{0,i}[V_n(s\wt{u},Y_n,Z_n^i,\vr_i)|y,z^i,\vr_i]\\
&\qquad+q_{n-1}(s,y)e^{-\gamma_n^i(\phi^{i,*}_{n-1} d-\theta_i \mu_n((\bs\wt{d})^n,\by))}
\ex^{0,i}[V_n(s\wt{d},Y_n,Z_n^i,\vr_i)|y,z^i,\vr_i]\Bigr\} \nn
\end{split}
\ee
for each realization $(x^i, \bs,\by,z^i,\vr_i)$ of $ (X_{n-1}^{i,*}, \bS^{n-1},\bY^{n-1},Z_{n-1}^i,\vr_i)$
with the conventions $(s,y)=(s_{n-1},y_{n-1})$. 
Let $V_{n-1}$ be defined in $(\ref{single-Vnm1})$, which again satisfies  the uniform bounds
$c_{n-1}\leq V_{n-1}\leq C_{n-1}$ on its domain with some positive constants $c_{n-1}$ and $C_{n-1}$.
By invoking  $(\ref{single-rp-mfe})$, the value function at $t_{n-1}$ can now be expressed as
\be
-\exp\bigl(-\gamma_{n-1}^i(x^i-\theta_i \mu_{n-1}(\bs,\by^-))\bigr)V_{n-1}(s,y,z^i,\vr_i) \nn
\ee
yielding  a problem of the same form as in $(\ref{problem-single-tmp})$ used in Lemma~\ref{lemma-single-tmp}.
The boundedness condition on the mean-field terms $(\mu_n)_{n=1}^N$ thus reduces to that of $\mu_0:=\ex^1[\xi_1]$.
Since $\xi_i\in [\ul{\xi},\ol{\xi}]$ is bounded, the entire process $(\mu_n)_{n=0}^N$ is now bounded and becomes consistent with our assumption. 
Thus we conclude that the above constructed optimal strategy $\phi^{i,*}$
and the associated mean-field $\mu=(\mu_n)_{n=0}^N$ satisfy the fixed point condition $(\ref{def-single-mu})$, 
and hence $\mu$ and $\phi^{i,*}$ constitute a RP-MFE solution pair, which we denote by 
$(\wh{\mu}, \wh{\phi}^i)$ for the problem $(\ref{problem-single})$.

The uniqueness of RP-MFE follows immediately from the construction:  firstly, the 
optimal control in each interval by Lemma~\ref{lemma-single-tmp} for a given $\mu_n$ is unique;
secondly, the dynamics of the mean-field term $(\mu_n)_{n\geq 1}$ is 
uniquely determined by the evolution equations $(\ref{single-rp-mfe})$ with a given initial condition $\mu_0\in \mbb{R}$.
\end{proof}

\begin{remark}
We analyze three regimes defined by the value of the expected relative 
performance concerns $\mathbb{E}^1[\theta_1]:$ $\mathbb{E}^1[\theta_1] \leq 0$, 
$0<\mathbb{E}^1[\theta_1]<1$, 
and $\mathbb{E}^1[\theta_1]>1$.
First, we introduce the variable:
\be
\calt_n:=\frac{1}{1-\ex^1[\theta_1]}\ex^1\Bigl[\frac{1}{\gamma_n^1}\Bigr]. \nn
\ee
Observing the expression in $(\ref{single-optimal-exp})$, we can identify $\mathcal{T}_n$ 
as the aggregate effective risk tolerance. 
\begin{itemize}[noitemsep]
\item $\ex^1[\theta_1]\leq 0$:  
In this regime, on average, the lower the expected relative concern $\mathbb{E}^1[\theta_1]$ is, 
the smaller the aggregate risk tolerance $\mathcal{T}_n$ becomes, 
implying that the agents are, on average, more risk-averse. 
This is because the mean-field interaction term $\theta_i \mu_n(\mathbf{S}^n,\mathbf{Y}^{n-1})$ in the objective function
$(\ref{problem-single-tmp})$ aligns with the wealth of agent-$i$ and thus increases the exposure to the risk 
in the stock position. 
\item $0<\ex^1[\theta_1]<1$: 
In this regime, the agents are, on average, more risk-tolerant, and thus take larger positions as $\mathbb{E}^1[\theta_1]$ 
increases. In the limit $\mathbb{E}^1[\theta_1]\rightarrow 1$, the aggregate risk tolerance and hence the stock position diverge,
becoming a singular point within the current setup. This is caused by the divergence of the feedback loop 
of relative performance concerns, represented by the series $1+\ex^1[\theta_1]+(\ex^1[\theta_1])^2+\cdots$,
i.e., an  agent is concerned with the performance of peers, while each peer is, in turn,  concerned with the performance of others, \ldots.
\item $\ex^1[\theta_1]>1$: 
In this regime, the relative performance concern becomes too large to hedge in the conventional manner.
From $(\ref{single-optimal-exp})$, we observe that the sign of the average stock position flips.
Suppose, for example, the transition probability for the up-move, $p_{n-1}(s,y)$, is significantly higher than $q_{n-1}(s,y)$.
If we can neglect the contribution from $f_{n-1}$, then agents would typically take a long position in the stock in the first two regimes.
However, this is not the case here. Although a long strategy is likely to produce a positive gain, 
the gain of peers, when multiplied by a large $\theta>1$, would reduce the agents' utility significantly more.
Thus, agents, on average, take a short position, bearing a loss from the stock trade but deriving
a higher utility gain from the loss of their peers, which is multiplied by a large $\theta>1$. 
Consider an agent who deviates from the majority and takes a long position.
If the stock price rises, the agent would enjoy a positive gain from the stock position as well as a significant utility gain derived from the loss of peers. 
However, if the stock price falls, the agent would bear a loss from the stock position as well as a significant utility loss due to the gain of peers. 
Given the concavity of the utility function, such a deviation would not be optimal. 
In fact, the optimal strategy of each agent is characterized by $(\ref{single-optimal-tmp})$
regardless of the sign of $\Del_n$.
\end{itemize}
\end{remark}

\subsection{Market-clearing mean-field equilibrium}
\label{sec-single-mc}
In the previous subsection, assuming $\ex^1[\theta_1]\neq 1$, we established 
the unique existence of the RP-MFE and provided the explicit form of its solution pair $(\wh{\mu},\wh{\phi}^i)$. 
Importantly, the specific functional form of the transition probabilities $(p_{n-1}(s,y), q_{n-1}(s,y))_{n=1}^N$ 
did not play a decisive role. In this subsection, utilizing this degree of freedom, we show that it is possible to 
choose an appropriate functional form for the stock price transition probabilities satisfying 
Assumption~\ref{assumption-single-1} (vi) so that the market-clearing mean-field equilibrium (MC-MFE) 
is achieved while preserving the RP-MFE. Naturally, the market-clearing condition implies 
that the agents' stock positions must not diverge. Interestingly, this suggests that the appropriate choice 
of transition probabilities that clears the market may eliminate the singularity at $\ex^1[\theta_1]=1$ 
observed in the relative performance game. 
We shall show that this is indeed the case.

As in \cite{Fujii-Trees}, we incorporate an external stochastic order flow, $L_{n-1}(S_{n-1},Y_{n-1})$,
which represents the aggregate net stock supply per capita at each time $t=t_{n-1}$, for $1\leq n\leq N$.
The external order flow is intended to model the aggregate contribution from other populations. In particular, it can be used to represent the
aggregate net supply from individual investors, whose behavior is often difficult to model via rigorous  
optimization, or the supply from a major financial institution such as a central bank.

\begin{assumption}
\label{assumption-single-3}
For every $1\leq n\leq N$, $L_{n-1}:\cals_{n-1}\times \caly_{n-1}\rightarrow \mbb{R}$ is a bounded measurable function.
\end{assumption}

\begin{definition}
\label{def-single-MFE}
We say that the RP-MFE defined in Definition~\ref{def-single-rp-mfe} constitutes a market-clearing 
mean-field equilibrium (MC-MFE) if
\be
\lim_{N_p\rightarrow \infty}\frac{1}{N_p}\sum_{i=1}^{N_p} \wh{\phi}_{n-1}^i
=L_{n-1}(S_{n-1},Y_{n-1}), \nn
\ee
$\mbb{P}$-a.s. for every $1\leq n\leq N$. 
\end{definition}
\noindent
This condition implies that the excess demand/supply 
per capita converges to zero in the large population limit.

\begin{remark}
We often consider the baseline case $L_{n-1} \equiv 0$, which corresponds to a net zero position among the agents.
The special case $L_{n-1}(s,y)=N^\# s$, $1\leq n\leq N$,  corresponds to the situation 
where the supply is given by a constant number of shares $N^{\#}$ per agent. 
In a closed market setting, this implies that the net initial number of shares per capita is  $N^\#$.
Note that, in our formulation, $L$ represents the monetary value of the supply per capita.
\end{remark}

\begin{theorem}
\label{th-single-MC-MFE}
Let Assumptions~\ref{assumption-single-1}, \ref{assumption-single-2} and \ref{assumption-single-3} be in force.
Then there exists a unique MC-MFE. The associated equilibrium transition probabilities of the stock price are given by
\be
\label{single-MC-transition}
\begin{split}
&p_{n-1}(s,y):=\mbb{P}^0\Bigl(S_n=\wt{u}S_{n-1}|(S_{n-1},Y_{n-1})=(s,y)\Bigr)\\
&=(-d)\Big/\left\{u \exp\left(\frac{1}{\ex^1[1/\gamma_n^1]}\Bigl\{\ex^1\Bigl[\frac{\log f_{n-1}(s,y,Z^1_{n-1},\vr_1)}{\gamma_n^1}\Bigr]
-(1-\ex^1[\theta_1])(u-d)L_{n-1}(s,y)\Bigr\}\right)-d\right\}
\end{split}
\ee
for every $(s,y)\in \cals_{n-1}\times \caly_{n-1}$, $1\leq n\leq N$. 
Here, the functions $f_{n-1}:\cals_{n-1}\times\caly_{n-1}\times \calz_{n-1}\times \Gamma\rightarrow \mbb{R}$, $1\leq n\leq N$
are measurable functions satisfying the uniform bounds $0<\ol{c}_n\leq f_{n-1}\leq \ol{C}_n<\infty$ 
on their respective domains with some positive constants $\ol{c}_n$ and $\ol{C}_n$.
They are determined via the backward induction in Theorem~\ref{th-single-1},  with transition probabilities replaced by those given above at each step.
Under the equilibrium transition probabilities, 
the optimal strategy $\wh{\phi}^i_{n-1}$ of agent-$i$ is given by, for each $(s,y,z^i,\vr_i)\in \cals_{n-1}\times \caly_{n-1}\times
\calz_{n-1}\times \Gamma$, 
\be
\label{single-MC-position}
\begin{split}
\wh{\phi}^i_{n-1}(s,y,z^i,\vr_i)&=\frac{1}{\ex^1[1/\gamma_n^1]}\Bigl(\frac{1-\ex^1[\theta_1]}{\gamma_n^i}+\theta_i\ex^1
\Bigl[\frac{1}{\gamma_n^1}\Bigr]\Bigr)L_{n-1}(s,y) \\
&+\frac{1}{u-d}\Bigl\{\frac{\log f_{n-1}(s,y,z^i,\vr_i)}{\gamma_n^i}-\frac{1}{\ex^1[1/\gamma_n^1]}\frac{1}{\gamma_n^i}
\ex^1\Bigl[\frac{\log f_{n-1}(s,y,Z_{n-1}^1,\vr_1)}{\gamma_n^1}\Bigr]\Bigr\}.
\end{split}
\ee
Moreover, there exists a positive constant $\calc_{n-1}$ such that
\be
\ex\Bigl| \frac{1}{N_p}\sum_{i=1}^{N_p}\wh{\phi}^i_{n-1}(S_{n-1},Y_{n-1},Z_{n-1}^i,\vr_i)-L_{n-1}(S_{n-1}, Y_{n-1})\Bigr|^2\leq \frac{\calc_{n-1}}{N_p} \nn
\ee
for every $1\leq n\leq  N$, which establishes the convergence rate in the large population limit.
\end{theorem}
\begin{remark*}
Recall that the definition  $\vr_i:=(\gamma_i,\theta_i)$ that incorporates the $\theta_i$-dependence.
\end{remark*}
\begin{proof}
{\bf(Step 1)}: We first assume that $\ex^1[\theta_1]\neq 1$. \\
In this case, we can directly use the result of Theorem~\ref{th-single-1}.
Since the sequence  $\{(Z^i, \vr_i), i\in \mbb{N}\}$ is i.i.d. and also independent of the process $(S,Y)$, the market-clearing condition 
in Definition~\ref{def-single-MFE} is equivalently given by
\be
\label{single-MC-MFE-eq}
\ex^1\bigl[\wh{\phi}_{n-1}^1(s,y,Z_{n-1}^1,\vr_1)\bigr]=L_{n-1}(s,y)
\ee
for every $(s,y)\in \cals_{n-1}\times \caly_{n-1}$, $1\leq n\leq N$. 
The expression for the transition probabilities $(\ref{single-MC-transition})$
is a direct consequence of $(\ref{single-optimal-exp})$ in Theorem~\ref{th-single-1}
and the market-clearing condition $(\ref{single-MC-MFE-eq})$. By substituting the resulting expression for $p_{n-1}(s,y)$
(and $q_{n-1}(s,y)$) into $(\ref{single-optimal})$, we obtain $(\ref{single-MC-position})$.

To establish the first claim, it suffices to verify that the family of transition probabilities $(p_{n-1}(s,y))_{n=1}^N$ 
defined in $(\ref{single-MC-transition})$ satisfies the bound specified in  $(vi)$ in Assumption~\ref{assumption-single-1},
while updating the functions $(f_{n-1})$ via the backward induction process described in Theorem~\ref{th-single-1}.
At $t=t_{N-1}$, $f_{N-1}$ is a measurable function satisfying the uniform bounds $\ol{c}_N\leq f_{N-1}\leq \ol{C}_N$
on its domain for some positive constants $\ol{c}_N$ and $\ol{C}_N$ due to the boundedness assumption on $F$. 
Combined with  $d<0<u$  and the boundedness of other variables, such as $(\gamma_i,\theta_i, L)$,
we can confirm that $p_{N-1}$ (and hence $q_{N-1}$) given by $(\ref{single-MC-transition})$ satisfies $0<p_{N-1}(s,y), q_{N-1}(s,y)<1$
for every $(s,y)\in \cals_{N-1}\times \caly_{N-1}$, and is thus consistent with the condition.
Moreover,  $\wh{\phi}^i_{N-1}$ in $(\ref{single-MC-position})$ is a bounded function on its domain 
due to the uniform bounds of $f_{N-1}$. It follows that $V_{N-1}$ defined by $(\ref{single-Vnm1})$
is a measurable function satisfying the uniform bounds $c_{N-1}\leq V_{N-1}\leq C_{N-1}$
on its domain for some positive constants $c_{N-1}$ and $C_{N-1}$.
This,  in turn,  ensures that $f_{N-2}$ once again satisfies the desired uniform bounds, and so do $(p_{N-2}(s,y), q_{N-2}(s,y))$,
$(s,y)\in \cals_{N-2}\times \caly_{N-2}$. Proceeding in this way, by backward induction, we establish the desired 
consistency for every time step.

To establish the second claim, it is enough to show that, for $1\leq n\leq N$, there exists a positive constant $\calc_{n-1}$ such that
the inequality 
\be
\ex \Bigl|\frac{1}{N_p}\sum_{i=1}^{N_p}\wh{\phi}^i_{n-1}(s,y,Z_{n-1}^i,\vr_i)-L_{n-1}(s,y)\Bigr|^2\leq \frac{\calc_{n-1}}{N_p}\nn
\ee
holds uniformly for every $(s,y)\in \cals_{n-1}\times\caly_{n-1}$.
Using the i.i.d.\ property of $(\gamma_n^i, \theta_i, Z^i)$ as well as the  boundedness of $(1/\gamma_n^i, \theta_i, f_{n-1}, L_{n-1})$
and $\ex^1[\theta_1]\neq 1$, we can show that there exists some constant $\calc_{n-1}$ 
by rearranging the expression $(\ref{single-MC-position})$  as follows:
\be
\begin{split}
&\ex\Bigl|\frac{1}{N_p}\sum_{i=1}^{N_p}\wh{\phi}^i_{n-1}(s,y,Z_{n-1}^i,\vr_i)-L_{n-1}(s,y)\Bigr|^2 \leq \frac{\calc_{n-1}}{N_p^2}\ex \Bigl[\Bigl|\sum_{i=1}^{N_p}\Bigl(\frac{1}{\gamma_n^i}-\ex^1\Bigl[\frac{1}{\gamma_n^1}\Bigr]\Bigr)\Bigr|^2
+\Bigl|\sum_{i=1}^{N_p}(\theta_i-\ex^1[\theta_1])\Bigr|^2\\
&\qquad+\Bigl|\sum_{i=1}^{N_p}\Bigl(\frac{\log f_{n-1}(s,y,Z_{n-1}^i,\vr_i)}{\gamma_n^i}-\ex^1\Bigl[
\frac{\log f_{n-1}(s,y,Z_{n-1}^1,\vr_1)}{\gamma_n^1}\Bigr]\Bigr)\Bigr|^2\Bigr] \\
&\leq \frac{\calc_{n-1}}{N_p}
\ex^1\Bigl[\Bigl|\frac{1}{\gamma_n^1}-\ex^1\Bigl[\frac{1}{\gamma_n^1}\Bigr]\Bigr|^2 +\bigl|\theta_1-\ex^1[\theta_1]\bigr|^2
+\Bigl| \frac{\log f_{n-1}(s,y,Z^1_{n-1},\vr_1)}{\gamma_n^1}
-\ex^1\Bigl[\frac{\log f_{n-1}(s,y,Z_{n-1}^1,\vr_1)}{\gamma_n^1}\Bigr]\Bigr|^2\Bigr]. \nn
\end{split}
\ee
Since the last variance terms are finite,  this estimate establishes the desired result. 
Note that the cross terms appearing in the second line vanish 
due to the mutual independence of $(\gamma_n^i, \theta_i, Z^i)_{i\in \mbb{N}}$. 

\bigskip
\noindent
{\bf (Step 2)}: We now consider the special case $\ex^1[\theta_1]=1$. \\
In this case, the equation $(\ref{single-Deln-consistency})$, which is the consistency (i.e., fixed point) condition for the RP-MFE, 
can be satisfied if and only if
\be
\label{theta-1-condition}
\Bigl\{\log\Bigl(-\frac{p_{n-1}(s,y)u}{q_{n-1}(s,y)d}\Bigr)\ex^1\Bigl[\frac{1}{\gamma_n^1}\Bigr]
+\ex^1\Bigl[\frac{\log f_{n-1}(s,y,Z^1_{n-1},\vr_1)}{\gamma_n^1}\Bigr]\Bigr\}=0 
\ee
for every $(s,y)\in \cals_{n-1}\times \caly_{n-1}$. 
This equality determines the transition probabilities uniquely, and they are 
given by the same expression $(\ref{single-MC-transition})$ with $\ex^1[\theta_1]=1$.

If the above equality holds, then $\Del_n(\bs,\by)$ and hence also 
the optimal control $\phi^{i,*}_{n-1}$ in $(\ref{single-optimal-tmp})$ would
remain undetermined if we were to consider the problem solely within the RP-MFE framework.
However, if we impose the market-clearing condition, the optimal control $\wh{\phi}^i_{n-1}$ and $\Del_n(\bs,\by)$
are determined uniquely. Specifically, under the condition $(\ref{theta-1-condition})$, 
the equations $(\ref{single-MC-MFE-eq})$ and $(\ref{single-optimal-tmp})$ imply 
\be
\Del_n(\bs,\by)=(u-d)L_{n-1}(s,y), \nn
\ee
which only depends on the last elements of $(\bs,\by)$, i.e., $(s,y)\in \cals_{n-1}\times \caly_{n-1}$.
Thus, once again by $(\ref{single-optimal-tmp})$, we obtain 
\be
\wh{\phi}^i_{n-1}(s,y,z^i,\vr_i)=\theta_i L_{n-1}(s,y)+\frac{1}{\gamma_n^i (u-d)}
\Bigl\{\log\Bigl(-\frac{p_{n-1}(s,y) u}{q_{n-1}(s,y)d}\Bigr)+\log f_{n-1}(s, y, z^i, \vr_i)\Bigr\}
\ee
with the transition probabilities derived above.  This is also consistent with the expression in $(\ref{single-MC-position})$ when $\ex^1[\theta_1]=1$.
Given that $V_n$ satisfies the uniform bounds $0<c_n\leq V_n\leq C_n<\infty$ on its domain for some positive constants $c_n$ and $C_n$, 
the control $\wh{\phi}^i_{n-1}$ is bounded and measurable. Thus, under the assumption that 
$\ex^{0,1}[X_{n-1}^{1,*}|\bs,\by]=\mu_{n-1}(\bs,\by^-)$ with some bounded measurable
function $\mu_{n-1}$, $(\ref{single-MC-position})$ and $(\ref{single-rp-mfe})$ 
with transition probabilities $(\ref{single-MC-transition})$ with $(\ex^1[\theta_1]=1)$
provide a unique candidate for the solution to the  MC-MFE (and hence the RP-MFE) for the interval $[t_{n-1},t_n]$.

In order to complete the proof, it suffices to show that this procedure can be repeated backward from the last interval $[t_{N-1},t_N]$
to the first one $[t_0,t_1]$. The value function $V_{n-1}$ for agent-$i$ at $t_{n-1}$, which serves as the objective function for the 
optimization for the period $[t_{n-2}, t_{n-1}]$, is given by $(\ref{single-Vnm1})$ with $\wh{\phi}^i$ obtained above.
It  once again satisfies the desired uniform bounds for some positive constants $c_{n-1}$ and $C_{n-1}$. 
Thus we can proceed with the backward induction one step further. 
The condition for $\mu_{n-1}$ is reduced to the initial condition $\mu_0:=\ex^1[\xi_1]$ 
and is satisfied trivially. The second claim on the convergence rate can be shown as in (Step 1).
\end{proof}

\begin{remark}
\label{remark-single-theta}
Theorem~\ref{th-single-MC-MFE} shows that there is  no singularity at $\ex^1[\theta_1]=1$
and the three regimes $(\ex^1[\theta_1]\leq 0, 0<\ex^1[\theta_1]<1, \ex^1[\theta_1]>1)$
are continuously connected. Let us give some remarks on the behavior of the equilibrium transition probabilities.
The economic interpretation when there is no relative performance concern $\theta\equiv 0$
is detailed in Fujii~\cite{Fujii-Trees}[Section 2.4], in particular, on the effects from the liabilities (or negative of the endowments). 
Recall that the transition probability of the up-move under the risk-neutral measure $\mbb{Q}$
is given by $p^{\mbb{Q}}=(-d)/(u-d)$. Thus $p_{n-1}(s,y)>p^{\mbb{Q}}$ denotes 
a positive excess return (at the node $(s,y)\in \cals_{n-1}\times \caly_{n-1}$).
Suppose, for simplicity, the hedge needs for the effective liability are negligible, i.e., $\log f_{n-1}\approx 0$. 
In this case,  when $\ex^1 [\theta_1]<1$, the positive stock supply $L_{n-1}(s,y)$ implies that a positive excess
return is required by the agents to compensate for their risk of taking long positions for market clearing.
As $\ex^1[\theta_1]$ approaches  $1$, the effective risk-tolerance $\calt_n$ becomes  larger, which
reduces the required excess return. The agents become totally indifferent to the stock size they absorb
at the critical point $\ex^1[\theta_1]=1$.
Conversely, when $\ex^1[\theta_1]>1$, the effective risk aversion of the agents becomes negative due to extremely
strong relative performance concerns. In order to clear the market, the excess return turns negative $p_{n-1}(s,y)<p^{\mbb{Q}}$. 
With the negative effective risk aversion, a negative excess return is required to induce agents to hold a long position 
to balance the positive market supply.
\end{remark}

\section{A network of relative performance concerns among multiple populations}
\label{sec-multi-population}
\subsection{The setup and notation}

A primary limitation of the previous framework lies in their restriction to a single homogeneous population, 
where all agents share identical liability functions $F$, distributions of i.i.d. idiosyncratic shocks $Z^i$, initial wealth $\xi_i$, 
risk aversion $\gamma_i$, and relative performance concern $\theta_i$. Crucially, the relative performance concern $\theta_i$ was constrained to 
reference solely the aggregate wealth of the entire population. This simplified structure fails to capture more realistic scenarios 
where agents benchmark their performance against specific peer groups or competitors rather than the market average as a whole.

To address this limitation, we consider a more generalized setting where 
agents (i.e., financial firms) belong to distinct sectors or groups $p=1,2,\ldots, m$. 
Agents in different populations are assumed to have distinct liability functions $F^p$ and 
distributions of idiosyncratic factors, characterized by $\xi_i^p, \gamma_i^p$, and $Z^{i,p}$. 
Most importantly, we introduce a heterogeneous network of relative performance concerns 
represented by $\theta^i_{p,k}$, which denotes the sensitivity 
of agent-$i$ in population $p$ (denoted by agent-$(i,p)$ hereafter) relative to the performance of population $k$, for $k=1,\ldots, m$.
Moreover, we allow the liability function $F^p$ to be path-dependent on the stock price,
which is a natural extension of the previous case. As we will see, this generalization necessitates that the stock price transition probabilities also 
become path-dependent to achieve the MC-MFE. 
Note that, without loss of generality, we can assume that the liability depends only on the terminal values of $(Y_N,Z_N^{i,p})$, unlike the stock price. 
This simplification is justified because any path-dependent effects involving these factors can always be incorporated 
by lifting the underlying processes to higher dimensions.

Let us start by preparing the appropriate probability spaces. We use the same setup and notation as in Section~\ref{sec-setup-single},  
for the complete filtered probability space $(\Omega^0,\calf^0,(\calf^0_{t_n})_{n=0}^N, \mbb{P}^0)$.
The filtration $(\calf^0_{t_n})_{n=0}^N$ is generated by the stock price process $S:=(S_n:=S(t_n))_{n=0}^N$ 
and the common noise process $Y:=(Y_n:=Y(t_n))_{n=0}^N$. The notations for their ranges $(\cals_n, \cals^n, \caly_n, \caly^n)$, 
$0\leq n\leq N$, remain identical to those defined in Section~\ref{sec-setup-single}. Specifically, we continue to assume 
that the trajectories of the stock price are confined to the binomial tree, and that the process $Y$ possesses a finite state space.
We also use the same notation $\wt{R}_n:=S_n/S_{n-1}$, $R_n:=\wt{R}_n-\exp(r\Del)$, $u:=\wt{u}-\exp(r\Del)$, $d:=\wt{d}-\exp(r\Del)$,
and $\beta:=\exp(r\Del)$ as before.  To model heterogeneity across the populations, we introduce a countably infinite number of 
complete filtered probability spaces $(\Omega^{i,p}, \calf^{i,p},(\calf^{i,p}_{t_n})_{n=0}^N, \mbb{P}^{i,p}), i=1,2,\ldots$
for each population $p=1,2,\ldots, m$.  Here, $(\Omega^{i,p}, \calf^{i,p},(\calf^{i,p}_{t_n})_{n=0}^N, \mbb{P}^{i,p})$
is the space modeling the idiosyncratic variables and shocks specific to agent-$(i,p)$:
$\xi_i^p$ is the initial wealth and $\gamma_i^p$  is the risk aversion of the agent,
both of which are $\calf^{i,p}_0$-measurable. We also define $\gamma_n^{i,p}:=(\beta^N/\beta^n)\gamma_i^p$, $0\leq n\leq N$, for notational simplicity;
$Z^{i,p}:=(Z^{i,p}_n:=Z^{i,p}(t_n))_{n=0}^N$ denotes the $d_{Z^p}$-dimensional $(\calf^{i,p}_{t_n})_{n=0}^N$-adapted idiosyncratic shock process to the agent.
We denote the range of $Z_n^{i,p}$ by $\calz_n^p~(\subset \mbb{R}^{d_{Z^p}})$.
To represent a network of relative performance concerns, we introduce $\calf^{i,p}_0$-measurable 
real random variables $\theta^i_{p,k}$, $k=1,2,\ldots,m$. Here, $\theta^i_{p,k}$ denotes
the relative concern of agent-$(i,p)$ in population $p$ regarding the average performance of population $k$.

By standard procedures, we define
\be
(\Omega,\calf,(\calf_{t_n})_{n=0}^N, \mbb{P}):=(\Omega^0,\calf^0,(\calf^0_{t_n})_{n=0}^N, \mbb{P}^0)
\otimes_{p=1}^m\otimes_{i=1}^\infty (\Omega^{i,p},\calf^{i,p},(\calf^{i,p}_{t_n})_{n=0}^N, \mbb{P}^{i,p}) \nn
\ee
as the complete filtered probability space describing the entire environment of the $m$-population model.
On the other hand, the relevant probability space for each agent-$(i,p)$ is given by
\be
(\Omega^{0,(i,p)}, \calf^{0,(i,p)}, (\calf_{t_n}^{0,(i,p)})_{n=0}^N, \mbb{P}^{0,(i,p)}):=
(\Omega^0, \calf^0, (\calf^0_{t_n})_{n=0}^N, \mbb{P}^0)\otimes (\Omega^{i,p},\calf^{i,p},(\calf^{i,p}_{t_n})_{n=0}^N, \mbb{P}^{i,p}).\nn
\ee
Expectations with respect to $\mbb{P}^{0}$, $\mbb{P}^{i,p}$, $\mbb{P}^{0,(i,p)}$ and $\mbb{P}$
are denoted by $\ex^0[\cdot]$, $\ex^{i,p}[\cdot]$, $\ex^{0,(i,p)}$ and $\ex[\cdot]$, respectively.
We use the same conventions for conditional expectations as in Section~\ref{sec-setup-single}.

\begin{assumption}
\label{assumption-multi-1}
{\rm (i):} $\wt{u}$ and $\wt{d}$ are real constants satisfying $0<\wt{d}<\exp(r\Del)<\wt{u}<\infty$.\\
{\rm (ii):} The variables $(\xi_i^p, \gamma_i^p, (\theta^i_{p,k})_{k=1}^m, Z^{i,p})$ are identically distributed across all agents $i=1,2,\ldots$
within each population $p=1,\ldots,m$. \\
{\rm (iii):} For each population $p=1,\ldots,m$, there exist real constants $\ul{\xi}^p, \ol{\xi}^p, \ul{\gamma}^p, \ol{\gamma}^p$,
and $\ul{\theta}^p, \ol{\theta}^p$ such that for every $i\in \mbb{N}$,
\be
\xi_i^p\in [\ul{\xi}^p,\ol{\xi}^p]\subset \mbb{R}, \qquad \vr_i^p:=(\gamma_i^p, (\theta^i_{p,k})_{k=1}^m)\in \Gamma^p:=
[\ul{\gamma}^p,\ol{\gamma}^p]\times [\ul{\theta}^p,\ol{\theta}^p]^m\subset (0,\infty)\times \mbb{R}^m. \nn
\ee
{\rm (iv):} For each $(i,p)$, $i \in \mbb{N}, p=1,\ldots, m$, the process $Z^{i,p}$ is Markovian, i.e., 
$\ex^{i,p}[f(Z^{i,p}_n)|\calf^{i,p}_{t_k}]=\ex^{i,p}[f(Z^{i,p}_n)|Z^{i,p}_k]$ for every bounded measurable function $f$ on $\calz_n^p$
and $k\leq n$. \\
{\rm (v):} The process $Y$ is Markovian i.e., $\ex^0[f(Y_n)|\calf^0_{t_k}]=\ex^0[f(Y_n)|Y_k]$ for every bounded measurable
function $f$ on $\caly_n$ and $k\leq n$. \\
{\rm (vi):} The transition probabilities of $S=(S_n)_{n=0}^N$ satisfy, for every $0\leq n\leq N-1$, a.s.,
\be
\begin{split}
&\mbb{P}^0(S_{n+1}=\wt{u}S_n|\calf^0_{t_n})=\mbb{P}^0(S_{n+1}=\wt{u}S_n|\bS^n,Y_n)=:p_n(\bS^n,Y_n), \\
&\mbb{P}^0(S_{n+1}=\wt{d}S_n|\calf^0_{t_n})=\mbb{P}^0(S_{n+1}=\wt{d}S_n|\bS^n,Y_n)=:q_n(\bS^n,Y_n), \nn
\end{split}
\ee
where $p_n, q_n~(:=1-p_n): \cals^n\times \caly_n\rightarrow \mbb{R}$, $0\leq n\leq N-1$ are bounded measurable 
functions satisfying
\be
0<p_n(\bs,y), q_n(\bs,y)<1 \nn
\ee
for every $(\bs,y)\in \cals^n\times\caly_n$.
\end{assumption}

The transition probabilities of the stock price are now allowed to depend on its past trajectory.
Under conditions (v) and (vi) above, $(S_{n+1},Y_{n+1})$ now satisfy the property:
\be
\ex^0[f(S_{n+1})g(Y_{n+1})|\calf^0_{t_n}]=\ex^0[f(S_{n+1})|\bS^n,Y_n]\ex^0[g(Y_{n+1})|Y_n]~{\rm a.s.,} \quad 0\leq n\leq N-1, \nn
\ee
for any bounded measurable functions $f:\cals_{n+1}\rightarrow \mbb{R}$ and $g:\caly_{n+1}\rightarrow \mbb{R}$.
\begin{remark}
The  risk-neutral measure $\mbb{Q}$ remains the same as in the previous section, in which 
the transition probability of the up-move is $p^{\mbb{Q}}=(-d)/(u-d)$ and the down-move $q^{\mbb{Q}}=u/(u-d)$.
The bound on the transition probabilities in Assumption~\ref{assumption-multi-1} (vi)
guarantees that the probability measures $\mbb{P}^0\circ S^{-1}$ and $\mbb{Q}\circ S^{-1}$ are equivalent. 
Hence our system is arbitrage free.
\end{remark}

\subsection{Optimization problem}

We now formulate the optimization problem for each agent. Each agent-$(i,p)$ in population $p$,  
with initial wealth $\xi_i^p$, engages in self-financing trading with the risk-free money market account and
the single risky stock. The agent adopts an $(\calf^{0,(i,p)}_{t_n})_{n=0}^N$-adapted trading strategy $(\phi^{i,p}_n)_{n=0}^{N-1}$,
representing the cash amount invested  in the stock at time $t_n$.
The associated wealth process of agent-$(i,p)$, $X^{i,p}:=(X_n^{i,p}:=X^{i,p}(t_n))_{n=0}^N$, follows the dynamics
\be
\begin{split}
X^{i,p}_{n+1}&=\exp(r\Del)(X_n^{i,p}-\phi^{i,p}_n)+\phi^{i,p}_n\wt{R}_{n+1} \\
&=\beta X_n^{i,p}+\phi^{i,p}_n R_{n+1}, \nn
\end{split}
\ee
with $X_0^{i,p}=\xi_i^p$. 

We suppose that each agent-$(i,p)$ solves the following optimization problem:
\be
\label{problem-multi}
\sup_{(\phi^{i,p}_n)_{n=0}^{N-1}\in \mbb{A}^{i,p}}\ex^{0,(i,p)}
\left[-\exp\Bigl(-\gamma_i^p\Bigl(X_N^{i,p}-\sum_{k=1}^m \theta^i_{p,k}\mu_N^k(\bS^N,\bY^{N-1})-F^p(\bS^N,Y_N,Z_N^{i,p})\Bigr)
\Bigr)\bigg|\calf_0^{0,(i,p)}\right], 
\ee
where
\be
\mbb{A}^{i,p}:=\{(\phi^{i,p}_n)_{n=0}^{N-1}: \phi^{i,p}_n~\text{is an $\calf^{0,(i,p)}_{t_n}$-measurable real-valued random variable}\} \nn
\ee
denotes the admissible control space.
$\mu^p_N:\cals^N\times \caly^{N-1}\rightarrow \mbb{R}$, $p=1,\ldots, m$, are  measurable functions
denoting the average wealth of the population $p$. We seek to find a fixed point $(\mu_n^p)_{n=0}^N$ with $\mu_0^p:=\ex^{i,p}[\xi_i^p]$ 
and $\mu_n^p:\cals^n\times \caly^{n-1}\rightarrow \mbb{R}$, $1\leq n\leq N$,  
satisfying
\be
\mu_n^p (\bS^n,\bY^{n-1})=\ex^{0,(i,p)}\bigl[X_n^{i,p}|\calf^0_{t_n}\bigr]~{\rm a.s.}\nn
\ee  
for every population $p=1,\ldots, m$.

For notational convenience in the subsequent analysis, let us introduce matrix notations.
We define the interaction matrix $\Theta \in \mbb{R}^{m\times m}$ representing a network of 
aggregate relative performance concerns among populations, whose $(p,k)$-th entry is given by the expected concern:
\be
\Theta_{p,k} := \ex^{i,p}[\theta^i_{p,k}], \quad 1\leq p, k\leq m, \nn
\ee
which is independent of the specific agent $i\in \mbb{N}$ due to the i.i.d. assumption.
For any $m$-dimensional vector $\bm{\mu}=(\mu^1, \ldots, \mu^m)^\top\in \mbb{R}^m$, 
we adopt the notation:
\be
\bigl(\theta^i\bm{\mu}\bigr)_p:=\sum_{k=1}^m \theta^i_{p,k}\mu^k, \quad \bigl(\Theta \bm{\mu}\bigr)_p:=\sum_{k=1}^m \Theta_{p,k}\mu^k. \nn
\ee

\begin{assumption}
\label{assumption-multi-2} 
For each $1\leq p\leq m$, the following conditions hold:\\
{\rm (i):} The function $F^p:\cals^N\times \caly_N\times \calz_N^p\rightarrow \mbb{R}$ is measurable and bounded. \\
{\rm (ii):} Every agent-$(i,p)$, $i=1,2,\ldots$, is a price-taker in the sense that they consider the stock price process (and 
hence its transition probabilities specified in Assumption~\ref{assumption-multi-1} (vi)) to be exogenously determined by 
the collective actions of the others and unaffected by the agent's own trading strategies. \\
{\rm (iii):} Every agent-$(i,p)$, $i=1,2,\ldots$, treats the mean-field terms $(\mu_n^k)_{n=0}^N, k=1,\ldots,m$,
where $\mu_n^k:\cals^n\times \caly^{n-1}\rightarrow \mbb{R}$, as exogenous bounded measurable functions, believing that 
they are determined by the collective actions of the agents in the population $k$ and unaffected by the agent's
own trading strategies.\footnote{As noted in Assumption~\ref{assumption-single-2}, since the domains are finite, 
the boundedness assumption is theoretically redundant but kept for clarity.}
\end{assumption}

Following the approach in the previous section, for $1\leq n\leq N$, we analyze the one-period problem at $t=t_{n-1}$
for the interval $[t_{n-1},t_n]$:
\be
\label{problem-multi-tmp}
\sup_{\phi^{i,p}_{n-1}}\ex^{0,(i,p)}\left[-\exp\Bigl(-\gamma_n^{i,p}\bigl(X_n^{i,p}-(\theta^i \bm{\mu}_n)_p (\bS^n,\bY^{n-1})\bigr)
\Bigr)V_n^p(\bS^n,Y_n,Z_n^{i,p},\vr_i^p)\bigg|\calf_{t_{n-1}}^{0,(i,p)}\right]
\ee
where the supremum is taken over the $\calf^{0,(i,p)}_{t_{n-1}}$-measurable real-valued random variables.
We recall that $\gamma_n^{i,p}:=(\beta^N/\beta^n)\gamma_i^p$ and $\vr_i^p:=(\gamma_i^p, (\theta^i_{p,k})_{k=1}^m)$.

\begin{lemma}
\label{lemma-multi-tmp}
Suppose that Assumption~\ref{assumption-multi-1} and Assumption~\ref{assumption-multi-2} (ii) and (iii) hold.
Furthermore, assume that $V_n^p:\cals^n\times \caly_n\times \calz_n^p\times \Gamma^p\rightarrow \mbb{R}$
is a measurable function satisfying the uniform bounds $0<c_n\leq V_n^p\leq C_n<\infty$ on its
domain with some positive constants $c_n$ and $C_n$.
Then, for every $p=1,\ldots, m$, the problem $(\ref{problem-multi-tmp})$ admits a unique optimal solution $\phi^{(i,p),*}_{n-1}$
given by a bounded measurable function $\phi^{(i,p),*}_{n-1}:\cals^{n-1}\times \caly^{n-1}\times\calz_{n-1}^p\times \Gamma^p
\rightarrow \mbb{R}$, such that $\phi^{(i,p),*}_{n-1}:=\phi^{(i,p),*}_{n-1}(\bS^{n-1}, \bY^{n-1}, Z_{n-1}^{i,p},\vr_i^p)$ a.s., where
\be
\label{multi-optimal-tmp}
\begin{split}
\phi^{(i,p),*}_{n-1}(\bs,\by,z^{i,p},\vr_i^p)&:=\frac{(\theta^i\bm{\Del}_n)_p(\bs,\by)}{u-d} 
+\frac{1}{\gamma_n^{i,p}(u-d)}\Bigl\{\log\Bigl(-\frac{p_{n-1}(\bs,y) u}{q_{n-1}(\bs,y)d}\Bigr)
+\log f_{n-1}^p(\bs,y,z^{i,p},\vr_i^p)\Bigr\}.
\end{split}
\ee
Here, $y=y_{n-1}\in \caly_{n-1}$ is the last element of $\by\in \caly^{n-1}$. 
Furthermore, $f_{n-1}^p:\cals^{n-1}\times \caly_{n-1}\times \calz_{n-1}^p\times \Gamma^p\rightarrow \mbb{R}$ 
is a measurable function satisfying the uniform bounds $0<\ol{c}_n\leq f_{n-1}^p\leq \ol{C}_n<\infty$
on its domain for some positive constants $\ol{c}_n$ and $\ol{C}_n$, and $\bm{\Del}_n:=(\Del_n^k)_{k=1}^m$, $\Del_n^k:\cals^{n-1}\times
\caly^{n-1}\rightarrow \mbb{R}$, $1\leq k\leq m$ are bounded measurable functions. They are defined respectively by
\be
\begin{split}
&f_{n-1}^p(\bs,y,z^{i,p},\vr^p_i):=\frac{\ex^{0, (i,p)}[V_n^p((\bs\wt{u})^n,Y_n,Z_n^{i,p},\vr_i^p)|y,z^{i,p},\vr_i^p]}
{\ex^{0,(i,p)}[V_n^p((\bs\wt{d})^n,Y_n,Z_n^{i,p},\vr_i^p)|y,z^{i,p}, \vr_i^p]}, \\
&\Del_n^k(\bs,\by):=\mu_n^k((\bs\wt{u})^n,\by)-\mu_n^k((\bs\wt{d})^n,\by). \nn
\end{split}
\ee
\end{lemma}
\begin{proof}
We solve the problem on each set $\{\omega^{0,(i,p)}\in \Omega^{0,(i,p)}:
(X_{n-1}^{i,p},\bS^{n-1},\bY^{n-1}, Z_{n-1}^{i,p},\vr_i^p)=(x^{i,p},\bs,\by,z^{i,p},\vr_i^p)\}$.
Here, with a slight abuse of notation, we use the same symbols for the realizations of $\calf^{i,p}_0$-measurable
random variables. We also set  $y=y_{n-1}$, i.e., the last element of $\by\in \caly^{n-1}$. 
The problem $(\ref{problem-multi-tmp})$ can be rewritten equivalently as
\be
\begin{split}
\label{multi-value-tmp}
&\hspace{-5mm}\inf_{\phi^{i,p}\in \mbb{R}}\ex^{0,(i,p)}\Bigl[\exp\Bigl(-\gamma_n^{i,p}(\beta x^{i,p}+\phi^{i,p} R_n-(\theta^i\bm{\mu}_n)_p(\bS^n,
\bY^{n-1})\Bigr)V_n^p(\bS^n,Y_n,Z_n^{i,p},\vr_i^p)|\bs,\by,z^{i,p},\vr_i^p\Bigr] \\
&=\exp\bigl(-\gamma_{n-1}^{i,p} x^{i,p}\bigr) \inf_{\phi^{i,p}} \Bigl\{ \\
&p_{n-1}(\bs,y)\exp\Bigl(-\gamma_n^{i,p}\bigl(\phi^{i,p} u-(\theta^i\bm{\mu}_n)_p((\bs\wt{u})^n,\by)\bigr)\Bigr)
\ex^{0,(i,p)}\bigl[V_n^p((\bs\wt{u})^n,Y_n,Z_n^{i,p},\vr_i^p)|y,z^{i,p},\vr_i^p\bigr]\\ 
&\hspace{-3mm}+q_{n-1}(\bs,y)\exp\Bigl(-\gamma_n^{i,p}\bigl(\phi^{i,p} d-(\theta^i\bm{\mu}_n)_p((\bs\wt{d})^n,\by)\bigr)\Bigr)
\ex^{0,(i,p)}\bigl[V_n^p((\bs\wt{d})^n,Y_n,Z_n^{i,p},\vr_i^p)|y,z^{i,p},\vr_i^p\bigr]\Bigr\},  
\end{split}
\ee
where we have used property (vi) in Assumption~\ref{assumption-multi-1}. Given that $\gamma_n^{i,p}>0$
and $d<0<u$,  the optimal trade position $\phi^{(i,p),*}$ is uniquely characterized by the first-order condition:
\be
\begin{split}
&p_{n-1}(\bs,y) u \exp\Bigl(-\gamma_n^{i,p}\bigl(\phi^{i,p} u-(\theta^i\bm{\mu}_n)_p((\bs\wt{u})^n,\by)\bigr)\Bigr)
\ex^{0,(i,p)}\bigl[V_n^p((\bs\wt{u})^n,Y_n,Z_n^{i,p},\vr_i^p)|y,z^{i,p},\vr_i^p\bigr]\\
&\quad +q_{n-1}(\bs,y)d \exp\Bigl(-\gamma_n^{i,p}\bigl(\phi^{i,p} d-(\theta^i\bm{\mu}_n)_p((\bs\wt{d})^n,\by)\bigr)\Bigr)
\ex^{0,(i,p)}\bigl[V_n^p((\bs\wt{d})^n,Y_n,Z_n^{i,p},\vr_i^p)|y,z^{i,p},\vr_i^p\bigr]=0, \nn
\end{split}
\ee
which yields the desired result. The existence of the uniform bounds on $f_{n-1}^p$ is
a direct consequence of its definition and the uniform bounds on $V_n^p$.
\end{proof}

\subsection{Relative performance mean-field equilibrium}

We now study the mean-field equilibrium in a network of relative performance concerns among the $m$ populations.
\begin{definition}
\label{def-multi-rp-mfe}
We say that the system is in the relative performance mean-field equilibrium (RP-MFE) if the problem $(\ref{problem-multi})$
admits an optimal solution $(\phi^{(i,p),*}_{n-1})_{n=1}^N$, $1\leq p\leq m$, $i=1,2,\ldots$,
for agents satisfying Assumptions~\ref{assumption-multi-2} (ii) and (iii), such that, 
with $\mu_0^p:=\ex^{i,p}[\xi_i^p]$, the bounded measurable functions
$\mu_n^p: \cals^n\times \caly^{n-1}\rightarrow \mbb{R}$, $1\leq n\leq N$,  satisfy
the fixed point condition:
\be
\label{def-multi-mu}
\mu_n^p(\bS^n,\bY^{n-1})=\ex^{0,(i,p)}\bigl[X_n^{(i,p),*}|\calf^0_{t_n}\bigr]~{\rm a.s.,} \quad  1\leq n\leq N,  
\ee
for every population $p=1,\ldots, m$. Here,  $X^{(i,p),*}$ denotes the wealth process of agent-$(i,p)$ associated with the optimal control $\phi^{(i,p),*}$.
Moreover, we denote the associated processes in the RP-MFE by $(\wh{\mu}^p, \wh{\phi}^{i,p})$, $1\leq p\leq m$, and refer to them
as the solution pairs of the RP-MFE. 
For each $p=1,\ldots, m$ and $i=1,2,\ldots$, we also use the symbol $\wh{X}^{i,p}$ to 
represent the wealth process of agent-$(i,p)$ associated with $\wh{\phi}^{i,p}$ and the initial condition $\xi_i^p$.
\end{definition}

Since we are dealing with a symmetric problem with i.i.d. variables and processes $(\xi_i^p,\gamma_i^p,\theta^i_{p,k}, Z^{i,p})$, $i\in \mbb{N}$,
the choice of  a representative agent-$(i,p)$ in each population $p=1,\ldots, m$ is arbitrary. 
Given that filtration $(\calf^0_{t_k})_{k=0}^N$ is generated by $(S,Y)$, the fixed-point condition $(\ref{def-multi-mu})$
can be equivalently represented by using agent-$(1,p)$ as the representative:
\be
\mu_n^p(\bs,\by^-)=\ex^{1,p}\bigl[X_n^{(1,p),*}|\bs,\by\bigr]=\ex^{1,p}\bigl[X_n^{(1,p), *}|\bs,\by^-\bigr], \nn
\ee
for every $(\bs,\by^-)\in \cals^n\times \caly^{n-1}$. Here, $\by=(\by^-,y_n)\in \caly^n$.
In the following, we prove the existence and uniqueness of the RP-MFE. We will show that the 
path-dependence of $\phi^{(i,p),*}_{n-1}$ on $\by \in \caly^{n-1}$ in Lemma~\ref{lemma-multi-tmp} reduces to a dependence 
on the current state $y=y_{n-1} \in \caly_{n-1}$ in the RP-MFE.

\begin{theorem}
\label{th-multi-rp}
Suppose that Assumptions~\ref{assumption-multi-1} and $\ref{assumption-multi-2}$ hold.
We also assume that  the $m\times m$ matrix $(I-\Theta)$, with entries defined by 
$\{( I-\Theta)_{p, k} := (\delta_{p,k}-\ex^{1,p}[\theta^1_{p,k}]), 1\leq p, k\leq m\}$, has a bounded inverse $(I-\Theta)^{-1}$. 
Then, the problem $(\ref{problem-multi})$ admits a unique RP-MFE with the solution 
pairs $(\wh{\mu}^p, \wh{\phi}^{i,p})_{p=1}^m$. For each population $p=1,\ldots, m$, the associated optimal strategy $(\wh{\phi}^{i,p}_{n-1})_{n=1}^N$
is given by the bounded measurable function $\wh{\phi}^{i,p}_{n-1}: \cals^{n-1}\times \caly_{n-1}\times \calz_{n-1}^p
\times \Gamma^p\rightarrow \mbb{R}$, $1\leq n\leq N$, such that
\be
\label{multi-optimal}
\begin{split}
&\wh{\phi}^{i,p}_{n-1}(\bs,y,z^{i,p},\vr_i^p):=\frac{1}{\gamma_n^{i,p}(u-d)}\Bigl\{\log\Bigl(-\frac{p_{n-1}(\bs,y)u}{q_{n-1}(\bs,y)d}\Bigr)
+\log f_{n-1}^p(\bs,y,z^{i,p},\vr_i^p)\Bigr\} \\
&\hspace{-8mm}+\frac{1}{u-d}\sum_{k=1}^m(\theta^i(I-\Theta)^{-1})_{p,k}\Bigl\{ \log\Bigl(-\frac{p_{n-1}(\bs,y) u}{q_{n-1}(\bs,y)d}\Bigr)
\ex^{1,k}\Bigl[\frac{1}{\gamma_n^{1,k}}\Bigr]+\ex^{1,k}\Bigl[\frac{\log f_{n-1}^k(\bs,y,Z_{n-1}^{1,k},\vr_i^k)}{\gamma_n^{1,k}}\Bigr]\Bigr\}
\end{split}
\ee
and the dynamics of the associated mean-field terms $(\wh{\mu}_n^p)_{n=1}^N$ is given by
\be
\label{multi-rp-mfe}
\begin{split}
&\wh{\mu}_n^p((\bs\wt{u})^n,\by)=\beta \wh{\mu}^p_{n-1}(\bs,\by^-)+u\ex^{1,p}\bigl[\wh{\phi}^{1,p}_{n-1}(\bs,y,Z_{n-1}^{1,p},\vr_1^p)\bigr], \\
&\wh{\mu}_n^p((\bs\wt{d})^n, \by)=\beta \wh{\mu}^p_{n-1}(\bs,\by^-)+d\ex^{1,p}\bigl[\wh{\phi}^{1,p}_{n-1}(\bs,y, Z_{n-1}^{1,p},\vr_1^p)\bigr], 
\end{split}
\ee
with the initial condition $\wh{\mu}^p_0:=\ex^{1,p}[\xi_1^p]$ and
\be
\label{multi-rp-exp}
\begin{split}
&\ex^{1,p}\bigl[\wh{\phi}^{1,p}_{n-1}(\bs,y,Z_{n-1}^{1,p},\vr_1^p)\bigr]\\
&\quad=\frac{1}{u-d}\sum_{k=1}^m (I-\Theta)^{-1}_{p,k}\Bigl\{\log\Bigl(-\frac{p_{n-1}(\bs,y)u}{q_{n-1}(\bs,y)d}\Bigr)
\ex^{1,k}\Bigl[\frac{1}{\gamma_n^{1,k}}\Bigr]+\ex^{1,k}\Bigl[\frac{\log f_{n-1}^k(\bs,y,Z_{n-1}^{1,k},\vr_1^k)}{\gamma_n^{1,k}}\Bigr]\Bigr\}. 
\end{split}
\ee
Here, $y:=y_{n-1}$ is the last element of $\by\in \caly^{n-1}$, 
and $\by^-$ denotes the history up to $t_{n-2}$ such that $(\by^{-}, y_{n-1})=\by$.
For each $p=1,\ldots, m$, the function $f_{n-1}^p:\cals^{n-1}\times \caly_{n-1}\times \calz_{n-1}^p\times \Gamma^p\rightarrow \mbb{R}$
is defined as in Lemma~\ref{lemma-multi-tmp}. It satisfies the uniform bounds $0<\ol{c}_n\leq f_{n-1}^p\leq \ol{C}_n<\infty$
on its domain for some positive constants $\ol{c}_n$ and $\ol{C}_n$, and is given by
\be
\label{multi-fnm1-update}
f_{n-1}^p(\bs,y,z^{i,p},\vr_i^p):=\frac{\ex^{0,(i,p)}[V_n^p((\bs\wt{u})^n,Y_n,Z_n^{i,p},\vr_i^p)|y,z^{i,p},\vr_i^p]}
{\ex^{0,(i,p)}[V_n^p((\bs\wt{d})^n,Y_n,Z_n^{i,p},\vr_i^p)|y,z^{i,p},\vr_i^p]}. 
\ee
Here, for each $p=1,\ldots, m$, $V_n^p:\cals^n\times \caly_n\times\calz_n^p\times \Gamma^p\rightarrow \mbb{R}$, $0\leq n\leq N$,
are measurable functions satisfying the uniform bounds $0<c_n\leq V_n^p\leq C_n<\infty$ on their respective domains
with some positive constants $c_n$ and $C_n$.
They are defined recursively by 
$
V_N^p(\bs,y,z^{i,p},\vr_i^p):=\exp\bigl(\gamma_i^p F^p(\bs,y,z^{i,p})\bigr), \nn
$
for each $(\bs,y,z^{i,p},\vr_i^p)\in \cals^N\times \caly_N\times \calz_N^p\times \Gamma^p$, and
\be
\label{multi-rp-Vnm1}
\begin{split}
V_{n-1}^p(\bs,y,z^{i,p},\vr_i^p)&=p_{n-1}(\bs,y)\exp\Bigl(-\gamma_n^{i,p} u \bigl(\wh{\phi}^{i,p}_{n-1}(\bs,y,z^{i,p},\vr_i^p)
-\sum_{k=1}^m \theta^i_{p,k}\ex^{1,k}[\wh{\phi}^{1,k}_{n-1}(\bs,y,Z_{n-1}^{1,k},\vr_1^k)]\bigr)\Bigr) \\
&\qquad\qquad \times  \ex^{0,(i,p)}\bigl[V_n^p((\bs\wt{u})^n,Y_n,Z_n^{i,p},\vr_i^p)|y,z^{i,p},\vr_i^p\bigr]\\
&+q_{n-1}(\bs,y)\exp\Bigl(-\gamma_n^{i,p} d \bigl(\wh{\phi}^{i,p}_{n-1}(\bs,y,z^{i,p},\vr_i^p)
-\sum_{k=1}^m \theta^i_{p,k}\ex^{1,k}[\wh{\phi}^{1,k}_{n-1}(\bs,y,Z_{n-1}^{1,k},\vr_1^k)]\bigr)\Bigr) \\
&\qquad\qquad \times  \ex^{0,(i,p)}\bigl[V_n^p((\bs\wt{d})^n,Y_n,Z_n^{i,p},\vr_i^p)|y,z^{i,p},\vr_i^p\bigr] 
\end{split}
\ee
for each $(\bs,y,z^{i,p},\vr_i^p)\in \cals^{n-1}\times\caly_{n-1}\times \calz_{n-1}^p\times \Gamma^p$, $1\leq n\leq N$.
\end{theorem}
\begin{remark*}
Note that $(1/\gamma_n^{i,p})\log V_n^p$ represents the effective liability for agent-$(i,p)$ at $t=t_n$.
\end{remark*}
\begin{proof}
Suppose that, for every $p=1,\ldots, m$,  the problem for agent-$(i,p)$ at $t_{n-1}$ for the period $[t_{n-1},t_n]$ takes the 
form $(\ref{problem-multi-tmp})$. To apply Lemma~\ref{lemma-multi-tmp}, we hypothesize that  $V_n^p:\cals^n\times \caly_n\times\calz_n^p\times\Gamma^p\rightarrow \mbb{R}$
is a measurable function satisfying the uniform bounds $c_n\leq V_n^p\leq C_n$ on its domain 
for some positive constants $c_n$ and $C_n$. This clearly holds at $t=t_{N-1}$ for the last interval $[t_{N-1},t_N]$
with  $V_N^p(\bs,y,z^{i,p},\vr_i^p):=\exp\bigl(\gamma_i^p F^p(\bs,y,z^{i,p})\bigr)$.

Using $(\ref{multi-optimal-tmp})$ in Lemma~\ref{lemma-multi-tmp}, the evolution of the wealth process of agent-$(i,p)$
under the optimal strategy $\phi^{(i,p),*}_{n-1}$ is given by
\be
X_n^{(i,p),*}=\beta X_{n-1}^{(i,p),*}+\phi^{(i,p),*}_{n-1}(\bS^{n-1},\bY^{n-1},Z_{n-1}^{i,p},\vr_i^p) R_n. \nn
\ee
Consider the problem conditioned on the event $\{(\bS^{n-1},\bY^{n-1})=(\bs,\by)\}$ in $\calf^0_{t_{n-1}}$.
Under the induction hypothesis that $\ex^{0,(1,p)}[X^{(1,p),*}_{n-1}|\bs,\by]$ is given by the form $\mu_{n-1}^p(\bs,\by^-)$
with $(\bs,\by^-)\in \cals^{n-1}\times \caly^{n-2}$ (when $n=1$, we simply set $\mu_0^p=\ex^{1,p}[\xi_1^p]$),
 and using the i.i.d. property of the variables $(\vr_i^p, Z^{i,p}), i=1,2,\ldots$, the fixed point condition  $(\ref{def-multi-mu})$ 
for the RP-MFE for the period $[t_{n-1},t_n]$ is equivalently given by the following 
evolution equations:
\be
\label{multi-RP-MFE-eq}
\begin{split}
&\mu_n^p((\bs\wt{u})^n,\by)=\beta \mu_{n-1}^p(\bs,\by^-)+u\ex^{1,p}\bigl[\phi^{(1,p),*}_{n-1}(\bs,\by,Z_{n-1}^{1,p},\vr_1^p)\bigr], \\
&\mu_n^p((\bs\wt{d})^n,\by)=\beta \mu_{n-1}^p(\bs,\by^-)+d\ex^{1,p}\bigl[\phi^{(1,p),*}_{n-1}(\bs,\by,Z_{n-1}^{1,p},\vr_1^p)\bigr], 
\end{split}
\ee
for $p=1,\ldots, m$. To solve the above equations, notice that $\phi^{(i,p),*}_{n-1}$ in $(\ref{multi-optimal-tmp})$
depends on the mean-field terms $\mu_n^k, k=1,\ldots m$ only through their differences $\Del_n^k(\bs,\by), k=1,\ldots, m$.
Taking the difference in $(\ref{multi-RP-MFE-eq})$, we obtain the simultaneous equations for $(\Del_n^p(\bs,\by))_{p=1}^m$ as
\be
\label{multi-rp-mfe-consistency}
\begin{split}
\Del_n^p(\bs,\by)&=\sum_{k=1}^m \Theta_{p,k}\Del_n^k(\bs,\by)\\
&+\Bigl\{ \log\Bigl(-\frac{p_{n-1}(\bs,y)u}{q_{n-1}(\bs,y)d}\Bigr)\ex^{1,p}\Bigl[\frac{1}{\gamma_n^{1,p}}\Bigr]
+\ex^{1,p}\Bigl[\frac{\log f_{n-1}^p(\bs,y,Z_{n-1}^{1,p},\vr_1^p)}{\gamma_n^{1,p}}\Bigr]\Bigr\}.
\end{split}
\ee
Since $(I-\Theta)$ is invertible by assumption, this equation determines $(\Del_n^p)_{p=1}^m$ uniquely as
\be
\label{multi-Deln}
\Del_n^p(\bs,y)=\sum_{k=1}^m (I-\Theta)^{-1}_{p,k}\Bigl\{ \log\Bigl(-\frac{p_{n-1}(\bs,y)u}{q_{n-1}(\bs,y)d}\Bigr)
\ex^{1,k}\Bigl[\frac{1}{\gamma_n^{1,k}}\Bigr]
+\ex^{1,k}\Bigl[\frac{\log f_{n-1}^k(\bs,y,Z_{n-1}^{1,k},\vr_1^k)}{\gamma_n^{1,k}}\Bigr]\Bigr\} 
\ee
for every $(\bs,y)\in \cals^{n-1}\times\caly_{n-1}$, $p=1,\ldots, m$. Note that $\bm{\Del}_n$ turns out to be independent of the 
entire trajectory of $Y$ up to time $t_{n-2}$, depending only on the current state $y=y_{n-1}$.
By substituting $\bm{\Del}_n(\bs,\by)$ in $(\ref{multi-optimal-tmp})$ with $\bm{\Del}_n(\bs,y)$ in $(\ref{multi-Deln})$, 
we obtain the desired expression $\phi^{(i,p),*}_{n-1}$ as in $(\ref{multi-optimal})$
and hence its expectation as in $(\ref{multi-rp-exp})$, which are also independent of the trajectory of $Y$ up to time $t_{n-2}$.
The measurable functions $f_{n-1}^p, p=1,\ldots m$ defined by $(\ref{multi-fnm1-update})$ satisfy
the uniform bounds $\ol{c}_n\leq f_{n-1}^p\leq \ol{C}_n$ for some positive constants $\ol{c}_n$ and $\ol{C}_n$
due to the uniform bounds on $(V_n^p)_{p=1}^m$.
Hence $\phi^{(i,p),*}_{n-1}$, $p=1,\ldots, m$ with expression of $(\ref{multi-optimal})$
are bounded and measurable,  and from $(\ref{multi-RP-MFE-eq})$,
$\mu_n^p$, $p=1,\ldots, m$ are also bounded and measurable as long as $\mu_{n-1}^p$, $p=1,\ldots, m$ are.
Thus, under the assumption that $\ex^{0, (1,p)}[X_{n-1}^{(1,p),*}|\bs,\by]=\mu_{n-1}^p(\bs,\by^-)$ with some bounded measurable
functions $\mu_{n-1}^p$, $(\ref{multi-optimal})$ and $(\ref{multi-rp-mfe})$ provide a unique candidate 
for the solution pairs of RP-MFE for the interval $[t_{n-1},t_n]$.

It remains to show that this procedure can be repeated backward from the last interval $[t_{N-1},t_N]$
to the first one $[t_0,t_1]$.  It follows from $(\ref{multi-value-tmp})$ that the value function for agent-$(i,p)$ at $t_{n-1}$ is 
\be
\begin{split}
&-\exp\bigl(-\gamma_{n-1}^{i,p}x^{i,p}\bigr)\Bigl\{
p_{n-1}(\bs,y)e^{-\gamma_n^{i,p}\bigl(\phi^{(i,p),*}_{n-1} u
-(\theta^i\bm{\mu}_n)_p((\bs\wt{u})^n,\by)\bigr)}\ex^{0,(i,p)}\bigl[V_n^p((\bs\wt{u})^n,Y_n,Z_n^{i,p},\vr_i^p)|y,z^{i,p},\vr_i^p\bigr]\\ 
&\qquad +q_{n-1}(\bs,y)e^{-\gamma_n^{i,p}\bigl(\phi^{(i,p),*}_{n-1} d-(\theta^i\bm{\mu}_n)_p((\bs\wt{d})^n,\by)\bigr)}
\ex^{0,(i,p)}\bigl[V_n^p((\bs\wt{d})^n,Y_n,Z_n^{i,p},\vr_i^p)|y,z^{i,p},\vr_i^p\bigr]\Bigr\}, \nn
\end{split}
\ee
for each realization $(x^{i,p},\bs,\by,z^{i,p},\vr_i^p)$ of $(X^{(i,p),*}_{n-1},\bS^{n-1}, \bY^{n-1}, Z^{i,p}_{n-1},\vr_i^p)$
with $y=y_{n-1}$. From $(\ref{multi-rp-mfe})$, the above value function becomes
\be
\label{objective-tnm1}
-\exp\Bigl(-\gamma_{n-1}^{i,p}\bigl(x^{i,p}-(\theta^i\bm{\mu}_{n-1})_p(\bs,\by^-)\bigr)\Bigr)V_{n-1}^p(\bs,y,z^{i,p},\vr_i^p), 
\ee
where $V_{n-1}^p$ is given by $(\ref{multi-rp-Vnm1})$, which once again satisfies the uniform bounds $c_{n-1}\leq V_{n-1}^p\leq C_{n-1}$
for some positive constants $c_{n-1}$ and $C_{n-1}$. Therefore, we reproduce the problem for $[t_{n-2}, t_{n-1}]$
of the same form as in $(\ref{problem-multi-tmp})$ used in Lemma~\ref{lemma-multi-tmp}.
Thus the boundedness condition on the mean-field terms $(\mu_n^p)_{n=1}^N$, $p=1,\ldots,m$
is reduced to that of $\mu_0^p:=\ex^{1,p}[\xi_1^p]$, $p=1,\ldots, m$.
Since $\xi_i^p\in [\ul{\xi}^p,\ol{\xi}^p], p=1,\ldots, m$ are bounded, all the processes $\mu^p, p=1,\ldots,m$
 are now bounded and become consistent with our assumption.
Thus we conclude that the above constructed mean-field $\mu^p$ and the optimal strategy $\phi^{(i,p),*}$
satisfy the fixed point condition $(\ref{def-multi-mu})$, and hence $(\mu^p)$ and $(\phi^{(i,p),*})$
constitute the RP-MFE solution pairs, which we denote by $(\wh{\mu}^p, \wh{\phi}^{i,p})$ for the problem $(\ref{problem-multi})$.

The uniqueness of RP-MFE immediately follows from construction: firstly, the optimal control in each interval by Lemma~\ref{lemma-multi-tmp}
for given $(\mu_n^p)$ is unique; secondly, the dynamics of the mean-field terms $(\mu_n^p)_{n\geq 1}, p=1,\ldots,m $
is uniquely determined by the evolution equations $(\ref{multi-rp-mfe})$
with a given set of initial conditions $\mu_0^p\in \mbb{R}$, $p=1,\ldots, m$.
\end{proof}

\subsection{Market-clearing mean-field equilibrium}
\label{sec-multi-mc}
As in Section~\ref{sec-single-mc}, we leverage the degrees of freedom in the transition probabilities $(p_{n-1}(\bs,y),q_{n-1}(\bs,y))$,
$1\leq n\leq N$, to establish the market-clearing mean-field equilibrium (MC-MFE).
This equilibrium analysis reveals how the inter-population network of relative performance concerns,
characterized by the matrix $\Theta$,
affects the equilibrium stock price transition probabilities, and more specifically, the equilibrium
excess return required by the agents to compensate for their risk.
In this section, we allow the external stochastic order flow at each $t=t_{n-1}$
to depend on the stock price trajectory up to $t_{n-1}$, taking the form $L_{n-1}(\bS^{n-1}, Y_{n-1})$.

\begin{assumption}
\label{assumption-multi-3}
For every $1\leq n\leq N$, $L_{n-1}:\cals^{n-1}\times\caly_{n-1}\rightarrow \mbb{R}$ is a bounded
measurable function.
\end{assumption}

As in Fujii~\cite{Fujii-Trees}, we study the mean-field market clearing as the 
large population limit while keeping the ratios of relative population size constant.
Let us denote the number of agents in population $p$ by $N_p$ and set 
$\caln:=N_1+\cdots+N_m$ as the total population size.
We use $w_p:=N_p/\caln$ to denote the relative size of population $p$.
Observe that we have the relation:
\be
\frac{1}{\caln}\sum_{p=1}^m \sum_{i=1}^{N_p}\wh{\phi}^{i,p}_{n-1}
=\sum_{p=1}^m w_p \Bigl(\frac{1}{N_p}\sum_{i=1}^{N_p}\wh{\phi}^{i,p}_{n-1}\Bigr). \nn
\ee

\begin{definition}
\label{def-multi-mc}
We say that the RP-MFE defined in Definition~\ref{def-multi-rp-mfe} constitutes a market-clearing 
mean-field equilibrium (MC-MFE) if
\be
\lim_{\caln\rightarrow \infty}\frac{1}{\caln}\sum_{p=1}^m \sum_{i=1}^{N_p}\wh{\phi}^{i,p}_{n-1}=L_{n-1}(\bS^{n-1}, Y_{n-1}), \nn
\ee
$\mbb{P}$-a.s. for every $1\leq n\leq N$, where the large population limit is taken with 
the population ratios $(w_p)_{p=1}^m$ kept constant.
\end{definition}

Let $f_{n-1}^p:\cals^{n-1}\times \caly_{n-1}\times \calz_{n-1}^p\times \Gamma^p\rightarrow \mbb{R}$, $p=1,\ldots, m$,
$1\leq n\leq N$ be the functions defined in Theorem~\ref{th-multi-rp}. 
To simplify the notation and facilitate the subsequent discussion, we introduce the following effective variables, 
assuming that the matrix $(I-\Theta)$ has a bounded inverse:
\be
\label{def-effective-variables}
\begin{split}
&\calt^{i,p}_n:=\frac{1}{\gamma_n^{i,p}}+\sum_{k=1}^m\bigl(\theta^i(I-\Theta)^{-1}\bigr)_{p,k}\ex^{1,k}\Bigl[\frac{1}{\gamma_n^{1,k}}\Bigr],  \\
&\calt_n:=\sum_{p,k=1}^m w_p (I-\Theta)^{-1}_{p,k}\ex^{1,k}\Bigl[\frac{1}{\gamma_n^{1,k}}\Bigr], \\
&\calv_{n-1}^{i,p}(\bs,y,z^{i,p},\vr_i^p):=\frac{\log f_{n-1}^p(\bs,y,z^{i,p},\vr_i^p)}{\gamma_n^{i,p}}
+\sum_{k=1}^m \bigl(\theta^i(I-\Theta)^{-1})_{p,k}\ex^{1,k}\Bigl[
\frac{\log f_{n-1}^k(\bs,y,Z^{1,k}_{n-1},\vr_1^k)}{\gamma_n^{1,k}}\Bigr], \\
&\calv_{n-1}(\bs,y):=\sum_{p,k=1}^m w_p(I-\Theta)^{-1}_{p,k}\ex^{1,k}\Bigl[
\frac{\log f_{n-1}^k(\bs,y,Z_{n-1}^{1,k},\vr_1^k)}{\gamma_n^{1,k}}\Bigr].
\end{split}
\ee
We recall that $\gamma_n^{i,p}:=(\beta^N/\beta^n)\gamma_i^p$ and $\vr_i^p:=(\gamma_i^p, (\theta^i_{p,k})_{k=1}^m)$.
Here, $\calt^{i,p}_n$, which is $\calf^{i,p}_0$-measurable, represents the \textit{effective risk tolerance} of agent-$(i,p)$ at $t_n$. 
Note that $\calt_n$ corresponds to the \textit{aggregate risk tolerance} of the market, satisfying the relation:
\be
\label{effective-conv-1}
\calt_n=\sum_{p=1}^m w_p\ex^{1,p}[\calt_n^{1,p}].
\ee
Similarly, $\calv_{n-1}^{i,p}:\cals^{n-1}\times \caly_{n-1}\times \calz_{n-1}^p\times \Gamma^p\rightarrow \mbb{R}$
is a  measurable function representing the sensitivity of the effective liability at $t_{n-1}$ for agent-$(i,p)$,
where the dependence on $\gamma_n^{i,p}$ is incorporated by the argument $\vr_i^p$. Thus, strictly speaking,
the superscript $i$ in $\calv^{i,p}_{n-1}$ is redundant, but we retain it to clearly associate the variable 
with the relevant agent.
Note also that  $\calv_{n-1}:\cals^{n-1}\times \caly_{n-1}\rightarrow \mbb{R}$ corresponds to the aggregate 
sensitivity of the effective liability, which satisfies
\be
\label{effective-conv-2}
\calv_{n-1}(\bs,y)=\sum_{p=1}^m w_p\ex^{1,p}\bigl[\calv_{n-1}^{1,p}(\bs,y,Z_{n-1}^{1,p},\vr_1^p)\bigr]. 
\ee
With these effective variables, the optimal control $(\ref{multi-optimal})$ in the RP-MFE
can be expressed as
\be
\label{optimal-with-effective}
\wh{\phi}^{i,p}_{n-1}(\bs,y,z^{i,p},\vr_i^p)=\frac{1}{u-d}\Bigl\{
\log\Bigl(-\frac{p_{n-1}(\bs,y)u}{q_{n-1}(\bs,y)d}\Bigr)\calt_n^{i,p}+\calv_{n-1}^{i,p}(\bs,y,z^{i,p},\vr_i^p)\Bigr\}. 
\ee

\begin{theorem}
\label{th-multi-mc-mfe}
Suppose that Assumptions~\ref{assumption-multi-1}, \ref{assumption-multi-2} and \ref{assumption-multi-3} hold.
Furthermore, we assume that the matrix $(I-\Theta)$ has a bounded inverse and that $\calt_N\neq 0$ (which implies $\calt_n\neq 0$ 
for all $1\leq n \leq N$). Then there exists a
unique MC-MFE. The associated equilibrium transition probabilities of the stock price are given by
\be
\label{multi-mc-transition}
\begin{split}
p_{n-1}(\bs,y)=(-d)\Big/\left\{ u \exp\left(
\frac{\displaystyle \calv_{n-1}(\bs,y)-(u-d)L_{n-1}(\bs,y)}{\displaystyle \calt_n}\right)-d\right\} 
\end{split}
\ee
for every $(\bs,y)\in \cals^{n-1}\times\caly_{n-1}$, $1\leq n\leq N$. Here, for each $1\leq n\leq N$,
the functions $f_{n-1}^p:\cals^{n-1}\times \caly_{n-1}\times \calz_{n-1}^p\times \Gamma^p\rightarrow \mbb{R}$, $p=1,\ldots,m$,
which are components of $(\calv_{n-1}^{i,k},\calv_{n-1})$,
are measurable functions satisfying the uniform bounds $0<\ol{c}_n\leq f_{n-1}^p\leq \ol{C}_n<\infty$ 
on their respective domains for some positive constants $\ol{c}_n$ and $\ol{C}_n$.
They are determined by the backward induction in Theorem~\ref{th-multi-rp},
with the transition probabilities replaced by those given above at each step. 
Under the equilibrium transition probabilities, the optimal strategy of agent-$(i,p)$ is given by, for each
$(\bs,y,z^{i,p},\vr_i^p)\in \cals^{n-1}\times\caly_{n-1}\times \calz_{n-1}^p\times \Gamma^p$,
\be
\label{multi-mc-optimal}
\wh{\phi}^{i,p}_{n-1}(\bs,y,z^{i,p},\vr_i^p)=\frac{\calt_n^{i,p}}{\calt_n}L_{n-1}(\bs,y)
+\frac{1}{u-d}\Bigl(\calv_{n-1}^{i,p}(\bs,y,z^{i,p},\vr_i^p)-\frac{\calt_n^{i,p}}{\calt_n}\calv_{n-1}(\bs,y)\Bigr). 
\ee
Moreover, there exists a positive constant $\calc_{n-1}$ such that
\be
\ex\Bigl|\frac{1}{\caln}\sum_{p=1}^m \sum_{i=1}^{N_p}\wh{\phi}^{i,p}_{n-1}(\bS^{n-1},Y_{n-1},Z_{n-1}^{i,p},\vr_i^p)
-L_{n-1}(\bS^{n-1}, Y_{n-1})\Bigr|^2\leq \frac{\calc_{n-1}}{\caln} \nn
\ee
for every $1\leq n\leq N$, which establishes the convergence rate in the large population limit.
\end{theorem}
\begin{proof}
Since, for each $p=1,\ldots, m$, $(Z^{i,p},\vr_i^p), i\in \mbb{N}$ are i.i.d. and also independent of $(S,Y)$, 
the market-clearing condition in Definition~\ref{def-multi-mc} is equivalently given by
\be
\label{multi-mc-condition}
\sum_{p=1}^m w_p \ex^{1,p}\bigl[\wh{\phi}^{1,p}_{n-1}(\bs,y,Z^{1,p}_{n-1},\vr_1^p)\bigr]=L_{n-1}(\bs,y)
\ee
for every $(\bs,y)\in \cals^{n-1}\times \caly_{n-1}$, $1\leq n\leq N$. The expression for the transition 
probabilities $(\ref{multi-mc-transition})$ is a direct consequence of $(\ref{multi-rp-exp})$ in Theorem~\ref{th-multi-rp}
and the market-clearing condition $(\ref{multi-mc-condition})$.
By substituting the resulting expression for $p_{n-1}(\bs,y)$ (and $q_{n-1}(\bs,y)$) into $(\ref{multi-optimal})$
(or equivalently $(\ref{optimal-with-effective})$), we obtain the desired equality $(\ref{multi-mc-optimal})$.

To establish the first claim, it suffices to verify that the family of transition probabilities 
$(p_{n-1}(\bs,y))_{n=1}^N$ defined in $(\ref{multi-mc-transition})$ satisfies the bound given by (vi) 
in Assumption~\ref{assumption-multi-1}, while updating the functions $(f_{n-1}^p)$ via the backward induction 
process described in Theorem~\ref{th-multi-rp}. At $t=t_{N-1}$, $f_{N-1}^p, p=1,\ldots,m$
are measurable functions satisfying the uniform bounds $\ol{c}_N\leq f_{N-1}^p\leq \ol{C}_N$
on their respective domains for some positive constants $\ol{c}_N$ and $\ol{C}_N$ due to the 
boundedness assumption on $F^p$. This makes $\calv_{N-1}$ a bounded function on $\cals^{N-1}\times\caly_{N-1}$.
Moreover, $L_{N-1}$ and  $1/\calt_N$ are bounded. Combined with the fact $d<0<u$, we can confirm
that $p_{N-1}$ (and hence $q_{N-1}$) given by $(\ref{multi-mc-transition})$ satisfies 
$0<p_{N-1}(\bs,y),q_{N-1}(\bs,y)<1$ for every $(\bs,y)\in \cals^{N-1}\times \caly_{N-1}$, and is thus consistent with 
Assumption~\ref{assumption-multi-1} (vi).
Then it is easy to observe that $\wh{\phi}_{N-1}^{i,p}$ is bounded on its domain by the expression $(\ref{optimal-with-effective})$.
Then, using the formula $(\ref{multi-rp-Vnm1})$, $V_{N-1}^p, p=1,\ldots, m$ are shown to 
satisfy the uniform bounds $c_{N-1}\leq V_{N-1}^p\leq C_{N-1}$ on their respective domains
for some positive constants $c_{N-1}$ and $C_{N-1}$.
This in turn ensures that $f_{N-2}^p, p=1,\ldots, m$ once again satisfy the desired properties, and so do
$(p_{N-2}(\bs,y),q_{N-2}(\bs,y))$, $(\bs,y)\in \cals^{N-2}\times\caly_{N-2}$. Proceeding in this way,
by backward induction, we establish the desired consistency for every time step.

To establish the second claim, it suffices to prove that 
\be
\ex\Bigl|\frac{1}{\caln}\sum_{p=1}^m \sum_{i=1}^{N_p}\wh{\phi}^{i,p}_{n-1}(\bs,y,Z_{n-1}^{i,p},\vr_i^p)
-L_{n-1}(\bs,y)\Bigr|^2\leq \frac{\calc_{n-1}}{\caln} \nn
\ee
holds uniformly for every $(\bs,y)\in  \cals^{n-1}\times \caly_{n-1}$. 
Using the equilibrium condition $(\ref{multi-mc-condition})$, it is useful to observe that
\be
\begin{split}
&\frac{1}{\caln}\sum_{p=1}^m \sum_{i=1}^{N_p}\wh{\phi}^{i,p}_{n-1}(\bs,y,Z^{i,p}_{n-1},\vr_i^p)-L_{n-1}(\bs,y) \\
&= \sum_{p=1}^m w_p \left( \frac{1}{N_p}\sum_{i=1}^{N_p} \wh{\phi}^{i,p}_{n-1}(\bs,y,Z^{i,p}_{n-1},\vr_i^p) 
- \ex^{1,p}\bigl[\wh{\phi}^{1,p}_{n-1}(\bs,y,Z_{n-1}^{1,p},\vr_1^p)\bigr] \right). \nn
\end{split}
\ee
Since the term inside the parenthesis for each $p$ is the average of centered i.i.d.~random variables with finite variance 
(guaranteed by the boundedness of the effective variables), the desired convergence rate follows from 
the standard law of large numbers arguments used in Theorem~\ref{th-single-MC-MFE}.  Equivalently, one can also use the expression 
$(\ref{multi-mc-optimal})$ with the relations $(\ref{effective-conv-1})$ and $(\ref{effective-conv-2})$. 
\end{proof}

\begin{remark}[Economic Interpretation of the Equilibrium Strategy]
The expression $(\ref{multi-mc-optimal})$ provides a clear economic interpretation of the equilibrium strategy. 
It can be decomposed into two distinct components:
\be
\wh{\phi}^{i,p}_{n-1} = \underbrace{\frac{\calt_n^{i,p}}{\calt_n} L_{n-1}}_{\text{(I) Supply Distribution}} 
+ \underbrace{\frac{1}{u-d}\Bigl( \calv_{n-1}^{i,p}
- \frac{\calt_n^{i,p}}{\calt_n}\calv_{n-1} \Bigr)}_{\text{(II) Hedge Distribution}}. \nn
\ee
\bi
\item[(I)] The first term represents the sharing of the external supply $L_{n-1}$. 
Each agent absorbs a portion of the supply proportional to their effective risk tolerance $\calt_n^{i,p}$ 
relative to the market aggregate $\calt_n$. Note that the network effect $\Theta$ modifies the 
effective tolerance.
\item[(II)] The second term corresponds to the hedging demand against the effective liability at $t_n$. 
Since $\calv_{n-1}^{i,p}$ represents the sensitivity of the liability (normalized by risk aversion), 
the term $\calv_{n-1}^{i,p}/(u-d)$ essentially corresponds to the Delta hedge required for agent-$(i,p)$. 
However, since the market must clear, agents cannot simply hold their desired hedge. 
Instead, the aggregate hedging demand of the market, $\calv_{n-1}/(u-d)$, is redistributed back to the agents 
according to their risk tolerance shares. 
\ei
This decomposition of the equilibrium strategy is universal: one can confirm that the same relation holds
for $(\ref{single-MC-position})$ in Theorem~\ref{th-single-MC-MFE}, although the $\theta$-dependent terms in (II)
cancel out in the single population case.
\end{remark}

\subsection{Market-clearing mean-field equilibrium when $\mathrm{dim}~\mathrm{Ker}(I-\Theta)=1$}
From the definition of the effective variables in $(\ref{def-effective-variables})$, if some eigenvalues 
$(\lambda_k, 1\leq k\leq m)$ of $(I-\Theta)$ approach zero, we can reasonably expect that the effective 
risk tolerance diverges, rendering the relative performance game ill-defined.
However, as observed in the single-population case (Theorem~\ref{th-single-MC-MFE}), 
we may recover the equilibrium with appropriate transition probabilities of the stock price 
by imposing the market-clearing condition, which precludes the divergence of the agents' positions. 
In this subsection, we demonstrate that this is indeed the case when ${\rm dim}~{\rm Ker}(I-\Theta)=1$.
This single dimension of the kernel corresponds to the existence of only one closed loop 
(e.g., population-$a \to$ population-$b \to \dots \to$ population-$a$) 
where the feedback effect of relative performance concerns diverges.
We also show that there is generally no RP-MFE when ${\rm dim}~{\rm Ker}(I-\Theta)\geq 2$.
Let us recall the definition of the pseudo inverse, $(I-\Theta)^\dagger$:
\be
(I-\Theta)^\dagger:=\Bigl((I-\Theta)\bigr|_{\mathrm{Ker}(I-\Theta)^\perp}\Bigr)^{-1}: 
\mathrm{Im}(I-\Theta)\rightarrow \mathrm{Ker}(I-\Theta)^\perp. \nn
\ee
In general, for a matrix $A$ and $\bm{x}\in \mathrm{Im}(A)$, $A^\dagger \bm{x}$ is the minimal
norm solution to the equation $A\bm{y}=\bm{x}$ and we clearly have $AA^\dagger x=x$.
On the other hand, $A^\dagger A\bm{x}$ gives the projection of $\bm{x}$ onto $\mathrm{Ker}(A)^\perp$ for any $\bm{x}$.
For general properties of pseudo inverse operators, see, e.g., \cite{Liu-Rockner}[Appendix C].

We introduce the following notation for $1\leq p\leq m$ and $i\in \mbb{N}$,  which is slightly different from that defined in 
$(\ref{def-effective-variables})$:
\be
\label{def-effective-2}
\begin{split}
&\mr{\calt}_n^{i,p}:=\frac{1}{\gamma_n^{i,p}}, \hspace{20mm}\mr{\calt}_n^p:=\ex^{1,p}\Bigl[\frac{1}{\gamma_n^{1,p}}\Bigr],\\
&\mr{\calv}_{n-1}^{i,p}(\bs,y,z^{i,p},\vr_i^p):=\frac{\log f_{n-1}^p(\bs,y,z^{i,p},\vr_i^p)}{\gamma_n^{i,p}}, 
\quad \mr{\calv}_{n-1}^p(\bs,y):=\ex^{1,p}\Bigl[\frac{\log f_{n-1}^p(\bs,y,Z_{n-1}^{1,p},\vr_1^p)}{\gamma_n^{1,p}}\Bigr], 
\end{split}
\ee
for each $(\bs,y,z^{i,p},\vr_i^p)\in \cals^{n-1}\times\caly_{n-1}\times \calz_{n-1}^p\times \Gamma^p$.
The functions $f_{n-1}^p:\cals^{n-1}\times\caly_{n-1}\times \calz_{n-1}^p\times \Gamma^p\rightarrow \mbb{R}$, $p=1,\ldots, m$
are defined as in Lemma~\ref{lemma-multi-tmp}. Their complete determination by backward induction will be 
detailed in the following theorem. The functions $\mr{\calv}_{n-1}:\cals^{n-1}\times\caly_{n-1}\rightarrow \mbb{R}$
and $\mr{\calv}_{n-1}^{i,p}:\cals^{n-1}\times \caly_{n-1}\times \calz_{n-1}^p\times \Gamma^p\rightarrow \mbb{R}$
are bounded measurable functions. Although the superscript $i$ in $\mr{\calv}_{n-1}^{i,p}$ is  redundant, 
since the dependence on $\gamma_n^{i,p}$ is incorporated by the argument $\vr_i^p:=(\gamma_i^p, (\theta^i_{p,k})_{k=1}^m)$,
we retain it to clearly associate the variable with the relevant agent as before.
We also introduce the vector notations $\mr{\bm{\calt}}_n:=(\mr{\calt}_n^p)_{p=1}^m$ and 
$\mr{\bm{\calv}}_{n-1}(\mathbf{s},y):=(\mr{\calv}_{n-1}^p(\mathbf{s},y))_{p=1}^m$. 
In addition to the matrix notation explained in the previous subsection, we use the notation $(\bm{u},\bm{v})$ 
to denote the inner product between any $m$-dimensional vectors $\bm{u}$ and $\bm{v}$.
We use $\bm{w}=(w_p)_{p=1}^m\in \mbb{R}^m$ to denote the vector of relative population size.

\begin{theorem}
\label{th-mc-mfe-kernel-1}
Suppose that Assumptions~\ref{assumption-multi-1}, \ref{assumption-multi-2} and \ref{assumption-multi-3} hold.
Furthermore, we assume that $\dim \mathrm{Ker}(I-\Theta)=1$. 
Let $\bm{v}$ and $\bm{\kappa}$ be unit vectors such that $\bm{v}\in {\rm Ker}((I-\Theta)^\top)$ and $\bm{\kappa}\in 
\mathrm{Ker}(I-\Theta)$, respectively.
We assume that they satisfy the non-degeneracy conditions $(\bm{v}, \mr{\bm{\calt}}_N)\neq 0$ and $(\bm{w},\bm{\kappa})\neq 0$.
Then there exists a unique MC-MFE. The associated equilibrium transition probabilities of the stock price are given by
\be
\label{mc-transition-kernel}
p_{n-1}(\bs,y)=(-d)\Big/\left\{u\exp\left(\frac{\bigl(\bm{v},\mr{\bm{\calv}}_{n-1}(\bs,y)\bigr)}{\bigl(\bm{v},\mr{\bm{\calt}}_n\bigr)}\right)-d\right\} 
\ee
for every $(\bs,y)\in \cals^{n-1}\times \caly_{n-1}$, $1\leq n\leq N$.
For each population $p=1,\ldots, m$, the equilibrium strategy of agent-$(i,p)$, $(\wh{\phi}^{i,p}_{n-1})_{n=1}^N$,
is given by the bounded measurable function $\wh{\phi}^{i,p}:\cals^{n-1}\times\caly_{n-1}\times\calz_{n-1}^p\times 
\Gamma^p\rightarrow \mbb{R}$, $1\leq n\leq N$, such that
\be
\label{mc-optimal-kernel-1}
\begin{split}
&\wh{\phi}^{i,p}_{n-1}(\bs,y,z^{i,p},\vr_i^p)=\frac{(\theta^i\bm{\kappa})_p}{(\bm{w},\bm{\kappa})}L_{n-1}(\bs,y) \\
&+\frac{1}{u-d}\Bigl\{\calu^{i,p}_{n-1}(\bs,y,z^{i,p},\vr_i^p)+\bigl(\theta^i(I-\Theta)^\dagger\bm{\calu}_{n-1}(\bs,y)\bigr)_p
-\frac{(\theta^i\bm{\kappa})_p}{(\bm{w},\bm{\kappa})}\bm{w}^\top (I-\Theta)^\dagger \bm{\calu}_{n-1}(\bs,y)\Bigr\}, 
\end{split}
\ee
where $\bm{w}=(w_p)_{p=1}^m\in \mbb{R}^m$ is the vector of relative population size. The bounded measurable 
functions $\calu_{n-1}^{i,p}:\cals^{n-1}\times \caly_{n-1}\times \calz_{n-1}^p\times \Gamma^p\rightarrow \mbb{R}$
and $\calu_{n-1}^p:\cals^{n-1}\times \caly_{n-1} \rightarrow \mbb{R}$, $p=1,\ldots, m$  are defined by
\be
\begin{split}
\calu_{n-1}^{i,p}(\bs,y,z^{i,p},\vr_i^p)&:=\mr{\calv}_{n-1}^{i,p}(\bs,y,z^{i,p},\vr_i^p)-\frac{(\bm{v},\mr{\bm{\calv}}_{n-1}(\bs,y))}{(\bm{v},\mr{\bm{\calt}}_n)}\mr{\calt}_n^{i,p}, \\
\calu_{n-1}^p(\bs,y)&:=\mr{\calv}_{n-1}^p(\bs,y)-\frac{(\bm{v},\mr{\bm{\calv}}_{n-1}(\bs,y))}{(\bm{v},\mr{\bm{\calt}}_n)}\mr{\calt}_n^{p}.\nn
\end{split}
\ee
We denote the vector formed by the latter components as $\bm{\mathcal{U}}_{n-1}(\mathbf{s},y):=(\mathcal{U}_{n-1}^p(\mathbf{s},y))_{p=1}^m$.
Here, for each $1\leq n\leq N$, the functions $f_{n-1}^p:\cals^{n-1}\times \caly_{n-1}\times \calz_{n-1}^p\times \Gamma^p\rightarrow \mbb{R}$, 
$p=1,\ldots,m$, which are  components of $(\mr{\calv}^{i,k}_{n-1}, \mr{\calv}^k_{n-1}, \calu_{n-1}^{i,k},\calu_{n-1}^k)$,
are measurable functions satisfying the  uniform bounds $0<\ol{c}_{n}\leq f_{n-1}^p\leq \ol{C}_{n}<\infty$
for some positive constants $\ol{c}_n$ and $\ol{C}_n$. They are determined by the backward induction as in Theorem~\ref{th-multi-rp}:
We use the same terminal condition $V_N^p(\bs,y,z^{i,p},\vr_i^p):=\exp\bigl(\gamma_i^p F^p(\bs,y,z^{i,p})\bigr)$ 
and the formulas $(\ref{multi-fnm1-update})$ and $(\ref{multi-rp-Vnm1})$, 
replacing the transition probabilities $(p_{n-1}(\bs,y), q_{n-1}(\bs,y))_{n=1}^N$ and the optimal control $(\wh{\phi}^{i,p}_{n-1})_{n=1}^N$ 
with those given above at each step.
The dynamics of the associated mean-field terms $(\wh{\mu}_n^p)_{n=1}^N$ is given by 
\be
\label{mc-mu-kernel-1}
\begin{split}
&\wh{\mu}_n^p((\bs\wt{u})^n,\by)=\beta \wh{\mu}^p_{n-1}(\bs,\by^-)+u\ex^{1,p}\bigl[\wh{\phi}^{1,p}_{n-1}(\bs,y,Z_{n-1}^{1,p},\vr_1^p)\bigr], \\
&\wh{\mu}_n^p((\bs\wt{d})^n, \by)=\beta \wh{\mu}^p_{n-1}(\bs,\by^-)+d\ex^{1,p}\bigl[\wh{\phi}^{1,p}_{n-1}(\bs,y, Z_{n-1}^{1,p},\vr_1^p)\bigr], 
\end{split}
\ee
with
\be
\begin{split}
&\ex^{1,p}\bigl[\wh{\phi}^{1,p}(\bs,y,Z_{n-1}^{1,p},\vr_1^p)\bigr]\\
&\quad =\frac{\kappa_p}{(\bm{w},\bm{\kappa})}L_{n-1}(\bs,y)
+\frac{1}{u-d}\Bigl\{\bigl((I-\Theta)^\dagger \calu_{n-1}(\bs,y)\bigr)_p-\frac{\kappa_p}{(\bm{w},\bm{\kappa})}
\bm{w}^\top (I-\Theta)^\dagger \calu_{n-1}(\bs,y)\Bigr\}. \nn
\end{split}
\ee
under the initial condition $\mu_0^p:=\ex^{1,p}[\xi_1^p]$.
Moreover, there exists a positive constant $\calc_{n-1}$ such that
\be
\ex\Bigl|\frac{1}{\caln}\sum_{p=1}^m \sum_{i=1}^{N_p}\wh{\phi}^{i,p}_{n-1}(\bS^{n-1},Y_{n-1},Z_{n-1}^{i,p},\vr_i^p)
-L_{n-1}(\bS^{n-1}, Y_{n-1})\Bigr|^2\leq \frac{\calc_{n-1}}{\caln} \nn
\ee
for every $1\leq n\leq N$, which establishes the convergence rate in the large population limit.
\end{theorem}
\begin{proof}
We proceed as in the proof for Theorem~\ref{th-multi-rp}.
We suppose that, for every $p=1,\ldots, m$,  the problem for agent-$(i,p)$ at $t_{n-1}$ for the period $[t_{n-1},t_n]$ is given by the 
form $(\ref{problem-multi-tmp})$. To apply Lemma~\ref{lemma-multi-tmp}, we hypothesize 
that the measurable function $V_n^p:\cals^n\times \caly_n\times\calz_n^p\times\Gamma^p\rightarrow \mbb{R}$
satisfies the uniform bounds $c_n\leq V_n^p\leq C_n$ for some positive constants $c_n$ and $C_n$.
This ensures that the function $f_{n-1}^p$ also satisfies the desired uniform bounds.  
This clearly holds at $t=t_{N-1}$ for the last interval $[t_{N-1},t_N]$
with  $V_N^p(\bs,y,z^{i,p},\vr_i^p):=\exp\bigl(\gamma_i^p F^p(\bs,y,z^{i,p})\bigr)$.

Using $(\ref{multi-optimal-tmp})$ in Lemma~\ref{lemma-multi-tmp}, the evolution of the wealth process of agent-$(i,p)$
under the optimal strategy $\phi^{(i,p),*}_{n-1}$ is given by
\be
X_n^{(i,p),*}=\beta X_{n-1}^{(i,p),*}+\phi^{(i,p),*}_{n-1}(\bS^{n-1},\bY^{n-1},Z_{n-1}^{i,p},\vr_i^p) R_n. \nn
\ee
Consider the problem conditioned on the event $\{(\bS^{n-1},\bY^{n-1})=(\bs,\by)\}$ in $\calf^0_{t_{n-1}}$.
Under the induction hypothesis that $\ex^{0,(1,p)}[X^{(1,p),*}_{n-1}|\bs,\by]$ is given by the form $\mu_{n-1}^p(\bs,\by^-)$
with $(\bs,\by^-)\in \cals^{n-1}\times \caly^{n-2}$ (when $n=1$, we simply set $\mu_0^p=\ex^{1,p}[\xi_1^p]$),
using the i.i.d. property of the variables $(\vr_i^p, Z^{i,p}), i=1,2,\ldots$, the fixed point condition $(\ref{def-multi-mu})$ 
for the RP-MFE for the period $[t_{n-1},t_n]$ is given by
\be
\label{mu-rp-dynamics}
\begin{split}
&\mu_n^p((\bs\wt{u})^n,\by)=\beta \mu_{n-1}^p(\bs,\by^-)+u\ex^{1,p}\bigl[\phi^{(1,p),*}_{n-1}(\bs,\by,Z_{n-1}^{1,p},\vr_1^p)\bigr], \\
&\mu_n^p((\bs\wt{d})^n,\by)=\beta \mu_{n-1}^p(\bs,\by^-)+d\ex^{1,p}\bigl[\phi^{(1,p),*}_{n-1}(\bs,\by,Z_{n-1}^{1,p},\vr_1^p)\bigr], 
\end{split}
\ee
which is equal to $(\ref{multi-RP-MFE-eq})$. 
Taking the difference on both sides, 
and defining the vectors $\mr{\bm{\calt}}_n:=(\mr{\calt}_n^p)_{p=1}^m$ 
and $\mr{\bm{\calv}}_{n-1}(\mathbf{s},y):=(\mr{\calv}_{n-1}^p(\mathbf{s},y))_{p=1}^m$,
we obtain $(\ref{multi-rp-mfe-consistency})$, or equivalently, for the relative performance equilibrium, the equality
\be
\label{eq-rp-condition}
(I-\Theta)\bm{\Del}_n(\bs,\by)=\log\Bigl(-\frac{p_{n-1}(\bs,y)u}{q_{n-1}(\bs,y)d}\Bigr)\mr{\bm{\calt}}_n+\mr{\bm{\calv}}_{n-1}(\bs,y)  
\ee
must hold for every $(\bs,\by)\in \cals^{n-1}\times \caly^{n-1}$. 
For this equation to have a solution, the right-hand side must be in $\mathrm{Im}(I-\Theta)$.
By the Fredholm theorem, since $\mathrm{Im}(I-\Theta)=\bigl(\mathrm{Ker}(I-\Theta)^\top\bigr)^\perp$, 
this is the case if and only if
\be
\log\Bigl(-\frac{p_{n-1}(\bs,y)u}{q_{n-1}(\bs,y)d}\Bigr)(\bm{v}, \mr{\bm{\calt}}_n)
+(\bm{v},\mr{\bm{\calv}}_{n-1}(\bs,y))=0  \nn
\ee
for a vector $\bm{v}\in \mathrm{Ker}(I-\Theta)^\top$. We normalize it as $\|\bm{v}\|=1$.
Note that $\dim \mathrm{Ker}(I-\Theta)^\top=\dim \mathrm{Ker}(I-\Theta)=1$ since $(I-\Theta)$ is a square matrix.
This uniquely fixes the form of $p_{n-1}(\bs,y)$ (and hence $q_{n-1}(\bs,y)$) as in $(\ref{mc-transition-kernel})$.
Note that $(\bm{v},\mr{\bm{\calt}}_n)\neq 0$ for every $n$. The uniform bounds $0<\ol{c}_n\leq f_{n-1}^p\leq \ol{C}_n<\infty$ ensure that
$0<p_{n-1}(\bs,y),q_{n-1}(\bs,y)<1$ for every $(\bs,y)\in \cals^{n-1}\times\caly_{n-1}$.
In this case, a general solution to $(\ref{eq-rp-condition})$ can be written in the following form:
\be
\bm{\Del}_n(\bs,\by)=(I-\Theta)^\dagger \Bigl\{
\mr{\bm{\calv}}_{n-1}(\bs,y)-\frac{(\bm{v},\mr{\bm{\calv}}_{n-1}(\bs,y))}{(\bm{v}, \mr{\bm{\calt}}_n)}\mr{\bm{\calt}}_n\Bigr\}+\del_n(\bs,\by)\bm{\kappa}, \nn
\ee
where $\del_n:\cals^{n-1}\times \caly^{n-1}\rightarrow \mbb{R}$ is an arbitrary (bounded) measurable function
and $\bm{\kappa}\in \mbb{R}^m$ is a unit vector with  $\bm{\kappa}\in \mathrm{Ker}(I-\Theta)$.
Substituting these results into the optimal strategy $(\ref{multi-optimal-tmp})$  in Lemma~\ref{lemma-multi-tmp}, we obtain
\be
\label{optimal-kernel-1-tmp}
\begin{split}
&\phi^{(i,p),*}_{n-1}(\bs,\by,z^{i,p},\vr_i^p)=\frac{1}{u-d} 
\left(\theta^i(I-\Theta)^\dagger\Bigl\{
\mr{\bm{\calv}}_{n-1}(\bs,y)-\frac{(\bm{v},\mr{\bm{\calv}}_{n-1}(\bs,y))}{(\bm{v}, \mr{\bm{\calt}}_n)}\mr{\bm{\calt}}_n\Bigr\}\right)_p \\
&\quad+\frac{1}{u-d}\Bigl\{\mr{\calv}_{n-1}^{i,p}(\bs,y,z^{i,p},\vr_i^p)
-\frac{(\bm{v},\mr{\bm{\calv}}_{n-1}(\bs,y))}{(\bm{v}, \mr{\bm{\calt}}_n)}\mr{\calt}^{i,p}_n\Bigr\}+\del_n(\bs,\by)\frac{(\theta^i \bm{\kappa})_p}{u-d}. 
\end{split}
\ee
Hence, with the transition probabilities given by $(\ref{mc-transition-kernel})$,
the solution to the relative performance game for the period $[t_{n-1},t_n]$ exists but it is not unique.
We also observe that the path-dependence $\by\in \caly^{n-1}$ in $\bm{\Del}_n$ and $\phi^{(i,p),*}_{n-1}$ appears only through 
the function $\del_n(\bs,\by)$. 

The additional degree of freedom in the choice of $\del_n$ is uniquely fixed by 
imposing the market-clearing condition:
\be
\label{mc-kernel-1-tmp}
\sum_{p=1}^m w_p \ex^{1,p}\bigl[\phi^{(1,p),*}_{n-1}(\bs,y,Z^{1,p}_{n-1},\vr_1^p)\bigr]=L_{n-1}(\bs,y).
\ee
From $(\ref{optimal-kernel-1-tmp})$ and $(\ref{mc-kernel-1-tmp})$, it is straightforward to obtain
\be
\begin{split}
\del_n(\bs,y)=\frac{1}{(\bm{w},\bm{\kappa})}\left\{ (u-d)L_{n-1}(\bs,y)-\bm{w}^\top (I-\Theta)^\dagger
\Bigl(\mr{\bm{\calv}}_{n-1}(\bs,y)-\frac{(\bm{v},\mr{\bm{\calv}}_{n-1}(\bs,y))}{(\bm{v}, \mr{\bm{\calt}}_n)}\mr{\bm{\calt}}_n\Bigr)\right\}, \nn
\end{split}
\ee
which only depends on the last element $y\in \caly_{n-1}$ in $\by\in \caly^{n-1}$.
Substituting this expression for $\delta_n$ into  $(\ref{optimal-kernel-1-tmp})$, 
we obtain $\phi^{(i,p),*}_{n-1}(\bs,y,z^{i,p},\vr_i^p)$ in the form given in $(\ref{mc-optimal-kernel-1})$.
Since $(f_{n-1}^p)_{p=1}^m$ satisfies the uniform bounds $\ol{c}_n\leq f_{n-1}^p\leq \ol{C}_n$ 
due to the induction hypothesis on $(V_n^p)_{p=1}^m$,
the optimal control $\phi^{(i,p),*}_{n-1}$ and hence the mean-field term $\mu_n^p$  are also given by bounded measurable functions.
Therefore, with a given bounded initial condition $(\mu_{n-1}^p(\bs,\by^-))_{p=1}^m$,
the transition probabilities $(\ref{mc-transition-kernel})$, the control $(\ref{mc-optimal-kernel-1})$,
and the dynamics of the mean-field term $(\ref{mc-mu-kernel-1})$ provide the unique solution 
for MC-MFE for this period $[t_{n-1},t_n]$. 

In order to complete the proof, it suffices to show that this procedure can be repeated backward from the last interval $[t_{N-1},t_N]$
to the first one $[t_0,t_1]$. The value function for the agent-$(i,p)$ at $t_{n-1}$, which is the objective function 
for the optimization for the period $[t_{n-2},t_{n-1}]$, is given by the same formula $(\ref{objective-tnm1})$
with $V_{n-1}^p$ in $(\ref{multi-rp-Vnm1})$ with transition probabilities and $\wh{\phi}^{i,p}$ replaced by 
those in $(\ref{mc-transition-kernel})$ and $(\ref{mc-optimal-kernel-1})$ derived above.
It is easy to confirm that $V_{n-1}^p$ once again satisfies the uniform bounds
$c_{n-1}\leq V_{n-1}^p\leq C_{n-1}$ on its domain for some positive constants $c_{n-1}$ and $C_{n-1}$.
Hence we succeed in recovering the same form of problem as in Lemma~\ref{lemma-multi-tmp}.
Hence we can proceed with the backward induction by one time step. The induction hypothesis of bounded measurability  for $\mu_{n-1}^p$ is reduced
to the constant $\mu_0^p:=\ex^{1,p}[\xi_1^p]$ and is satisfied trivially.
The last claim on the convergence rate can be proved in the same way as in Theorem~\ref{th-multi-mc-mfe}.
\end{proof}

\begin{remark}
From $(\ref{eq-rp-condition})$, one can observe that there is generally no equilibrium for 
$d_{K}:=\dim \mathrm{Ker}(I-\Theta)\geq 2$. Specifically, in order to have a mean-field equilibrium, 
we need to satisfy the solvability conditions for every basis vector $\bm{v}_k \in \mathrm{Ker}((I-\Theta)^\top)$, $1\leq k\leq d_K$.
Since we only have a single scalar degree of freedom (the transition probability of the stock price) 
to satisfy these $d_K$ independent constraints simultaneously, the system is overdetermined.
Thus, a solution exists only in highly degenerate cases where the vectors $(\mr{\bm{\calv}}_{n-1}(\bs,y))$ 
coincidentally lie in the orthogonal complement of the kernel.
For instance, a trivial case is given where the liabilities $(F^p)$ and the external order flow $(L_n)$
are independent of the stock price process. In this case, $(f^p_{n-1})$ and hence $\mr{\bm{\calv}}_{n-1}$ vanish,
and the transition probabilities become equal to those in the risk-neutral measure $\mbb{Q}$, i.e.,
\be
\log\Bigl(-\frac{p_{n-1}(\bs,y)u}{q_{n-1}(\bs,y)d}\Bigr)=0. \nn
\ee
In this special situation, relative performance concerns become irrelevant due to the
independence of the mean-field term $\wh{\mu}$ from the stock price.
\end{remark}

\begin{remark}
\label{remark-path-dependence}
If we assume that the liabilities and external order flow depend only on the current state variables, i.e., take the form
$F^p(S_N,Y_N,Z_N^{i,p})$ and $L_{n-1}(S_{n-1},Y_{n-1})$, then we can show that the equilibrium transition probabilities
and the controls in Theorems~\ref{th-multi-rp}, \ref{th-multi-mc-mfe}, and \ref{th-mc-mfe-kernel-1} 
become Markovian (path-independent) as shown in Theorems~\ref{th-single-1} and \ref{th-single-MC-MFE}.
\end{remark}

\subsection{Perturbation of the inverse $(I-\Theta(\ep))^{-1}$ around $\ep=0$}
\label{sec-perturbation}
As observed in Theorem~\ref{th-mc-mfe-kernel-1}, a unique MC-MFE 
exists when $\mathrm{dim}~\mathrm{Ker}(I-\Theta)=1$. 
It is natural to expect that, similarly to the single-population case (Theorem~\ref{th-single-MC-MFE}), 
the market-clearing condition ensures that both the equilibrium transition probabilities $(p_{n-1}(\bs,y))$ 
and the equilibrium controls $(\wh{\phi}^{i,p}_{n-1})$ exhibit continuous dependence
on the interaction matrix as it approaches the singular limit.
We assume that the unperturbed interaction matrix satisfies $\mathrm{dim}~\mathrm{Ker}(I-\Theta)=1$. 
For simplicity, we consider a perturbation $\ep\in \mbb{R}$ of the form:
\be
\label{ep-Theta}
\theta^i_{p,k}(\ep):=\theta^i_{p,k}-\ep \del_{p,k}, \quad \Theta(\ep):=\Theta-\ep I 
\ee
for every $p,k=1,\ldots, m$ and $i\in \mbb{N}$, where $\del_{p,k}$ denotes the Kronecker delta . We need the following result:
\begin{assumption}
\label{assumption-1st-pole}
$\mathrm{dim}~\mathrm{Ker}(I-\Theta)=1$ and the pole of the resolvent $(I-\Theta(\ep))^{-1}$ has a simple pole at $\ep=0$.
\end{assumption}

\begin{lemma}
\label{lemma-resolvent-expansion}
Let Assumption~\ref{assumption-1st-pole} be in force.
Let $\bm{v}$ and $\bm{\kappa}$ be $m$-dimensional unit vectors satisfying $\bm{v}\in \mathrm{Ker}(I-\Theta)^\top$
and $\bm{\kappa}\in \mathrm{Ker}(I-\Theta)$.
Define the projection matrix $P$ and the pseudo inverse $G$ by
\be
P:=I-\frac{\bm{\kappa}\bm{v}^\top}{(\bm{v},\bm{\kappa})}, \quad G:=(I-\Theta)^\dagger. \nn
\ee
Then, for any $\ep$ satisfying $0 < |\ep| < \|PG\|^{-1}$ and such that 
$(1+\ep)$ is in the resolvent set of $\Theta$, the following expansion holds:
\be
(I-\Theta(\ep))^{-1} = \frac{1}{\ep}\frac{\bm{\kappa}\bm{v}^\top}{(\bm{v},\bm{\kappa})}+\sum_{n=0}^\infty (-\ep)^n(PG)^{n+1}P.
\ee
\end{lemma}
\begin{remark*}
Note that Assumption~\ref{assumption-1st-pole} implies $(\bm{v},\bm{\kappa})\neq 0$.
\end{remark*}
\begin{proof}
Under Assumption~\ref{assumption-1st-pole}, the resolvent has a simple pole at $\ep=0$.
Thus, we can postulate a Laurent expansion of the form:
\be
(I-\Theta(\ep))^{-1} = \frac{1}{\ep}R_{-1} + R_0 + \ep R_1 + \cdots. \nn
\ee
We determine the coefficients $(R_i)_{i=-1}^\infty$ using the identity $(I-\Theta(\ep))(I-\Theta(\ep))^{-1}=I$. 
Recall that $I-\Theta(\ep) = (I-\Theta) + \ep I$. Comparing terms of order $\ep^{-1}$, we get $(I-\Theta)R_{-1}=0$, 
which implies $R_{-1}=\bm{\kappa}\bm{c}^\top$ for some vector $\bm{c}\in \mbb{R}^m$. Comparing terms of order $\ep^0$, 
we have $(I-\Theta)R_0 + R_{-1} = I$, or $(I-\Theta)R_0=I-R_{-1}$. Multiplying by $\bm{v}^\top$ from the left, 
and noting that $\bm{v}^\top(I-\Theta)=0$, we obtain $0 = \bm{v}^\top - \bm{v}^\top \bm{\kappa}\bm{c}^\top$.
This determines $\bm{c}^\top=\bm{v}^\top/(\bm{v},\bm{\kappa})$, yielding the residue $R_{-1}=(\bm{\kappa}\bm{v}^\top)/(\bm{v},\bm{\kappa})$. 
Note that $I-R_{-1}=P$. Next, since $R_0$ is a solution to $(I-\Theta)R_0 = P$, it must take the form $R_0=GP+\bm{\kappa}\bm{c}_1^\top$ 
for some $\bm{c}_1\in \mbb{R}^m$. To determine $\bm{c}_1$, we compare terms of order $\ep^1$. We have $(I-\Theta)R_1 + R_0 = 0$, 
or $(I-\Theta)R_1 = -R_0$. Multiplying by $\bm{v}^\top$ from the left yields the condition 
$\bm{v}^\top R_0 = 0$. Substituting the expression for $R_0$, we obtain
$\bm{v}^\top (GP + \bm{\kappa}\bm{c}_1^\top) = 0$.
This uniquely determines $\bm{c}_1^\top$ as $\bm{c}_1^\top = - \frac{\bm{v}^\top GP}{(\bm{v},\bm{\kappa})}$. 
Substituting this back into the expression for $R_0$, we obtain
\be 
R_0 = GP - \frac{\bm{\kappa}\bm{v}^\top}{(\bm{v},\bm{\kappa})} GP = \Bigl( I-\frac{\bm{\kappa}\bm{v}^\top}{(\bm{v},\bm{\kappa})} 
\Bigr) GP = PGP. \nn
\ee
Proceeding inductively, requiring the existence of higher-order terms leads to the relation $R_n = (-1)^n (PG)^{n+1}P$. 
The series converges in the operator norm under the condition $|\ep|\|PG\| < 1$.
\end{proof}

\noindent
Note that, for any $\bm{x}\in \mbb{R}^m$, we have $(\bm{v},P\bm{x})=0$. 
Since $\mathrm{Im}(I-\Theta) = (\mathrm{Ker}(I-\Theta)^\top)^\perp = \{\bm{y}\in \mbb{R}^m \mid (\bm{v}, \bm{y})=0\}$, 
the matrix $P$ serves as the projection operator onto the image space $\mathrm{Im}(I-\Theta)$ 
along the kernel direction $\bm{\kappa}$.

\begin{theorem}
\label{th-pole-convergence}
Suppose that the assumptions of Theorem~\ref{th-mc-mfe-kernel-1} hold.
Furthermore, let Assumption~\ref{assumption-1st-pole} be in force.
Choose $\eta_0>0$ sufficiently small such that, for all $|\ep| \in(0,\eta_0)$, the matrix $(I-\Theta(\ep))$ is invertible 
and $\calt_n(\ep) \neq 0$, where $\calt_n(\ep)$ is obtained from $(\ref{def-effective-variables})$ by replacing $\Theta$ with $\Theta(\ep)$.
For $|\ep|\in (0, \eta_0)$, let us denote the unique solution for the MC-MFE associated with 
the perturbed relative performance concerns $(\theta^i_{p,k}(\ep), i\in \mbb{N})_{p,k=1}^m$
by $\wh{\phi}^{i,p}_{n-1}(\bs,y,z^{i,p},\vr_i^p)(\ep)$ and $p_{n-1}(\bs,y)(\ep)$
for each  $(\bs,y,z^{i,p},\vr_i^p)\in \cals^{n-1}\times \caly_{n-1}\times \calz_{n-1}^p\times \Gamma^p$, $1\leq n\leq N$.
These solutions are determined by Theorems~\ref{th-multi-rp} and \ref{th-multi-mc-mfe} with $(\theta^i_{p,k}, \Theta, \vr_i^p)$
replaced by $(\theta^i_{p,k}(\ep), \Theta(\ep), \vr_i^p(\ep))$, where we define $\vr_i^p(\ep):=(\gamma_i^p, (\theta^i_{p,k}(\ep))_{k=1}^m)$.
Then we have the convergence
\be
\begin{split}
&\lim_{\ep\rightarrow 0} p_{n-1}(\bs,y)(\ep)=p_{n-1}(\bs,y), \\
&\lim_{\ep\rightarrow 0} \wh{\phi}^{i,p}_{n-1}(\bs,y,z^{i,p},\vr_i^p)(\ep)=\wh{\phi}^{i,p}_{n-1}(\bs,y,z^{i,p},\vr_i^p), \quad i\in \mbb{N} \nn
\end{split}
\ee
for every $(\bs,y,z^{i,p},\vr_i^p)\in \cals^{n-1}\times \caly_{n-1}\times \calz_{n-1}^p\times \Gamma^p$, $1\leq n\leq N$.
Here, $p_{n-1}(\bs,y)$ and $\wh{\phi}^{i,p}_{n-1}(\bs,y,z^{i,p},\vr_i^p)$ are the solutions 
established  in Theorem~\ref{th-mc-mfe-kernel-1} for the singular case.
\end{theorem}
\begin{proof}
We proceed with backward induction. We define the following variables and functions 
(See $(\ref{def-effective-variables})$ and $(\ref{def-effective-2})$):
\be
\begin{split}
&\mr{\calt}_n^{i,p}:=\frac{1}{\gamma_n^{i,p}}, \quad \mr{\calt}_n^p:=\ex^{1,p}\Bigl[\frac{1}{\gamma_n^{1,p}}\Bigr], \\
&\mr{\calv}_{n-1}^{i,p}(\bs,y,z^{i,p},\vr_i^p)(\ep):=\frac{\log f_{n-1}^p(\bs,y,z^{i,p},\vr_i^p)(\ep)}{\gamma_n^{i,p}}, \quad
\mr{\calv}_{n-1}^{p}(\bs,y)(\ep):=\ex^{1,p}\bigl[\mr{\calv}_{n-1}^{1,p}(\bs,y,Z_{n-1}^{1,p},\vr_1^p)(\ep)\bigr], \\
&\calt_n^{i,p}(\ep):=\mr{\calt}_n^{i,p}+\bigl(\theta^i(\ep)(I-\Theta(\ep))^{-1}\mr{\bm{\calt}}_n\bigr)_p, \quad \calt_n(\ep):=
\bm{w}^\top (I-\Theta(\ep))^{-1}\mr{\bm{\calt}}_n, \\
&\calv_{n-1}^{i,p}(\bs,y,z^{i,p},\vr_i^p)(\ep):=\mr{\calv}_{n-1}^{i,p}(\bs,y,z^{i,p},\vr_i^p)(\ep)+
\Bigl(\theta^i(\ep)(I-\Theta(\ep))^{-1}\mr{\bm{\calv}}_{n-1}(\bs,y)(\ep)\Bigr)_p, \\
&\calv_{n-1}(\bs,y)(\ep):=\bm{w}^\top (I-\Theta(\ep))^{-1}\mr{\bm{\calv}}_{n-1}(\bs,y)(\ep), \nn
\end{split}
\ee
for each  $(\bs,y,z^{i,p},\vr_i^p)\in \cals^{n-1}\times \caly_{n-1}\times \calz_{n-1}^p\times \Gamma^p$, $i\in \mbb{N}$.
Here, $\bm{w}=(w_p)_{p=1}^m$ is the relative weight vector of populations.
Furthermore, the functions $(f_{n-1}^p(\cdot)(\ep), p=1,\ldots, m)$ are those determined by Theorem~\ref{th-multi-mc-mfe}
with perturbed interaction $\theta^i(\ep)$. We also employ vector notation, such as
$\mr{\bm{\calt}}_n:=(\mr{\calt}_n^p)_{p=1}^m$ and $\mr{\bm{\calv}}_{n-1}(\bs,y)(\ep):=(\mr{\calv}_{n-1}^p(\bs,y)(\ep))_{p=1}^m$.
Let us also introduce the notation $V_n^p(\bs,y,z^{i,p},\vr_i^p)(\ep), 1\leq n \leq N$ to represent the exponential effective 
liability for the $\ep$-perturbed setup defined by $(\ref{multi-rp-Vnm1})$  with $\theta^i(\ep), p_{n-1}(\cdot)(\ep), 
\wh{\phi}^{i,p}(\cdot)(\ep)$, etc. 

With these variables, 
the equilibrium control for $\theta^i(\ep)$ in Theorem~\ref{th-multi-mc-mfe} can be written as
\be
\label{eq-phi-ep}
\wh{\phi}^{i,p}_{n-1}(\bs,y,z^{i,p},\vr_i^p)(\ep)=\frac{\calt_n^{i,p}(\ep)}{\calt_n(\ep)}L_{n-1}(\bs,y) 
+\frac{1}{u-d}\Bigl(\calv_{n-1}^{i,p}(\bs,y,z^{i,p},\vr_i^p)(\ep)-\frac{\calt_n^{i,p}(\ep)}{\calt_n(\ep)}\calv_{n-1}(\bs,y)(\ep)\Bigr).  
\ee
As an induction hypothesis, we assume
\be
\begin{split}
&\lim_{\ep\rightarrow 0} V_n^p(\bs^n,y_n,z_n^{i,p},\vr_i^p)(\ep)=V_n^p(\bs^n,y_n,z_n^{i,p},\vr_i^p) \\
&\lim_{\ep\rightarrow 0}\mr{\calv}_{n-1}^{i,p}(\bs^{n-1},y_{n-1},z_{n-1}^{i,p},\vr_i^p)(\ep)
=\mr{\calv}_{n-1}^{i,p}(\bs^{n-1},y_{n-1},z_{n-1}^{i,p},\vr_i^p), \nn
\end{split}
\ee
for every $(\bs^n,\bs^{n-1})\in \cals^n\times \cals^{n-1}$, $(y_n,y_{n-1})\in \caly_n\times\caly_{n-1}$,
$(z_n^{i,p},z_{n-1}^{i,p})\in \calz_n^p\times \calz_{n-1}^p$, $\vr_i^p\in \Gamma^p$, 
where the limits $V_n^p(\bs^n,y_n,z_n^{i,p},\vr_i^p)$ and $\mr{\calv}_{n-1}^{i,p}(\bs^{n-1},y_{n-1},z_{n-1}^{i,p},\vr_i^p)$ in the right-hand sides 
correspond to the functions defined in Theorem~\ref{th-mc-mfe-kernel-1}. The convergence of $\mr{\calv}_{n-1}^{i,p}(\cdot)(\ep)$ also implies 
the convergence of $\mr{\bm{\calv}}_{n-1}(\bs,y)(\ep)~(:=(\mr{\calv}^p_{n-1}(\bs,y)(\ep))_{p=1}^m)\to \mr{\bm{\calv}}_{n-1}(\bs,y)$
as $\ep\to 0$ by the dominated convergence theorem. In particular, this implies that they remain bounded as $\ep\rightarrow 0$.
These convergences hold trivially at $n=N$ since the function $F^p$ is $\ep$-independent.
For notational simplicity, we fix the values of $(\bs,y,z^{i,p},\vr_i^p)\in \cals^{n-1}\times \caly_{n-1}\times\calz_{n-1}^p\times \Gamma^p$
and suppress these arguments hereafter  to focus on the dependence on $\ep$.

Using Lemma~\ref{lemma-resolvent-expansion}, let $R_{-1}:=\bm{\kappa}\bm{v}^\top/(\bm{v},\bm{\kappa})$ 
and $R_0:=PGP$. We then have the expansion:
\be
(\theta^i(\ep)(I-\Theta(\ep))^{-1})=\frac{1}{\ep}\theta^i R_{-1}+(\theta^i R_0-R_{-1})+\calo(\ep). \nn
\ee
Similarly, we obtain the expansions as
\be
\begin{split}
\calt_n^{i,p}(\ep)&=\frac{1}{\ep}(\theta^i R_{-1}\mr{\bm{\calt}}_n)_p+\mr{\calt}_n^{i,p}+\bigl((\theta^i R_0-R_{-1})\mr{\bm{\calt}}_n\bigr)_p+\calo(\ep), \\
\calt_n(\ep)&=\frac{1}{\ep}\bm{w}^\top R_{-1}\mr{\bm{\calt}}_n+\bm{w}^\top R_0 \mr{\bm{\calt}}_n+\calo(\ep), \\
\calv_{n-1}^{i,p}(\ep) &=\mr{\calv}_{n-1}^{i,p}(\ep)+\Bigl(\Bigl(\frac{1}{\ep}\theta^iR_{-1}+(\theta^i R_0-R_{-1})+\calo(\ep)\Bigr)\mr{\bm{\calv}}_{n-1}(\ep)\Bigr)_p\\
\calv_{n-1}(\ep)&=\bm{w}^\top\Bigl(\frac{1}{\ep}R_{-1}+R_0+\calo(\ep)\Bigr)\mr{\bm{\calv}}_{n-1}(\ep). \nn
\end{split}
\ee

Firstly, it is easy to confirm that
\be
\begin{split}
\frac{\calt_n^{i,p}(\ep)}{\calt_n(\ep)}=\rho^{i,p}-\ep\frac{1}{(\bm{w},\bm{\kappa})}\Bigl\{\kappa_p
-\frac{(\bm{v},\bm{\kappa})}{(\bm{v},\mr{\bm{\calt}}_n)}\bigl(\mr{\calt}_n^{i,p}+(\theta^i R_0\mr{\bm{\calt}}_n)_p
-\rho^{i,p}\bm{w}^\top R_0\mr{\bm{\calt}}_n\bigr)\Bigr\}+\calo(\ep^2),  \nn
\end{split}
\ee
where $\rho^{i,p}:=(\theta^i\bm{\kappa})_p/(\bm{w},\bm{\kappa})$.
This establishes the desired convergence of the first term in $(\ref{eq-phi-ep})$.

Next, we consider the convergence for the second term of $\wh{\phi}^{i,p}_{n-1}(\ep)$. We have
\be
\label{ep-calv}
\begin{split}
&\calv_{n-1}^{i,p}(\ep)-\frac{\calt_n^{i,p}(\ep)}{\calt_n(\ep)}\calv_{n-1}(\ep)
=\frac{1}{\ep}\Bigl\{(\theta^i R_{-1}\mr{\bm{\calv}}_{n-1}(\ep))_p-\rho^{i,p}\bm{w}^\top R_{-1}\mr{\bm{\calv}}_{n-1}(\ep)\Bigr\}\\
&\qquad+\mr{\calv}_{n-1}^{i,p}(\ep)+\bigl((\theta^iR_0-R_{-1})\mr{\bm{\calv}}_{n-1}(\ep)\bigr)_p-\rho^{i,p}\bm{w}^\top R_0\mr{\bm{\calv}}_{n-1}(\ep)\\
&\qquad+\frac{1}{(\bm{w},\bm{\kappa})}\Bigl\{\kappa_p-\frac{(\bm{v},\bm{\kappa})}{(\bm{v},\mr{\bm{\calt}}_n)}
\bigl(\mr{\calt}_n^{i,p}+(\theta^i R_0\mr{\bm{\calt}}_n)_p-\rho^{i,p}\bm{w}^\top R_0\mr{\bm{\calt}}_n\bigr)\Bigr\}\bm{w}^\top R_{-1}
\mr{\bm{\calv}}_{n-1}(\ep)+\calo(\ep). 
\end{split}
\ee
Note that the potentially diverging terms as $\ep\to 0$ given by the first line vanish completely by
the equality
\be
\bigl(\theta^iR_{-1}\mr{\bm{\calv}}_{n-1}(\ep)\bigr)_p-\rho^{i,p}\bm{w}^\top R_{-1}\mr{\bm{\calv}}_{n-1}(\ep)\equiv 0. \nn
\ee
By expanding the left matrix $P$ in $R_0:=PGP$, $(\ref{ep-calv})$ can be simplified, due to significant cancellations,  to
\be
\begin{split}
\calv_{n-1}^{i,p}(\ep)-\frac{\calt_n^{i,p}(\ep)}{\calt_n(\ep)}\calv_{n-1}(\ep)&=\mr{\calv}_{n-1}^{i,p}(\ep)+(\theta^i G P\mr{\bm{\calv}}_{n-1}(\ep))_p-\rho^{i,p}\bm{w}^\top GP\mr{\bm{\calv}}_{n-1}(\ep) \\
&-\frac{(\bm{v},\mr{\bm{\calv}}_{n-1}(\ep))}{(\bm{v},\mr{\bm{\calt}}_n)}\Bigl(\mr{\calt}_n^{i,p}+
(\theta^i GP\mr{\bm{\calt}}_n)_p-\rho^{i,p}\bm{w}^\top GP\mr{\bm{\calt}}_n\Bigr) +\calo(\ep). \nn
\end{split}
\ee
Finally, using the induction hypothesis on the convergence of $\mr{\calv}_{n-1}^{i,p}(\ep)$ and $\mr{\bm{\calv}}_{n-1}(\ep)$, we obtain
\be
\begin{split}
\lim_{\ep\rightarrow 0}\Bigl(\calv_{n-1}^{i,p}(\ep)-\frac{\calt_n^{i,p}(\ep)}{\calt_n(\ep)}\calv_{n-1}(\ep)\Bigr)
=\calu_{n-1}^{i,p}+(\theta^i G\mr{\bm{\calu}}_{n-1})_p-\rho^{i,p}\bm{w}^\top G\mr{\bm{\calu}}_{n-1}. \nn
\end{split}
\ee
Here, we have recalled the definition:
\be
\begin{split}
\calu_{n-1}^{i,p}=\mr{\calv}_{n-1}^{i,p}-\frac{(\bm{v},\mr{\bm{\calv}}_{n-1})}{(\bm{v},\mr{\bm{\calt}}_n)}\mr{\calt}_n^{i,p} 
\quad \calu_{n-1}^p=\mr{\calv}_{n-1}^{p}-\frac{(\bm{v},\mr{\bm{\calv}}_{n-1})}{(\bm{v},\mr{\bm{\calt}}_n)}\mr{\calt}_n^{p},  \nn
\end{split}
\ee
and used the fact that $P\mr{\bm{\calu}}_{n-1}=\mr{\bm{\calu}}_{n-1}$, which holds because $\mr{\bm{\calu}}_{n-1}\in \mathrm{Im}(I-\Theta)$.

Consequently,  $(\ref{eq-phi-ep})$ can be expressed as
\be
\wh{\phi}^{i,p}_{n-1}(\ep)=\rho^{i,p}L_{n-1}+\frac{1}{u-d}\Bigl(\calu_{n-1}^{i,p}+(\theta^i G\mr{\bm{\calu}}_{n-1})_p
-\rho^{i,p}\bm{w}^\top G\mr{\bm{\calu}}_{n-1}\Bigr)+\calo(\ep).  \nn
\ee
This establishes the desired convergence of $\wh{\phi}^{i,p}_{n-1}$.
The convergence of the transition probabilities is straightforward to verify by noting that
\be
\frac{\calv_{n-1}(\ep)}{\calt_n(\ep)}=\frac{(\bm{v},\mr{\bm{\calv}}_{n-1}(\ep))}{(\bm{v},\mr{\bm{\calt}}_n)}+\calo(\ep) \nn
\ee
and the boundedness of the external order flow $L_{n-1}$.

From $(\ref{multi-rp-Vnm1})$ and the dominated convergence theorem, the convergence established above 
as well as the induction hypothesis $V_n^p(\ep)\to V_n^p$ imply that
\be
\lim_{\ep\to 0}V_{n-1}^p(\bs,y,z^{i,p},\vr_i^p)(\ep)\rightarrow  V_{n-1}^p(\bs,y,z^{i,p},\vr_i^p) 
\ee
for every $(\bs,y,z^{i,p},\vr_i^p)\in \cals^{n-1}\times \caly_{n-1}\times\calz_{n-1}^p\times \Gamma^p$.
Since $(V_{n-1}^p)_{p=1}^m$ satisfy the uniform bounds $0<c_{n-1}\leq V_{n-1}^p\leq C_{n-1}<\infty$,
the above convergence of $V_{n-1}^p(\ep)$ also implies the convergence of $f_{n-2}^p(\ep)$ and hence $\calv_{n-2}^{i,p}(\ep)$.
Therefore,  we recover the induction hypothesis for the previous time $t_{n-2}$,  which establishes the claim.
\end{proof}

\begin{remark}
The continuity established in Theorem~\ref{th-pole-convergence} relies on the assumption that $(I-\Theta(\ep))^{-1}$ 
has a simple pole. Note, however, that the equilibrium existence result in Theorem~\ref{th-mc-mfe-kernel-1} 
does not require this condition. 
The asymptotic behavior for higher-order poles and more general forms of perturbation $\Theta(\ep)$ 
can possibly be analyzed using the methods described in Kato~\cite{Kato}.
\end{remark}

\section{Numerical examples}
\label{sec-numerical}
In this section, we provide numerical examples for the models studied in 
Sections~\ref{sec-single-population} and \ref{sec-multi-population}. 
To reduce computational cost, following Fujii~\cite{Fujii-Trees},  we assume that $Y$ and $Z^{i,p}$ are one-dimensional discrete processes 
taking values on binomial trees, and that $\theta^i_{p,k}$ are 
independent of $i$, i.e., they are constants common 
across all the agents
in each population $p$. Moreover, we  assume that $\calf^{i,p}_0$-measurable random variables $\gamma_i^p$ are uniformly 
distributed over finite sets. Since the cases without relative performance concerns (i.e., $\theta^i\equiv 0$) have been studied in 
\cite{Fujii-Trees} in detail, we focus here on providing illustrative examples 
that highlight  the impact of $\theta$ on the equilibrium price distribution and the optimal trading position size.

\subsection{Single population}

We first consider the model discussed in Section~\ref{sec-single-population}.
$\gamma_i$ is assumed to be uniformly distributed over the $(N_\gamma+1)$ discrete values given by
\be
\gamma_i(k_\gamma):=\ul{\gamma}+(\ol{\gamma}-\ul{\gamma})k_\gamma/N_\gamma, \quad k_\gamma=0, \ldots, N_\gamma. \nn
\ee
We assume that the coefficient of relative performance concerns $\theta^i$ is common across all the agents, i.e., 
$\theta^i=\theta\in \mbb{R}, ~\forall i\in \mbb{N}$.
The process $(Z_n^i)_{n=0}^N$ is modeled as a one-dimensional binomial process defined by
\be 
Z_{n+1}^i=Z_n^i R_{n+1}^i, \nn
\ee
where $(R_{n}^i)$ is an $(\calf^i_{t_n})$-adapted process taking values either $u_z$ or $d_z$.
Specifically,  $R_{n}^i=u_z$ occurs with probability $p_z$ and $R_{n}^i=d_z$ with $q_z:=1-p_z$.
We take $u_z=(d_z)^{-1}=\exp(\sigma_z \sqrt{\Del})$. We also assume $Z_0^i=z_0 \in (0,\infty)$ is common for all the agents to reduce
computational cost.  We model the process $(Y_n)_{n=0}^N$ similarly but assume it follows an approximate Gaussian process:
\be
Y_{n+1}=Y_n +R_{n+1}^y, \nn
\ee 
where $(R_n^y)$ is an $(\calf^0_{t_n})$-adapted process taking values of either $u_y$ or $d_y$. 
Specifically,  $R_{n}^y$ takes the value $u_y$ with probability $p_y$ and $d_y$ with probability $q_y:=1-p_y$.
We take $u_y=(-d_y)=\sigma_y \sqrt{\Del}$.
Finally, for the stock-price process $(S_n)$, we set $\wt{u}=(\wt{d})^{-1}=\exp(\sigma \sqrt{\Del})$ and $S_0=1.0$.
From Theorems~\ref{th-single-1} and \ref{th-single-MC-MFE}, one can see that there is no need to track the evolution of the path-dependent mean-field term $\mu_n$ for obtaining the equilibrium price distributions $(p_{n-1}(s,y))$ and the optimal strategies $\wh{\phi}_{n-1}^i$. We only need its difference, i.e., $\Del_n$, which is fixed by the market-clearing condition.

\begin{figure}[H]
\vspace{0mm}
    \centering
    \begin{minipage}[t]{0.48\textwidth}
        \centering
        \includegraphics[width=\linewidth]{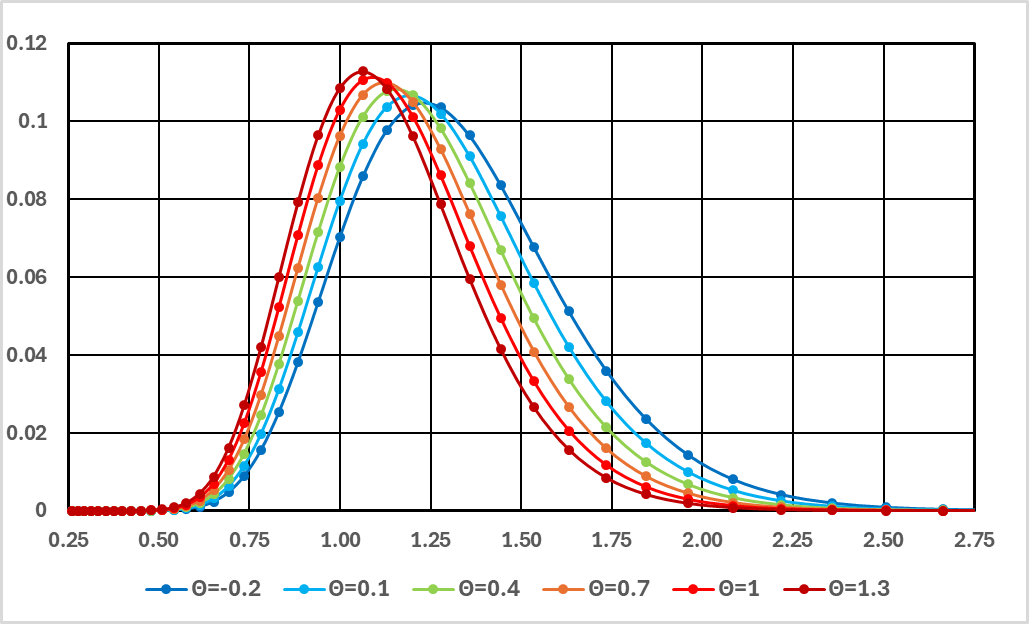}
	\caption{\footnotesize The equilibrium stock price distribution at the 2-year horizon for six different values of $\theta$. }
	 \label{fig-single-1}
    \end{minipage}
    \hspace{0\textwidth} 
    \begin{minipage}[t]{0.48\textwidth}
        \centering
        \includegraphics[width=\linewidth]{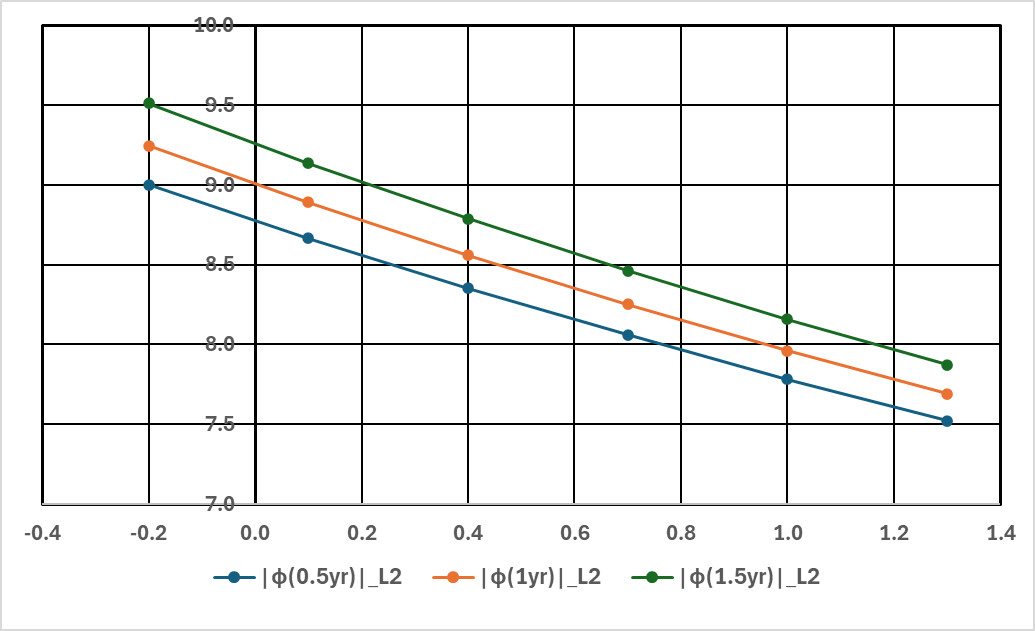}
	\caption{\footnotesize The root mean square of the equilibrium optimal strategy 
	$\ex[|\wh{\phi}^1(t_{n})|^2]^\frac{1}{2}$ as a function of $\theta$ at $t_n \in \{0.5, 1.0, 1.5\}$.}
	\label{fig-single-2}
    \end{minipage}
\end{figure}

As an example for a terminal liability, we adopt the parameterization
\be
F(S_N,Y_N,Z_N^i):=C- f_a S_N Y_N Z_N^i,  \nn
\ee
where $C\in \mbb{R}$ is an arbitrary real constant. Since the result is invariant under a constant shift, one may adjust the constant $C$, if necessary,
to make the liability positive. $f_a\in \mbb{R}$ is a parameter determining the sensitivity of the liability.
We model the external order flow as
\be
\label{numerical-L}
L_{n-1}(S_{n-1},Y_{n-1})=l_a (1+l_b Y_{n-1})S_{n-1}, 
\ee
with some real constants $l_a, l_b\in \mbb{R}$.
Table~\ref{tab-param-1}  summarizes the parameters to be used in Figures~\ref{fig-single-1} and \ref{fig-single-2}.  
We recall that the initial wealth $\xi_i$ is irrelevant to our analysis.
\begin{table}[h]
    \footnotesize
    \centering
    \begin{tabular}{c c c c c c c c c c c c c c c c c c} 
        \toprule
        \text{parameter} &  $\ul{\gamma}$ &  $\ol{\gamma}$ &  $N_\gamma$ & $z_0$ & $\sigma_z$ & $p_z$ &  $Y_0$ & $\sigma_y$ & $p_y$ 
		& $S_0$ & $\sigma$  & $r$ & $T$ & $N$ & $f_a$ & $l_a$ & $l_b$ \\
        \midrule
	  \text{value} & 0.5 & 1.5 & 4  & 1.0 & 12\% & 0.5 & 1.0 & 12\% & 0.5 &1.0 & 15\%&  2.5\% & 2yr &48 & 1.5 & 1 & 1\\
        \bottomrule
    \end{tabular}
\vspace{-2mm}
    \caption{ parameter values }
  \label{tab-param-1}
\end{table}

In Figure~\ref{fig-single-1}, we present the equilibrium distribution of the stock price at the 2-year horizon 
for six different values of $\theta \in \{-0.2, 0.1, 0.4, 0.7, 1.0, 1.3\}$. One can observe that the price distribution 
changes smoothly. More specifically, it monotonically shifts leftward as $\theta$ increases, which implies that a lower excess 
return is demanded when there exists a strong relative performance concern. 
In Figure~\ref{fig-single-2}, we plot $\ex[|\wh{\phi}^1(t_{n})|^2]^\frac{1}{2}$ for the corresponding values of 
$\theta$ at three different times $t_n \in \{0.5, 1.0, 1.5\}$ years. As expected, there is no irregularity around $\theta=1$.

\begin{figure}[H]
  \centering
  \includegraphics[width=0.48\textwidth]{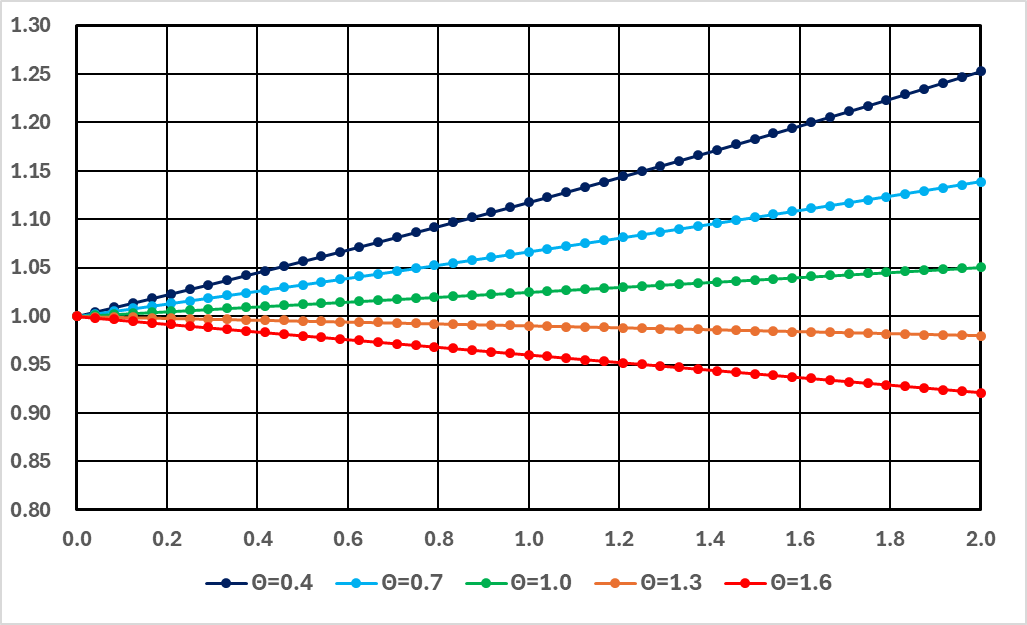}
  \caption{\footnotesize The evolution of $\ex[S(t)]$ for $\theta\in \{0.4, 0.7, 1.0, 1.3, 1.6\}$ with parameters in Table~\ref{tab-param-2}.}
  \label{fig-single-3}
\end{figure}

In the next example, we set $f_a=0$ to eliminate the terminal liability and use a larger value for $l_a$ to emphasize 
the effect of $\theta$ on the equilibrium price distribution.
The parameters are summarized in Table~\ref{tab-param-2}.
\begin{table}[h]
    \footnotesize
    \centering
    \begin{tabular}{c c c c c c c c c c c c c c c c c c} 
        \toprule
        \text{parameter} &  $\ul{\gamma}$ &  $\ol{\gamma}$ &  $N_\gamma$ & $z_0$ & $\sigma_z$ & $p_z$ &  $Y_0$ & $\sigma_y$ & $p_y$ 
		& $S_0$ & $\sigma$  & $r$ & $T$ & $N$ & $f_a$ & $l_a$ & $l_b$ \\
        \midrule
	  \text{value} & 0.5 & 1.5 & 4  & 1.0 & 12\% & 0.5 & 1.0 & 12\% & 0.5 &1.0 & 15\%&  2.5\% & 2yr &48 & 0 & 3 & 1\\
        \bottomrule
    \end{tabular}
\vspace{-2mm}
    \caption{ parameter values }
  \label{tab-param-2}
\end{table}
In Figure~\ref{fig-single-3}, we plot the time evolution of the expected stock price $\ex[S(t_n)]$ in equilibrium
for five different values of $\theta\in \{0.4, 0.7, 1.0, 1.3, 1.6\}$, based on the parameters in Table~\ref{tab-param-2}.
The risk-free interest rate is $r=2.5\%$, which coincides  with the expected growth rate of the stock  when $\theta=1.0$.
This is consistent with the discussion in Remark~\ref{remark-single-theta}. 
We can also observe that cases with $\theta>1$ yield a negative excess return.
It is interesting to note that, when $\theta=1.0$, the equilibrium price distribution coincides exactly with 
the distribution under the risk-neutral measure. Using $(\ref{single-MC-position})$ and $(\ref{single-Vnm1})$,
a simple induction argument shows that $f_{n-1}\equiv 1$ and $\wh{\phi}^i_{n-1}(s,y)=L_{n-1}(s,y)$ for all $1\leq n\leq N$.
Since the effective risk tolerance is now $+\infty$, there is no required risk premium regardless
of the size of $L_{n-1}$.

\subsection{Multiple populations}
We next provide numerical examples for the multi-population model studied in Section~\ref{sec-multi-population}.
For simplicity, we focus on the two-population model $(m=2)$ with relative population weights $w_p, p\in\{1,2\}$.
As in the previous case, $\gamma_i^p$ is assumed to be uniformly  distributed over the $(N_\gamma^p+1)$ discrete values:
\be
\gamma_i^p(k_\gamma)=\ul{\gamma}^p+(\ol{\gamma}^p-\ul{\gamma}^p) k_\gamma/N_\gamma^p, 
\quad p\in\{1,2\}, \quad k_\gamma=0,\ldots, N_\gamma^p. \nn
\ee
As mentioned at the beginning of this section, we assume that $\{(\theta^i_{p,k}), p,k\in \{1,2\}\}$ are  independent of the agent-$i$, i.e., 
$\theta^i_{p,k}=\Theta_{p,k}$ for every $p,k\in\{1, 2\}$. 
The processes $(Z_n^{i,p})_{n=0}^N$ for $p\in\{1,2\}$ are modeled in the same way as before:
\be 
Z_{n+1}^{i,p}=Z_n^{i,p} R_{n+1}^{i,p}, \nn
\ee
where $(R_{n}^{i,p})$ is an $(\calf^{i,p}_{t_n})$-adapted process taking  values in  $\{u_z^p, d_z^p\}$.
Specifically, $R_{n}^{i,p}$ takes the value $u_z^p$ with probability $p_z^p$ and $d_z^p$ with probability 
$q_z^p:=1-p_z^p$.
We take $u_z^p=(d_z^p)^{-1}=\exp(\sigma_z^p \sqrt{\Del})$. We also assume that $Z_0^{i,p}=z_0^p \in (0,\infty)$ is common to all the agents
in each population $p$. The dynamics of the stock price process $(S_n)$ as well as the common noise process $(Y_n)$
are exactly the same as in the previous example. 

For each population $p$,  the terminal liability is assumed to be 
\be
\label{numerical-F-multi}
F^p(S_N, Y_N, Z_N^{i,p}):=C^p-f_a^p S_N Y_N Z_N^{i,p}, \quad p\in\{1,2\}, 
\ee
where $C^p$ is an arbitrary real constant and $f_a^p$ is a parameter determining the level of sensitivity. 
The external order flow $(L_{n-1})$ is modeled by $(\ref{numerical-L})$ as before.
As mentioned in Remark~\ref{remark-path-dependence}, since both  the terminal liability 
and the external order flow are independent of the past history of the stock price, 
the equilibrium transition probabilities and the associated optimal controls also become path-independent.
In this case, the solutions given in Theorems~\ref{th-multi-rp}, \ref{th-multi-mc-mfe}, and \ref{th-mc-mfe-kernel-1} 
hold by replacing $\bs\in \cals^{n-1}$ with $s\in \cals_{n-1}$, the last element of $\bs$.

In the first example, we consider the matrix of relative performance concerns with the following parametrization:
\be
\label{example-theta-1}
\Theta(a):=\begin{pmatrix} a  & 0.4 \\ 0.4 & a \end{pmatrix}, 
\ee
with  $a$ chosen around the value $0.6$,  which makes $(I-\Theta)$ singular.
When $a=0.6$, we have
\be
(I-\Theta(0.6))^\dagger=\frac{5}{8}\begin{pmatrix} 1 & -1 \\ -1 & 1\end{pmatrix}, \bm{v}=\bm{\kappa}=\frac{1}{\sqrt{2}}(1,1)^\top. \nn
\ee
Recall that $(I-\Theta(0.6))^\dagger$ denotes the pseudo inverse defined on the $\mathrm{span}\{(1,-1)^\top\}$
and $\bm{v}$ and $\bm{\kappa}$ are unit vectors satisfying $\bm{v}\in \mathrm{Ker}(I-\Theta(0.6))^\top$ and $\bm{\kappa}\in \mathrm{Ker}(I-\Theta(0.6))$. 
It is easy to see that $a=0.6$ is a first-order pole. In fact, setting $a=0.6-\ep$, we have
\be
\label{theta-ep-1}
(I-\Theta(a))^{-1}=\frac{1}{\ep(0.8+\ep)}\begin{pmatrix} 0.4+\ep & 0.4 \\ 0.4 & 0.4+\ep \end{pmatrix}. 
\ee
The remaining parameters for this example are summarized in Table~\ref{tab-param-3}.
\begin{table}[h]
    \footnotesize
    \centering
    \begin{tabular}{c *{14}{c}}
        \toprule
        parameter &$w_1$ &$w_2$ &$\ul{\gamma}^1$ & $\ol{\gamma}^1$ & $\ul{\gamma}^2$ & 
	$\ol{\gamma}^2$ & $N_\gamma^1$ & $N_\gamma^2$ & $z_0^1$ & $z_0^2$ & $\sigma_z^1$ & $\sigma_z^2$ & $p_z^1$ & $p_z^2$ \\
        \midrule
        value & 0.3 & 0.7& 0.5 & 1.5 & 0.2 & 1.2 & 4 & 4 & 1.0 & 1.0 & 12\% & 12\% & 0.5 & 0.5 \\
        \bottomrule
    \end{tabular}
    \vspace{3mm} 
    \begin{tabular}{c *{12}{c}}
        \toprule
        parameter & $Y_0$ & $\sigma_y$ & $p_y$ & $S_0$ & $\sigma$ & $r$ & $T$ & $N$ & $f_a^1$ & $f_a^2$ & $l_a$ & $l_b$ \\
        \midrule
        value & 1.0 & 12\% & 0.5 & 1.0 & 15\% & 2.5\% & 2yr & 48 & 1.2 & 2.4 & 1.5 & 1.5 \\
        \bottomrule
    \end{tabular}
\vspace{-3mm}
    \caption{parameter values}
    \label{tab-param-3}
\end{table}

\begin{figure}[H]
\vspace{2mm}
    \centering
    \begin{minipage}[t]{0.48\textwidth}
        \centering
        \includegraphics[width=\linewidth]{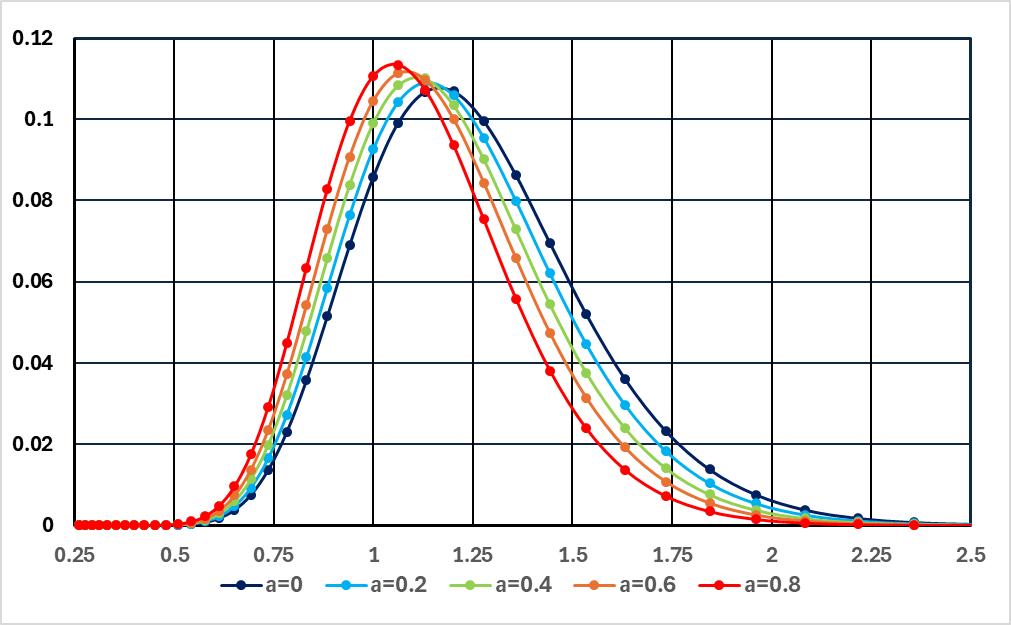}
	\caption{\footnotesize 
The equilibrium stock price distributions at the 2-year horizon for five different values of $a$.}
	 \label{fig-multi-1}
    \end{minipage}
    \hspace{-0.005\textwidth} 
    \begin{minipage}[t]{0.48\textwidth}
        \centering
        \includegraphics[width=\linewidth]{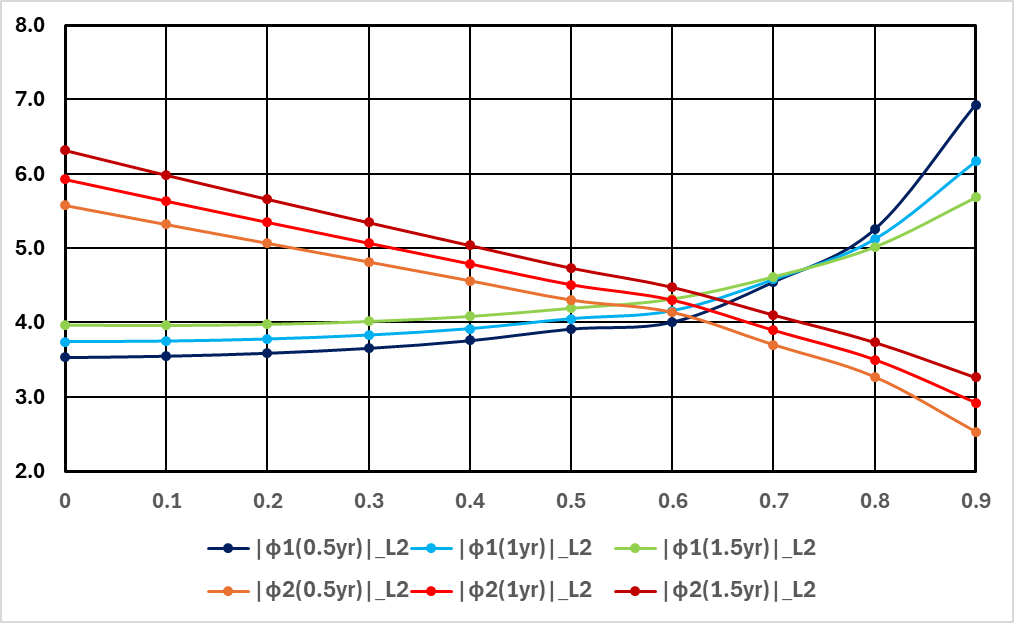}
	\caption{\footnotesize The root mean square of the equilibrium optimal strategy $\ex[|\wh{\phi}^p(t_n)|^2]^\frac{1}{2}$ for $p\in\{1,2\}$
	as a function of $a$ at $t_n\in \{0.5, 1.0, 1.5\}$.}
	\label{fig-multi-2}
    \end{minipage}
\end{figure}
In Figure~\ref{fig-multi-1}, we plot the equilibrium stock price distributions at the 2-year horizon
for $\Theta$ in $(\ref{example-theta-1})$ and the parameters in Table~\ref{tab-param-3}
with five different values of $a\in \{0,0.2, 0.4, 0.6, 0.8\}$. The distribution at the singular point $(a=0.6)$
is calculated using the results in Theorem~\ref{th-mc-mfe-kernel-1}, while the others are based on Theorem~\ref{th-multi-mc-mfe}.
In Figure~\ref{fig-multi-2}, we present the root mean square of the equilibrium optimal strategy $\ex[|\wh{\phi}^p(t_n)|^2]^\frac{1}{2}$
for each population $p\in\{1,2\}$  as a function of $a$ at $t_n\in \{0.5, 1.0, 1.5\}$ years.
We evaluate ten cases with $\{a=(0.1)k: k=0,1,\ldots, 9\}$. 
Consistent with the results in Section~\ref{sec-perturbation}, 
both the equilibrium price distributions and the optimal position sizes appear to change continuously with the parameter $a$.
Figure~\ref{fig-multi-2} shows that 
$\ex[|\wh{\phi}^2(t_n)|^2]^\frac{1}{2}>\ex[|\wh{\phi}^1(t_n)|^2]^\frac{1}{2}$ for small values of $a$, but this relationship reverses
for $a>0.6$.  
Since the second population has a larger risk tolerance ($\gamma^2 \leq \gamma^1$ on average)
and higher hedging needs ($f_a^1 \leq f_a^2$), it is natural that they hold larger positions 
than the first population when the effects of relative performance concerns are sufficiently small. 
The significant change around $a=0.6$ is attributable to the sign flip in $(I-\Theta(\epsilon))^{-1}$ 
from $(\ref{theta-ep-1})$ and the resulting impact on both the effective risk tolerance 
and the sensitivity of the effective liability defined in $(\ref{def-effective-variables})$.

\bigskip
In the second example, we consider the matrix of relative performance concerns 
of the following form:
\be
\Theta(a):=\begin{pmatrix} a & 0.1 \\ 0 & a\end{pmatrix} \nn
\ee
with $a\in \mbb{R}$ as a parameter. This asymmetric matrix has a second-order pole at $a=1$
and hence this model is not covered by the analysis in Section~\ref{sec-perturbation}.
By setting $a=1-\ep$, we obtain
\be
(I-\Theta(a))^{-1}=\begin{pmatrix}
1/\ep & 0.1/\ep^2 \\ 0 & 1/\ep
\end{pmatrix}. \nn
\ee
When $a=1$, it is not difficult to see that
\be
(I-\Theta(1))^\dagger=\begin{pmatrix} 0  & 0 \\ -10 & 0 \end{pmatrix}, 
\quad \bm{v}=(0,1)^\top, \quad \bm{\kappa}=(1, 0)^\top, \nn
\ee
where $(I-\Theta(1))^\dagger$ is the pseudo inverse defined on $\mathrm{span}\{(1, 0)^\top\}$
and $\bm{v}$ and $\bm{\kappa}$ are unit vectors satisfying $\bm{v}\in \mathrm{Ker}(I-\Theta(1))^\top$ and $\bm{\kappa}\in \mathrm{Ker}(I-\Theta(1))$. 
The remaining parameters for this example are summarized in Table~\ref{tab-param-4}.
\begin{table}[h]
    \footnotesize
    \centering
    \begin{tabular}{c *{14}{c}}
        \toprule
        parameter &$w_1$ &$w_2$ &$\ul{\gamma}^1$ & $\ol{\gamma}^1$ & $\ul{\gamma}^2$ & 
	$\ol{\gamma}^2$ & $N_\gamma^1$ & $N_\gamma^2$ & $z_0^1$ & $z_0^2$ & $\sigma_z^1$ & $\sigma_z^2$ & $p_z^1$ & $p_z^2$ \\
        \midrule
        value & 0.5 & 0.5& 0.5 & 1.5 & 0.5 & 1.5 & 4 & 4 & 1.0 & 1.0 & 12\% & 12\% & 0.5 & 0.5 \\
        \bottomrule
    \end{tabular}
    \vspace{3mm} 
    \begin{tabular}{c *{12}{c}}
        \toprule
        parameter & $Y_0$ & $\sigma_y$ & $p_y$ & $S_0$ & $\sigma$ & $r$ & $T$ & $N$ & $f_a^1$ & $f_a^2$ & $l_a$ & $l_b$ \\
        \midrule
        value & 1.0 & 12\% & 0.5 & 1.0 & 15\% & 2.5\% & 2yr & 48 & 0 & 0 & 2 & 2 \\
        \bottomrule
    \end{tabular}
\vspace{-3mm}
    \caption{parameter values}
    \label{tab-param-4}
\end{table}

\begin{figure}[H]
\vspace{2mm}
    \centering
    \begin{minipage}[t]{0.48\textwidth}
        \centering
        \includegraphics[width=\linewidth]{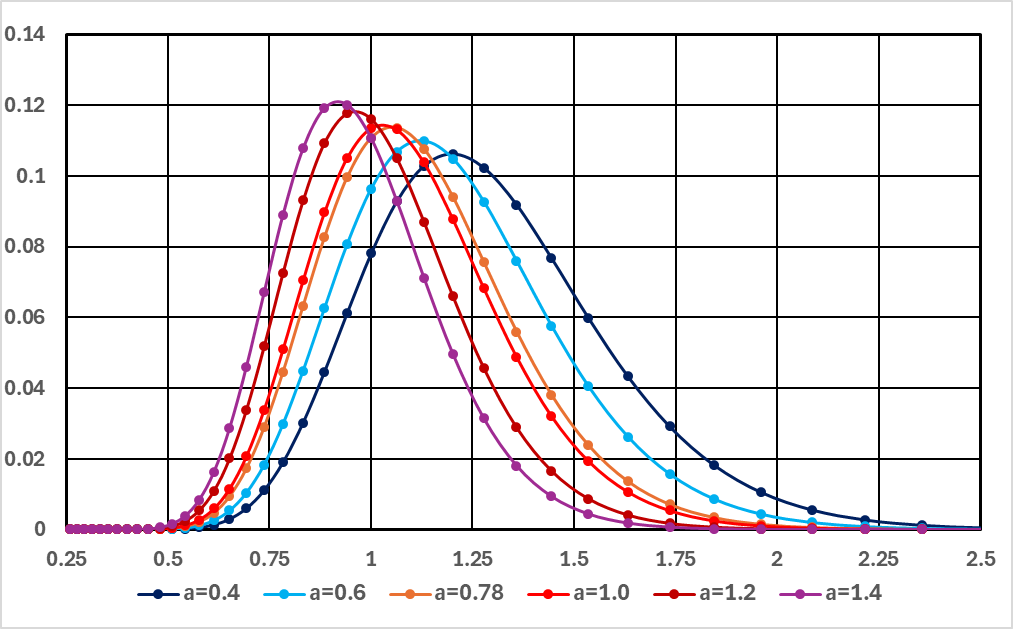}
	\caption{\footnotesize The equilibrium stock price distributions at the 2-year horizon for six different values of $a$.}
	 \label{fig-multi-3}
    \end{minipage}
    \hspace{-0.005\textwidth} 
    \begin{minipage}[t]{0.48\textwidth}
        \centering
        \includegraphics[width=\linewidth]{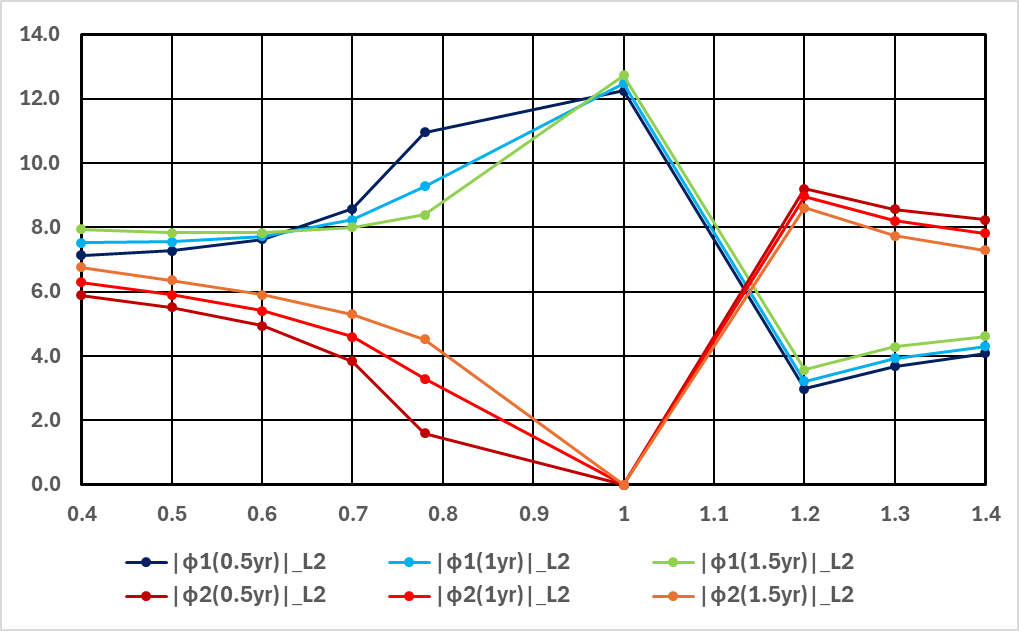}
	\caption{\footnotesize The root mean square of the equilibrium optimal strategy $\ex[|\wh{\phi}^p(t_n)|^2]^\frac{1}{2}$
for $p\in \{1,2\}$ as a function of $a$ at $t_n\in \{0.5, 1.0, 1.5\}$.}
	\label{fig-multi-4}
    \end{minipage}
\end{figure}

In Figure~\ref{fig-multi-3}, we plot the equilibrium stock price distributions at the 2-year horizon
for six different values of $a\in \{0.4, 0.6, 0.78, 1.0, 1.2, 1.4\}$, 
and in Figure~\ref{fig-multi-4}, we plot the root mean square of the equilibrium 
optimal strategy $\ex[|\wh{\phi}^{p}(t_n)|^2]^\frac{1}{2}$ for each population $p\in \{1,2\}$
at $t_n\in \{0.5, 1.0, 1.5\}$ years. For this figure, we evaluate nine different values of $a\in \{0.4, 0.5, 0.6, 0.7, 0.78, 1.0, 1.2, 1.3, 1.4\}$.
At the singular point $(a=1.0)$, the results in Theorem~\ref{th-mc-mfe-kernel-1} are used, 
while the others are based on Theorem~\ref{th-multi-mc-mfe}.
We could not obtain stable numerical results near the singular point $a=1.0$ using Theorem~\ref{th-multi-mc-mfe}
due to  numerical overflow arising from $(V_{n-1}^p)$, which are exponentials of the effective 
liabilities. Although the equilibrium price distributions still appear to shift monotonically leftward as $a$ increases,
the root mean square of the optimal strategies $\ex[|\wh{\phi}^p(t_n)|^2]^\frac{1}{2}$ exhibit distinctive behavior around the singular point.
In particular, the optimal strategy of the second population $\ex[|\wh{\phi}^2(t_n)|^2]^\frac{1}{2}$ decreases 
toward zero at $a=1.0$. Since there is no liability and $(\Theta \kappa)_2=0$, 
we can show, by a simple induction,  that $\wh{\phi}^2\equiv 0$ from $(\ref{mc-optimal-kernel-1})$ when $a=1.0$.
Furthermore, since an induction argument shows $\mr{\calv}^2 \equiv 0$ and we have 
$\bigl(\bm{v},\mr{\bm{\calv}}_{n-1}\bigr)=\mr{\calv}^2_{n-1}$, 
the equilibrium price distribution at $a=1.0$ coincides with that under the risk-neutral measure.

\section{Concluding remarks}
\label{sec-conclusion}
In this paper, we have investigated the mean-field equilibrium price formation in a discrete-time 
market populated by agents with exponential utility and relative performance concerns. By extending the tractable binomial tree framework, 
we were able to impose the market-clearing condition explicitly, thereby determining the asset price dynamics 
endogenously rather than treating it as exogenous. We established the existence and uniqueness of the Market-Clearing Mean-Field 
Equilibrium (MC-MFE) for both single-population and multi-population settings, characterizing how 
the network of relative concerns impacts the risk premium and trading strategies. 
Furthermore, using resolvent expansion techniques, 
we demonstrated that the equilibrium solutions depend continuously on the interaction matrix, even around the points
where $(I-\Theta)$ exhibits a first-order singularity.

\subsection*{Declarations of Interest and AI use}
{\footnotesize
This research did not receive any specific grant from funding agencies in the public, commercial, or not-for-profit sectors.
There is no competing interests to declare.
The author acknowledges the use of the large language model Gemini 
to refine the English clarity and style in the manuscript. After using this tool, the author reviewed and
edited the content as needed and takes full responsibility for the content of the published article.
}

\appendix
\section{Connection to the finite population game} 
\label{sec-A}

\setcounter{equation}{0}
\setcounter{theorem}{0}
\setcounter{proposition}{0}
\setcounter{lemma}{0}
\renewcommand{\theequation}{A.\arabic{equation}}
\renewcommand{\thetheorem}{A.\arabic{theorem}}
\renewcommand{\theproposition}{A.\arabic{proposition}}
\renewcommand{\thelemma}{A.\arabic{lemma}}

We briefly comment on the connection to the finite population game with agent-$i$, $1\leq i \leq N_p$.
Due to the symmetry of the problem, we focus on the optimization problem for the agent-$1$.
Let us define the empirical expectation of the wealth as
\be
\ol{\mu}^{N_p}_N:=\frac{1}{N_p-1}\sum_{j=2}^{N_p}\wh{X}_N^{j}, \nn
\ee
where $\wh{X}^{j}$ denotes the  wealth process of agent-$j$ associated with the mean-field optimal control 
$\wh{\phi}^{j}:=(\wh{\phi}^j_{n-1})_{n=1}^N$ given in Theorem~\ref{th-single-1}.

Suppose that agent-$1$ adopts an arbitrary admissible control $\phi^1:=(\phi^1_{n-1})_{n=1}^{N}$ in $\mbb{A}^1$ and set
\be
J^{N_p,1}(\phi^1, \wh{\phi}^{2},\ldots, \wh{\phi}^{N_p}):=
\ex\Bigl[-\exp\Bigl(-\gamma_1(X_N^1-\theta_1 \ol{\mu}_N^{N_p}-F(S_N,Y_N,Z_N^1)\Bigr)|\calf_0^{0,1}\Bigr], \nn
\ee 
where $X^1$ is the wealth process of agent-$1$ associated with the control $\phi^1$.
We also define the value function in the large population limit:
\be
\begin{split}
\calj^1(\phi^1,(\wh{\phi}^{j})_{j=2}^\infty):=\ex^{0,1}
\Bigl[-\exp\Bigl(-\gamma_1(X_N^1-\theta_1 \wh{\mu}_N(\bS^N,\bY^{N-1})-F(S_N,Y_N,Z_N^1)\Bigr)|\calf_0^{0,1}\Bigr], \nn
\end{split}
\ee
where $\wh{\mu}$ is the solution to the RP-MFE given in Theorem~\ref{th-single-1}.
Note that $\calj^1(\wh{\phi}^{1},(\wh{\phi}^{j})_{j=2}^\infty)$ corresponds to the value function in the RP-MFE 
studied in Theorem~\ref{th-single-1}.

Since $(\wh{X}^{j})_{j\geq 2}$ are $\calf^0$-conditionally i.i.d., we have 
\be
\label{single-LLN}
\begin{split}
&\mbb{E}\Bigl[\Bigl|\ol{\mu}^{N_p}-\wh{\mu}_N(\bS^N,\bY^{N-1})\Bigr|^2\Bigl|\bs,\by\Bigr]
=\frac{1}{N_p-1}\ex^{0,j}\Bigl[\Bigl|\wh{X}_N^{j}-\ex^{0,j}[\wh{X}_N^{j}|\bs,\by^-]\Bigr|^2\Bigr|\bs,\by^-\Bigr] \leq \frac{C}{N_p}, 
\end{split}
\ee
uniformly in  $(\bs,\by)\in \cals^{N}\times \caly^N$ with some positive constant $C$.
Note that the conditional variance is finite due to the boundedness of $\wh{X}^{j}$, which follows from 
the boundedness of $\xi_j$ and $\wh{\phi}^{j}$.

\begin{theorem}
\label{th-approx-Nash}
Let Assumptions~\ref{assumption-single-1} and \ref{assumption-single-2} be in force. We also assume that $\ex^1[\theta_1]\neq 1$.
For any admissible control $\phi^1\in \mbb{A}^1$, there exists $\ep_{N_p}>0$ such that 
\be
J^{N_p,1}(\phi^1,\wh{\phi}^{2},\ldots, \wh{\phi}^{N_p})\leq J^{N_p,1}(\wh{\phi}^{1},\wh{\phi}^{2},\ldots,\wh{\phi}^{N_p})+\ep_{N_p}, \nn
\ee
and $\ep_{N_p}\rightarrow 0$ as $N_p\rightarrow \infty$.
\end{theorem}
\begin{proof}
It is enough to constrain the admissible control space $\mbb{A}^1$ so that
$J^{N_p,1}(\phi^1,\wh{\phi}^{2},\ldots,\wh{\phi}^{N_p})\wedge \calj^1(\phi^1, (\wh{\phi}^{j})_{j=2}^\infty)>-M$
for sufficiently large $M>0$. Then we have
\be
\begin{split}
&|J^{N_p,1}(\phi^1,\wh{\phi}^{2},\ldots,\wh{\phi}^{N_p})-\calj^1(\phi^1, (\wh{\phi}^{j})_{j=2}^\infty)|\\
&\quad \leq M\ex\Bigl[\exp\Bigl(\gamma_1 |\theta_1|\bigl|\wh{\mu}_N(\bS^N,\bY^{N-1})-\ol{\mu}_N^{N_p}\bigr|\Bigr)-1\Bigr]
 \leq CM\ex\bigl[|\wh{\mu}_N(\bS^N,\bY^{N-1})-\ol{\mu}_N^{N_p}|\bigr]. \nn
\end{split}
\ee
where $C$ is some positive constant depending on the bounds of $\gamma_1,\theta_1, \wh{\mu}_N$, and $\ol{\mu}_N^{N_p}$.
Recall that the wealth process $\wh{X}^{j}$ associated with $\wh{\phi}^{j}$ is a bounded process, and so are $\wh{\mu}_N$ and $\ol{\mu}_N^{N_p}$.
In particular, $C$ can be taken independent of $N_p$. It thus follows from the estimate $(\ref{single-LLN})$ and Jensen's
inequality that
\be
\label{epsilon-Nash-tmp}
|J^{N_p,1}(\phi^1,\wh{\phi}^{2},\ldots,\wh{\phi}^{N_p})-\calj^1(\phi^1, (\wh{\phi}^{j})_{j=2}^\infty)|\leq C/\sqrt{N_p}
\ee
with some constant $C$ independent of $N_p$. 

On the other hand, by definition of $\wh{\phi}^{1}$, 
$\calj^1(\phi^1,(\wh{\phi}^{j})_{j=2}^\infty)\leq \calj^1(\wh{\phi}^{1},(\wh{\phi}^{j})_{j=2}^\infty)$
for any admissible control $\phi^1$. Therefore, using the estimate in $(\ref{epsilon-Nash-tmp})$,
\be
\begin{split}
&J^{N_p,1}(\phi^{1},\wh{\phi}^{2},\ldots,\wh{\phi}^{N_p})\leq  \calj^1(\phi^{1},(\wh{\phi}^{j})_{j=2}^\infty)+C/\sqrt{N_p}\\
&\quad \leq \calj^1(\wh{\phi}^{1},(\wh{\phi}^{j})_{j=2}^\infty)+C/\sqrt{N_p}  \leq J^{N_p,1}(\wh{\phi}^{1},\wh{\phi}^{2},\ldots,\wh{\phi}^{N_p})+2C/\sqrt{N_p}, \nn
\end{split}
\ee
which proves the claim with $\epsilon_{N_p}=2C/\sqrt{N_p}$.
\end{proof}
\begin{remark}
Theorem~\ref{th-approx-Nash} implies that the optimal controls $(\wh{\phi}^{i})_{i\geq 1}$ constitute
an (open loop) $\epsilon$-Nash equilibrium for the relative performance game.
\end{remark}



\end{document}

%% file: Fmacro-2015.tex

\newtheorem{definition}{Definition}[section]
\newtheorem{assumption}{Assumption}[section]
\newtheorem{condition}{$[$ C}
\newtheorem{lemma}{Lemma}[section]
\newtheorem{proposition}{Proposition}[section]
\newtheorem{theorem}{Theorem}[section]
\newtheorem{remark}{Remark}[section]
\newtheorem{example}{Example}[section]
\newtheorem{corollary}{Corollary}[section]
%
\def\cala{{\cal A}}
\def\calb{{\cal B}}
\def\calc{{\cal C}}
\def\cald{{\cal D}}
\def\cale{{\cal E}}
\def\calf{{\cal F}}
\def\calg{{\cal G}}
\def\calh{{\cal H}}
\def\cali{{\cal I}}
\def\calj{{\cal J}}
\def\calk{{\cal K}}
\def\call{{\cal L}}
\def\calm{{\cal M}}
\def\caln{{\cal N}}
\def\calo{{\cal O}}
\def\calp{{\cal P}}
\def\calq{{\cal Q}}
\def\calr{{\cal R}}
\def\cals{{\cal S}}
\def\calt{{\cal T}}
\def\calu{{\cal U}}
\def\calv{{\cal V}}
\def\calw{{\cal W}}
\def\calx{{\cal X}}
\def\caly{{\cal Y}}
\def\calz{{\cal Z}}
%
\def\sskip{\hspace{0.5cm}}
\def\simleq{ \raisebox{-.7ex}{\em $\stackrel{{\textstyle <}}{\sim}$} }
\def\leqsim{ \raisebox{-.7ex}{\em $\stackrel{{\textstyle <}}{\sim}$} }
\def\nn{\nonumber}
\def\be{\begin{equation}}
\def\ee{\end{equation}}
\def\bea{\begin{eqnarray}}
\def\eea{\end{eqnarray}}
%